\documentclass{aastex63}
\usepackage{amsmath}

\newcommand{\Eexc}{$E_{\rm exc}$}
\newcommand{\Teff}{$T_{\rm eff}$}  
\newcommand{\kms}{km\,s$^{-1}$}
\def\ione{\,{\sc i}}
\def\ii{\,{\sc ii}}
\def\iii{\,{\sc iii}}

\def\v{\,{\sc v}}

\received{}
\revised{}
\accepted{}
\submitjournal{ApJ}

\shorttitle{NLTE analysis for Y\ione\ and Y\ii\ in atmospheres of F-G-K stars }
\shortauthors{Alexeeva et al.}

\begin{document}


\title{NLTE analysis for Y\ione\ and Y\ii\ in atmospheres
of F-G-K stars}

\email{alexeeva@nao.cas.cn}
\author[0000-0002-8709-4665]{Sofya Alexeeva}
\affiliation{CAS Key Laboratory of Optical Astronomy, National Astronomical Observatories, Beijing, 100101, China \\}

\author[0000-0002-2448-3049]{Yu Wang}
\affiliation{School of Physics, Beijing Institute of Technology, Beijing, 100081, China \\}


\email{gzhao@nao.cas.cn}
\author[0000-0002-8980-945X]{Gang Zhao}
\affiliation{CAS Key Laboratory of Optical Astronomy, National Astronomical Observatories, Beijing, 100101, China \\}
\affiliation{School of Astronomy and Space Science, University of Chinese Academy of Sciences, Beijing, 100049, China\\}

\author[0000-0002-8514-4497]{Feng Wang}
\affiliation{School of Physics, Beijing Institute of Technology, Beijing, 100081, China \\}

\author[0000-0003-1874-9653]{Yong Wu}
\affiliation{Institute of Applied Physics and Computational Mathematics, Beijing 100088, China \\}
\affiliation{HEDPS, Center for Applied Physics and Technology, Peking University, Beijing 100084, China\\}

\author{Jianguo Wang}
\affiliation{Institute of Applied Physics and Computational Mathematics, Beijing 100088, China \\}

\author[0000-0002-8609-3599]{Hongliang Yan}
\affiliation{CAS Key Laboratory of Optical Astronomy, National Astronomical Observatories, Beijing, 100101, China \\}
\affiliation{School of Astronomy and Space Science, University of Chinese Academy of Sciences, Beijing, 100049, China\\}
\affiliation{Institute for Frontiers in Astronomy and Astrophysics, Beijing Normal University, Beijing, 102206, China\\}

\author[0000-0002-0349-7839]{Jianrong Shi}
\affiliation{CAS Key Laboratory of Optical Astronomy, National Astronomical Observatories, Beijing, 100101, China \\}
\affiliation{School of Astronomy and Space Science, University of Chinese Academy of Sciences, Beijing, 100049, China\\}

\begin{abstract}

 The non-local thermodynamical equilibrium (NLTE) line formation of Y\ione\ and Y\ii\ is considered in 1D LTE model atmospheres of F-G-K-type stars. The model atom was constructed with the most up-to-date atomic data, including quantum cross sections and rate coefficients for transitions in inelastic collisions of Y\ione\ and Y\ii\ with hydrogen atoms. For seven reference stars, we obtained an agreement between NLTE abundances inferred from the two ionization stages, while the difference in LTE abundance (Y\ione\ -- Y\ii) can reach up to -0.31 dex. In the atmospheres of F-G-K-type stars, for both Y\ione\ and Y\ii\ lines, the NLTE abundance corrections are positive. In solar metallicity stars, the NLTE abundance corrections for Y\ii\ lines do not exceed 0.12 dex, while in atmospheres of metal-poor stars they do not exceed 0.21 dex. For Y\ione\ lines, the NLTE abundance corrections can reach up to $\sim$0.5 dex. We determined the yttrium NLTE abundances for a sample of 65 F and G dwarfs and subgiants in the -2.62~$\leq$~[Fe/H]~$\leq$~+0.24 metallicity range, using high-resolution spectra. For stars with [Fe/H]~$\leq$~-1.5, [Y/Fe] versus [Fe/H] diagram reveals positive trend with an average value of [Y/Fe]~$\simeq$~0. For metal-poor stars, among Sr, Y, and Zr, the arrangement [Sr/Fe]~$<$~[Y/Fe]~$<$~[Zr/Fe] remains consistent. The current study is useful for the Galactic chemical evolution research. The model atom will be applied for NLTE yttrium abundance determination in very metal-poor stars studied with LAMOST and Subaru.

  
\end{abstract}


\section{Introduction} \label{sec:intro}

Yttrium (Y, Z = 39, A = 89) is one of the easily-observable light neutron-capture elements in B to K-type stars.
Yttrium can be produced by stars in three known types of neutron-capture reactions:
rapid (\textit{r}-) process, strong (or main) component of the slow (\textit{s}-) process, and the weak component of the \textit{s}-process \citep{1989RPPh...52..945K}. Synthesis of yttrium in main component of the s-process takes place in low- and intermediate-mass stars (1.5--4M$\odot$) during the double-shell burning stage as a result of thermal pulsations in the envelopes of asymptotic-branch giants (AGB stars) \citep{1998ApJ...497..388G, 2011RvMP...83..157K}. About 70-90\% of the solar yttrium was produced in main component. The weak component of the \textit{s}-process (forming elements up to A$\sim$90) takes place in the cores of massive (M $\geq$ 8M$\odot$) stars during convective helium core and carbon shell burning stages \citep{ 2010ApJ...710.1557P}. 
The astrophysical site for the \textit{r}-process remains a long-standing problem of nucleosynthesis \citep{2021RvMP...93a5002C}.
During the last several decades the \textit{r}-process was associated with type-II supernovae explosions \citep{1994ApJ...433..229W}. 
 However, the recent observations showed different scenarios, such as binary neutron star mergers \citep[see e.g.][]{2019Natur.574..497W} and magneto-rotationally driven supernovae \citep[e.g.][]{2016ApJ...816...79S} as the most probable astrophysical \textit{r}-process sources. One of them, neutron star mergers, was recently observed and related to the Gravitational Wave event GW170817, where the yttrium lines were successfully identified in optical spectra  \citep{2023ApJ...944..123V}. Since the Galactic chemical evolution (GCE) models underproduce the s-process component of the solar-system abundances of light neutron-capture elements (Sr-Y-Zr) by about 20--30\%, \citet{2004ApJ...601..864T} hypothesized another source of neutron-capture nucleosynthesis in the Galaxy defined as Light Element Primary Process (LEPP). The need of this additional LEPP of yet unknown origin for the light isotopes is still highly debated \citep{2007ApJ...671.1685M, 2010ApJ...710.1557P, 2014ApJ...787...10B, 2015ApJ...801...53C, 2020ApJ...900..179K}. The study of the Galactic trend of yttrium in stars uniformly distributed over the wide range of metallicity, from solar to  low-metallicity ([Fe/H] $<$ $-$2.0), may indicate the changes of relative contributions from the processes to the creation of the element as the Galaxy evolved. The presence of yttrium in stars of different ages and locations gives a good opportunity to restore the history of heavy elements enrichment of the interstellar medium and, thus, to provide constraints on the nucleosynthesis theories.

Many studies have been performed to establish [Y/Fe]--[Fe/H] Galactic trend, see e.g. \citet{1991A&A...244..425Z, 2002ApJ...579..616J, 2004ApJ...607..474H, 2007A&A...476..935F, 2011A&A...530A..15N, 2012A&A...545A..31H, 2014AJ....147..136R, 2014A&A...562A..71B, 2017A&A...608A..46R, 2021A&A...653A..67B,  2022ApJ...931..147L}. The most of the studies report on a moderate underabundance in [Y/Fe] in stars with [Fe/H]$<$ $-1$ and a large dispersion in the [Y/Fe] ratio with decreasing [Fe/H].  

All studies determined the yttrium abundance in stars are based on the assumption of the local thermodynamic equilibrium (LTE). The deviations from LTE for yttrium lines were not studied so far. However, departures from LTE are important, especially, for low surface gravities and metal poor atmospheres. The use of NLTE element abundances is a valuable approach that strengthens the reliability and accuracy of data interpretation within the context of the chemical evolution of the Galaxy.
NLTE calculations consider departures from the assumption of LTE and take into account the interactions between photons, electrons, and atoms in stellar atmospheres. Using NLTE element abundances in data interpretation improves credibility by offering a more realistic and comprehensive view of the Galaxy's chemical processes.

Accurate yttrium abundances in stars of different spectral types and populations are also important for various astrophysical purposes. For thin disc stars of different metallicities and solar-type stars accurate yttrium abundances combined with the other spectroscopic indicator, as magnesium, provide the method for the estimation of stellar ages, that was a crucial challenge since the advent of stellar astrophysics. While [Mg/Fe] increases with age, yttrium shows the opposite behavior meaning that [Y/Mg] can be used as a sensitive chronometer for Galactic evolution \citep{2012A&A...542A..84D, 2015A&A...579A..52N, 2016A&A...590A..32T, 2019A&A...622A..59T, 2022ApJ...936..100B}. Another combinaion of yttrium with europium, as [Y/Eu] abundance ratio, is a good indicator of the chemical evolution efficiency, because it characterizes the relative contribution of low- to intermediate-mass stars with respect to high-mass stars \citep{2021A&A...648A.108R}. First peak s-element yttrium was mainly produced by low- and intermediate mass AGB stars \citep{2018MNRAS.476.3432P}, while europium is produced by massive stars through rapid neutron captures \citep{2014ApJ...787...10B}.  In RR Lyrae variable stars, yttrium shows anomalous pattern, which cannot be explained in the framework of GCE. For example, \citet{1995AJ....110.2319C, 2013RAA....13.1307L, 2020AstBu..75..311G} detected anomalously low (beyond the errors) yttrium abundances in the atmospheres of the metal-rich RR Lyrae ([Fe/H] $>$ $-$1.0) stars.
\citet{1995AJ....110.2319C} attributed the anomalous abundances of Y to the departures from LTE. However, it is not clear why similar effect is not found in normal metal-poor stars.

There is a significant lack of information regarding NLTE effects for yttrium in the literature, highlighting the pressing need for precise spectroscopic analyses utilizing high-resolution observations. This study aims to construct a comprehensive model atom for Y\ione\ --Y\ii\ based on the most up-to-date atomic data available so far
electronic collisions, collisions with hydrogen atoms, and photoionization cross-section. The model atom is applicable to analysis of the yttrium lines in F, G, K spectral type stars. As a first application of the treated model atom, we obtain the yttrium abundances of the reference stars, with well-determined atmospheric parameters, using an extensive list of Y\ione\ and Y\ii\ lines. The simultaneous presence of lines of the two ionization stages, Y\ione\ and Y\ii, in FGK-type stars provides an opportunity to test our model atom. 

In order to draw more dependable conclusions regarding the origin and evolution of yttrium within the Galactic context, we determine NLTE yttrium abundances in a sample of 65 well-studied stars that includes F and G dwarfs and subgiants in a limited range of temperatures, gravities, and metallicities. For 51 star,
\citet{2015ApJ...808..148S} have diligently established precise stellar parameters. Their NLTE abundances for 17 chemical elements from Li to Eu were presented in 
\citet{2016ApJ...833..225Z}.

The article is organized as follows. Section\,\ref{Sect:atom} describes the model atom of Y\ione --Y\ii, adopted atomic data,  and the departures from LTE for Y\ione --Y\ii\ in model atmospheres with different stellar parameters. Analysis of the solar Y\ione --Y\ii\ lines is presented in Section\,\ref{subsec:sun}. 
 In Section\,\ref{Sect:Stars}, we determine the Y abundances of the reference FGK stars.  Galactic trend for 65 F and G dwarfs and subgiants is presented in Section\,\ref{sec:application}.
We summarize our conclusions in Section\,\ref{Sect:Conclusions}.

 \begin{figure}
\begin{center}
 \includegraphics[scale=0.4]{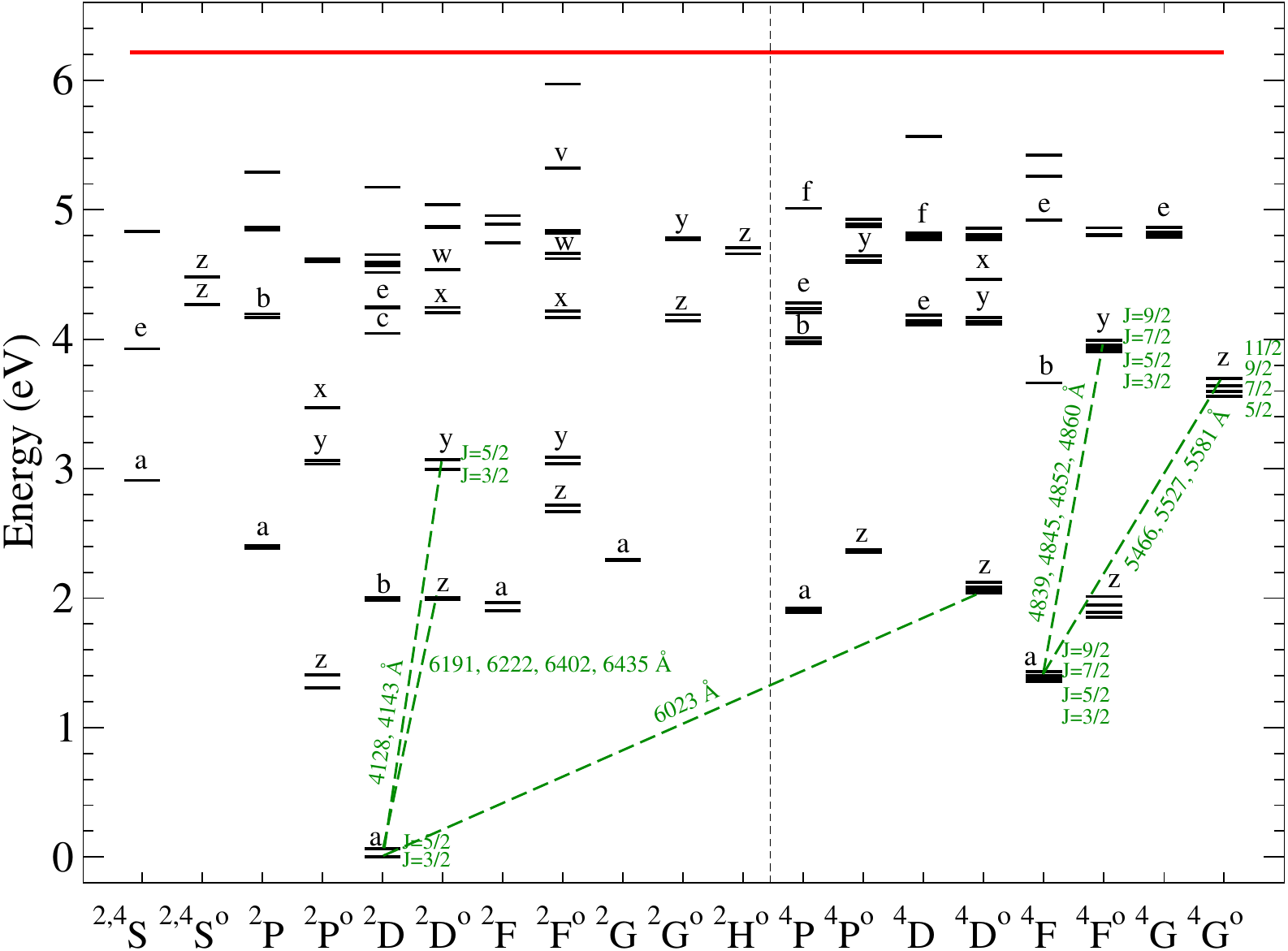}
 \caption{Grotrian term diagram for Y\ione. The ionization energy is shown by solid horizontal line. The dashed lines indicate the transitions, where the investigated spectral lines arise. }
 \label{Groteriane_Y1}
 \end{center}
 \end{figure}

  \begin{figure}
\begin{center}
 \includegraphics[scale=0.4]{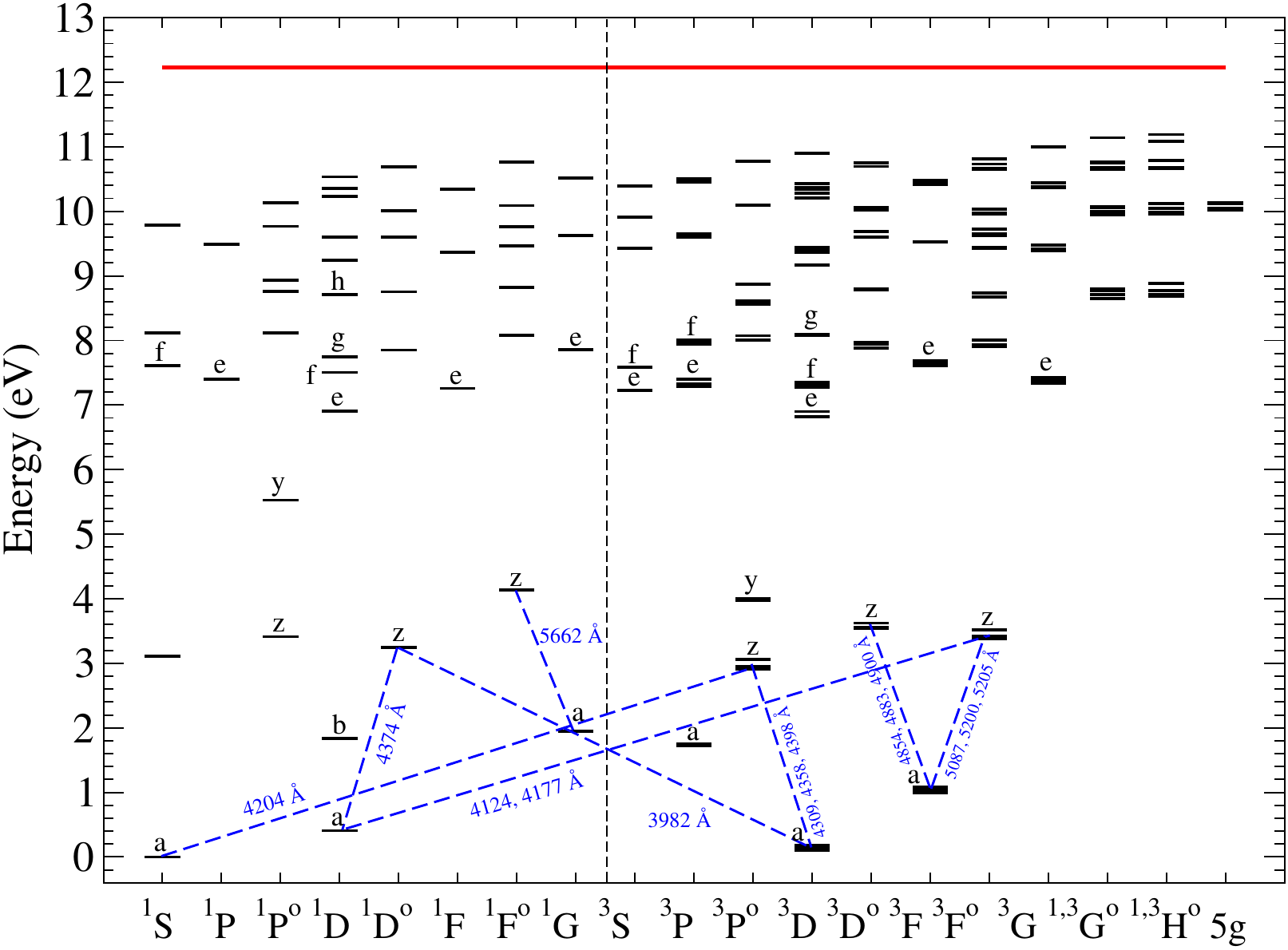}
 \caption{Grotrian term diagram for Y\ii. The ionization energy is shown by solid horizontal line. The dashed lines indicate the transitions, where the investigated spectral lines arise. }
 \label{Groteriane_Y2}
 \end{center}
 \end{figure}

\section{NLTE line formation for Y\ione\ -- Y\ii}\label{Sect:atom}

\subsection{Model atom and atomic data}
Yttrium is a transition element displaying chemical properties of rare-earth elements. Yttrium is a chemical analogue of scandium, and their spectra are similar. The natural abundance of yttrium is 100\% of stable isotope  $^{89}$Y \citep{2005JPCRD..34...57B}, this implies that there is no isotope shift seen in spectra. 
 
\subsubsection{Energy levels } 
 
\noindent {\bf Energy levels of Y\ione\ :}
Energy levels of Y\ione\ belong to doublet
terms of the 4d5s$^2$, 5s$^2$ $nl$ ($nl$ = 5p, 6s, 5d, 6p, 7s), 4d$^2$ 5$l$ ($l$ = s, p), 4d5s $nl$ ($nl$ = 5p, 6s, 5d, 6p),
4d$^3$, 5p$^2$ 5s, and 5s$^2$ $n$f ($n$ = 6--10) electronic configurations;
the quartet terms of 4d$^2$ $nl$ ($nl$ = 5s, 5p, 6s), 4d5s $nl$ ($nl$ = 5p, 6s, 5d, 6p, 7s), 4d$^3$, 5p$^2$ $nl$ ($nl$ = 5s, 4d).

\noindent {\bf Energy levels of Y\ii\ :}
Energy levels of Y\ii\ belong to singlet terms of the 5s$^2$, 
4d$nl$ ($n =$ 5--7, $l \le$ 3), 4d4f, 4d $n$s ($n =$ 8, 9), 4d$^2$,
5s $nl$ ($nl$ = 5p, 6s, 5d, 6p, 4f, 7s, 6d, 8s),
5p$^2$ electronic configurations, the triplet terms of 
4d$nl$ ($n =$ 5--7, $l \le$ 3), 4d4f, 4d $n$s ($n =$ 8, 9), 4d8d, 4d$^2$, 5s $nl$ ($nl$ = 5p, 5d, 6p, 4f, 6d, 5f),
5p$^2$, additional levels with configuration 4d5g, and the ground state of Y\iii.

Totally, we included 181 and 235 levels for Y\ione\ and Y\ii\ taking into account fine structure, respectively. 
In order to reduce computational time, we combine some fine structure levels with higher energies up to 135 and 193 levels for Y\ione\ and Y\ii, respectively.
The energy gaps between the highest levels and the ionization thresholds are 0.25 and 1.03 eV for Y\ione\ and Y\ii, respectively.  Fine-structure splitting was included everywhere for the most of the levels.

Energy levels were taken from the NIST database\footnote{\url{https://www.nist.gov/pml/atomic-spectra-database}} version 5.9 \citep{NIST_ASD}. Energy levels of Y\ione\ with 5s$^2$ $n$f ($n$ = 6--10) electronic configuration were taken from the Kurucz’s website\footnote{\url{http://kurucz.harvard.edu/atoms/3900}}.
 The Grotrian term diagrams for Y\ione\ and Y\ii\ in our model atom are shown in Fig.\,\ref{Groteriane_Y1} and Fig.\,\ref{Groteriane_Y2}. 
 
\subsubsection{Radiative data}  

The model atom includes 4748 allowed bound-bound ($b-b$) transitions for Y\ione\ and 6837 transitions for Y\ii. After combination of some fine structure levels, the amount of transitions were reduced to 2176 for Y\ione\ and 5023 for Y\ii. 
Semi-empirical oscillator strengths for 377 transitions in Y\ii\ were adopted from \citet{2017MNRAS.471..532P}. The remaining transition probabilities were taken from the NIST database (for Y\ione), and the Kurucz’s website. 

Accurate data on the photoionization cross-sections for Y\ione\ still remain undefined. Photo-ionization cross section for the ground state of Y\ione\ was taken from \citet{1987PhRvA..36.3187W}, who used many-body perturbation theory. For the remaining levels of Y\ione, we calculated photoionization cross-sections based on formula for inner-shell photoionization and non-hydrogen-like atoms and ions \citep{doi:10.1063/5.0022751}. 
 
Photo-ionization cross sections for the ground state and 29 lowest levels of Y\ii\ were provided by \citet{2020MNRAS.496.2558F}, and we employed a hydrogenic approximation with an effective principal quantum number for the higher excitation levels of Y\ii.

\subsubsection{Collisions with electrons}  
 
No theoretical calculations of the excitation cross sections were performed for both yttrium neutral atom and singly-ionized yttrium ion. 
Experiments based on observation of the optical radiation of excited atoms allow to directly measure the excitation cross sections (Q$_{ij}$) for spectral lines. We employed experimental values of the excitation cross sections of spectral lines of Y\ione\ and Y\ii\ presented in \citet{1984JApSp..40..368K, Smirnov2000, 2002OptSp..93..351S, Smirnov2001, 2002DokPh..47...34S}.
For almost all transitions, the dependence of excitation cross sections on the energy of incident electrons
is registered in the energy range of 0--200 eV. 
The error of measurement of relative values of the
cross sections varied from 10\% for the most intense
lines to 20\% for low-intensity ones.
For excitation from level $i$ to level $j$, an effective collision strenghts $\Upsilon_{ij}$ is given by

\begin{equation}
 \Upsilon_{ij} = g_i \dfrac{kT}{E_H^\infty} \int_{0}^{\infty} Q_{ij}(x)(x+x_{ij})\mathrm{e}^{-x} \,dx,  \label{eq:1}
\end{equation}

where $g_i$ is the statistical weight of the lower level, $k$ the Boltzmann's constant, 
$E_H^\infty$ the Rydberg unit of energy,
$Q_{ij}$ the collision cross-section expressed in unit of $\pi a_0^2$,
$x = E/kT$ the kinetic energy of electron after excitation and $x_{ij} = E_{ij}/kT$ the transition energy in unit of $kT$.
 $\Upsilon_{ij}$ is dimensionless and symmetric with respect to the transition ($\Upsilon_{ij}$ = $\Upsilon_{ji}$).
 
For NLTE calculations, we used the collision rate from state $i$ to state $j$ with electrons $C^e_{ij}$ in the unit of s$^{-1}$ defined as 

\begin{equation}
C^e_{ij} = n_e A \dfrac{\Upsilon_{ij}}{g_i \sqrt{T}} e^{\frac{-E_{ij}}{kT}},
\label{eq:2}
\end{equation}

where $n_e$ is the electron density, and $A$ the constant in unit of cm$^3$ s$^{-1}$ K$^{1/2}$

\begin{equation}
A = \pi a_0^2  \left(\frac{8 E_H^\infty}{\pi m_e}\right)^{1/2}\left(\frac{E_H^\infty}{k}\right)^{1/2} = 8.629 \times 10^{-6}, 
\label{eq:3}
\end{equation}

where $a_0^2$ is Bohr radius and $m_e$ the mass of electron. 

Totally, we take into account 175 transitions with available  experimental measurements,
namely, 9 quartet transitions of Y\ione\ from \citet{1984JApSp..40..368K}, 48 doublet transitions of Y\ione\ from \citet{Smirnov2000},   
41 high-lying triplet transitions of Y\ii\ from
\citet{2002OptSp..93..351S}, 
32 singlet transitions of Y\ii\ from \citet{Smirnov2001}, and
45 transitions of Y\ii\ from \citet{2002DokPh..47...34S}.  
The calculated effective collision strengths are given for thirteen temperatures in the range $T$=2000--20000 K for the states of Y\ione\ (Table \ref{ECS_Y1}) and fifteen temperatures in the range $T$=2000--28000 K for the states of Y\ii\ (Table \ref{ECS_Y2}).

For transitions with non-available experimental data, the electron-impact excitation was taken into account through the impact parameter method (IPM: \citet{1962PPS....79.1105S}) for the allowed transitions and the effective collision strength was assumed to be  $\Omega_{ij}$ = 1 for forbidden transitions.  Electron impact ionization cross-sections are computed with the formula from \citet{Seaton1962}.

\subsubsection{Collisions with hydrogen atoms}

For the first time the quantum cross sections and rate coefficients for bound-bound transitions in inelastic collisions of Y\ione\ and Y\ii\ with hydrogen atoms are used in this study. 
A simplified model presented by \citet{2017A&A...608A..33B} is used to estimate rate coefficients for inelastic processes in low-energy hydrogen collisions. 
This method relies on the asymptotic approach for electronic structure calculations and employs the Landau-Zener model for determining nonadiabatic transition probabilities. The method is physically reliable but less computationally expensive than a full quantum analysis.
The data for the rate coefficients are involving all levels and all possible bound-bound and bound-free transitions up to 4$d$ f$^2$P ($E$=5.697~eV) for Y\ione\ and 7$f$ $^1$H$^{\circ}$ ($E$=11.191~eV) for Y\ii. Totally, we included data for 311 bound-free and 20934 bound-bound transitions in the model. 
The more information about the calculated rate coefficients can be found in \citet{2023arXiv230807831W}.

The cross sections of inelastic collisions between atoms and hydrogen atoms could be estimated by Drawin’s formula \citep{1968ZPhy..211..404D}, which is an extension of the classical Thomson model. However, for low-energy atomic collisions, the Drawin formula compares poorly with the results of the available quantum mechanical calculations and cannot provide reliable results \citep{2011A&A...530A..94B}.

\subsection{Method of calculations}
 
The departure coefficients, $b_{\rm{i}}$ = $n_{\rm{NLTE}}$ / $n_{\rm{LTE}}$, where calculated with the code \textsc{DETAIL} \citep{detail} from the solution of the system of the radiative transfer and statistical equilibrium equations using the accelerated $\Lambda$-iteration method \citep{rh91}. Here, $n_{\rm{NLTE}}$ and $n_{\rm{LTE}}$ are the statistical equilibrium and thermal (Saha-Boltzmann) number densities, respectively. 
The \textsc{DETAIL} opacity package was updated by \citet{2011JPhCS.328a2015P, 2011A&A...528A..87M} by including bound-free opacities of neutral and ionized species for applications of the code to FGK and BA stars.
Theoretical NLTE spectra were calculated with the code \textsc{synthV\_NLTE} \citep{2019ASPC..518..247T} using the obtained departure coefficients. To examine theoretical stellar spectra and compare them to observations we employed 
widget program \textsc{binmag}\footnote{\url{http://www.astro.uu.se/~oleg/binmag.html}} \citep{2018ascl.soft05015K}. BinMag interfaces with mounted code \textsc{synthV\_NLTE} allows to obtain the best LTE and NLTE fits to the observed line profiles automatically to determine chemical abundances with high precision. 
  
We used classical (1D) plane-parallel model atmospheres from the MARCS model grid \citep{2008A&A...486..951G}. For given \Teff, log~$g$, and [Fe/H] the models were interpolated using a FORTRAN-based routine written by Thomas Masseron\footnote{\url{http://marcs.astro.uu.se/software.php}}.  
  
We selected the lines of Y\ione\ and Y\ii, which can be found in spectra of stars in visual wavelength range and can be suitable for abundance analyses. 
Table~\ref{tab1} presents the lines together with the adopted line data. 
The adopted oscillator strengths were obtained from laboratory measurements. 
For the most of the lines of Y\ione, the oscillator strengths were taken from \citet{NIST_ASD}. For three Y\ione\ lines at 6023.406, 6222.578, and 6402.007~\AA, we adopted the oscillator strengths from recent experimental work of \citet{2018JOSAB..35.2244L}. 
The most up-to-date semi-empirical oscillator strengths for Y\ii\ lines were adopted from \citet{2017MNRAS.471..532P}.

The splitting of the hyperfine components and isotopic shifts were neglected during the determination of Y abundances.
For the Y\ione\ and Y\ii\ lines, the effects due to isotopic splitting and hyperfine structure are insignificant, since there is only one stable isotope, and the hyperfine splitting factors are very small, typically less than 1 m\AA\ \citep{1982ApJ...261..736H, 1988JPhB...21..547G}.

For lines of Y\ione\ and Y\ii, Radiative, Stark, and Van der Waals damping constants are presented in Table~\ref{tab1}. They were adopted from VALD3\footnote{\url{http://vald.astro.uu.se/~vald3/php/vald.php}} database \citep{2015PhyS...90e4005R}.

\subsection{Departures from LTE for Y\ione\ - Y\ii\ in FGK-type Stars}
 
\begin{figure*}
   \begin{minipage}{170mm}
 \begin{center}
 \parbox{0.22\linewidth}{\includegraphics[scale=0.2]{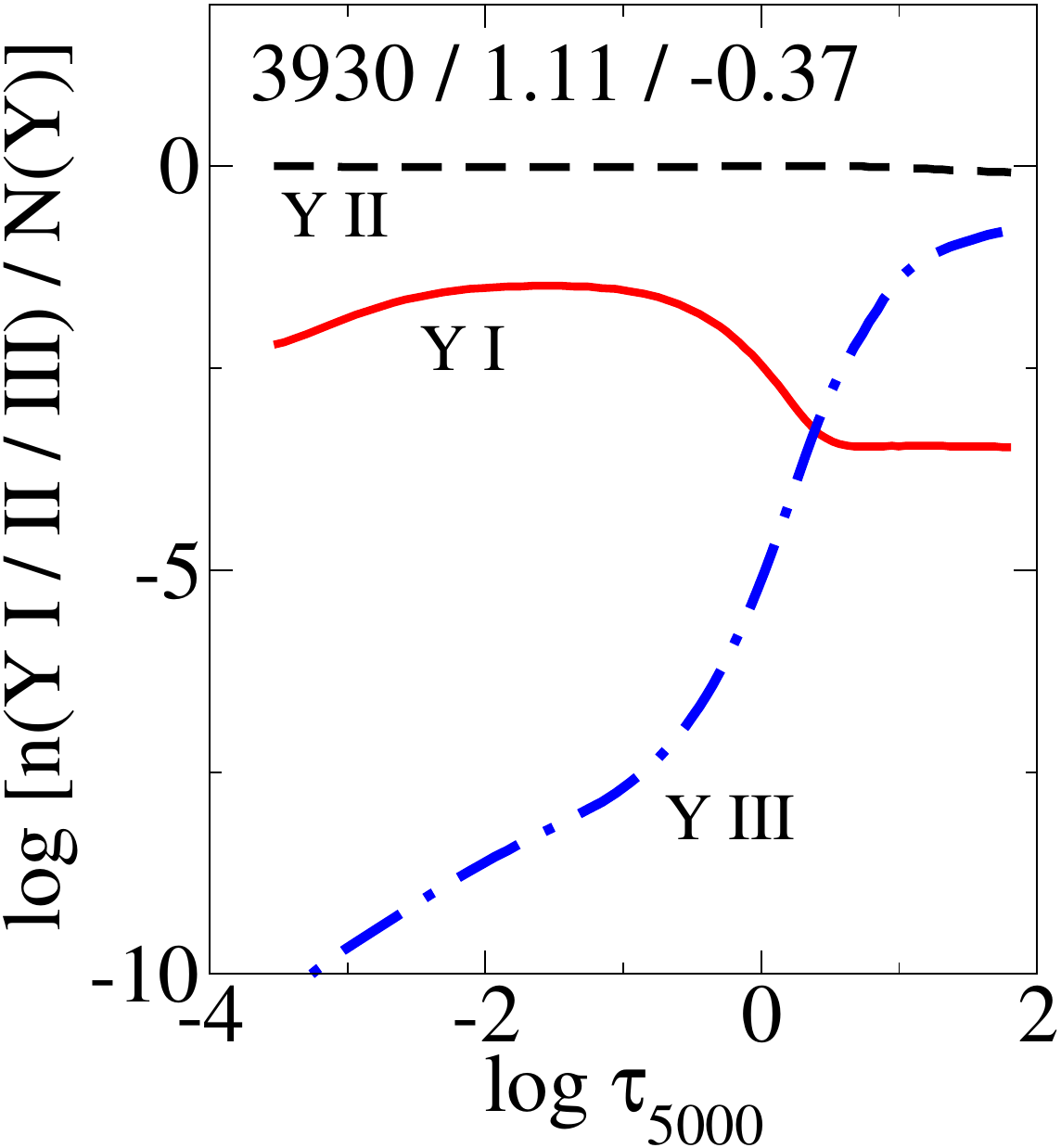}\\
 \centering}
 \parbox{0.22\linewidth}{\includegraphics[scale=0.2]{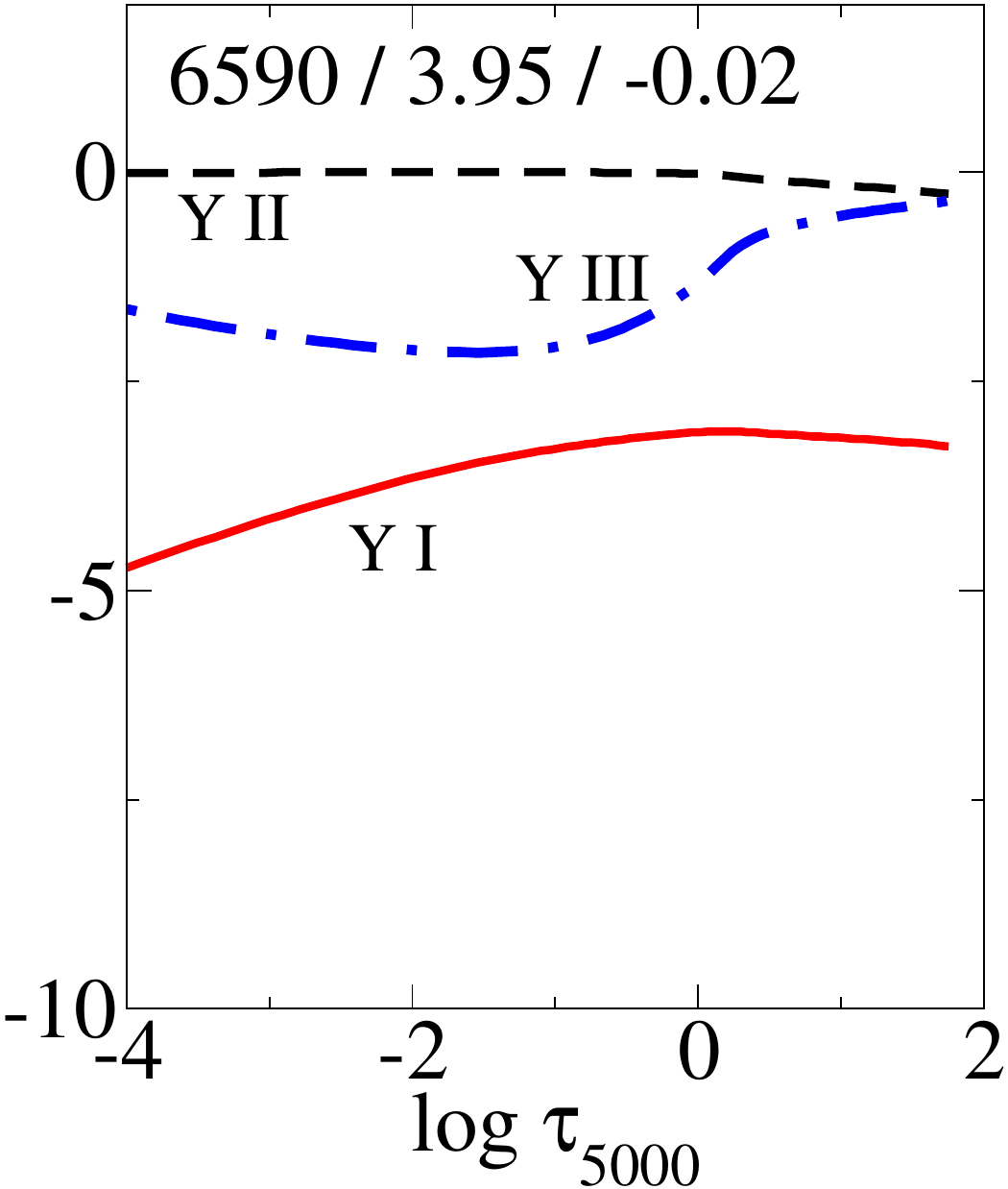}\\
 \centering}
 \hspace{1\linewidth}
 \hfill
 \\[0ex]
 \caption{The fractions of Y\ione, Y\ii, and Y\iii\  in the model atmospheres with different stellar parameters. }
 \label{balance}
 \end{center}
 \end{minipage}
 \end{figure*}

 {Figure}\,\ref{balance} displays fractions of Y\ione, Y\ii, and Y\iii\ in the model atmospheres with \Teff /log$g$/[Fe/H] = 3930/1.11/-0.37 and 6590/3.95/-0.02.
In the model 3930/1.11/-0.37, Y\ii\ is the dominant ionization stage with small admixtures of Y\ione\ (few parts in a thousand) in the line-formation region.
In hotter atmosphere, 6590/3.95/-0.02, Y\ii\ is still the dominant stage and the fraction of Y\iii\ increases, while Y\ione\ becomes the minority species.  

\begin{figure*}
  \begin{minipage}{170mm}
   \begin{center}
  \parbox{0.45\linewidth}{\includegraphics[scale=0.5]{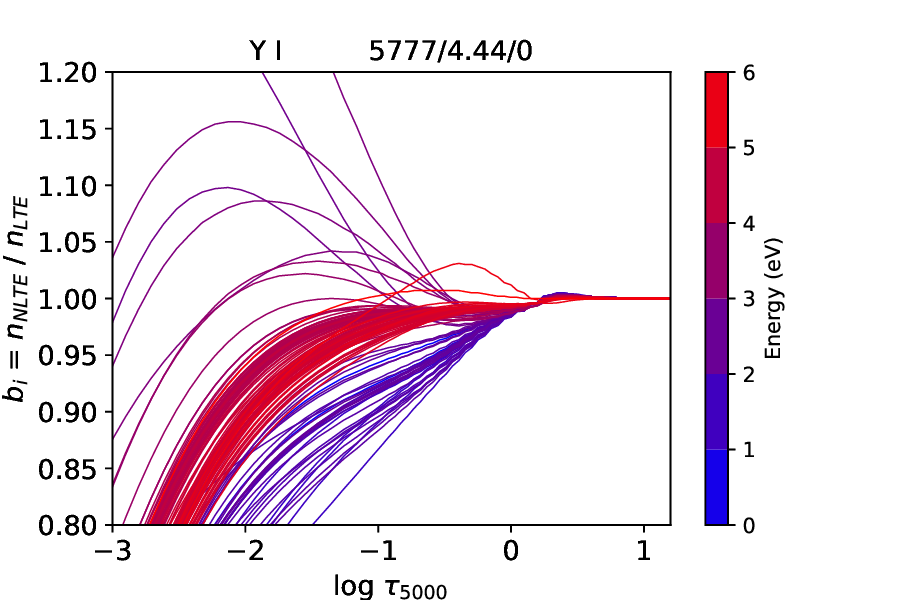}\\
  \centering}
  \parbox{0.45\linewidth}{\includegraphics[scale=0.5]{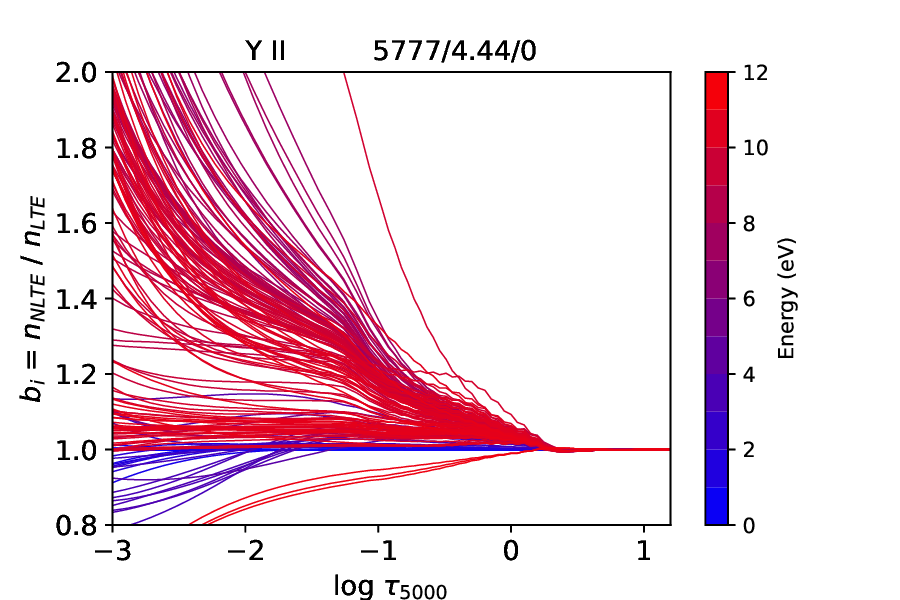}\\
  \centering}
  \hspace{1\linewidth}
  \hfill
  \\[0ex]
  \parbox{0.45\linewidth}{\includegraphics[scale=0.5]{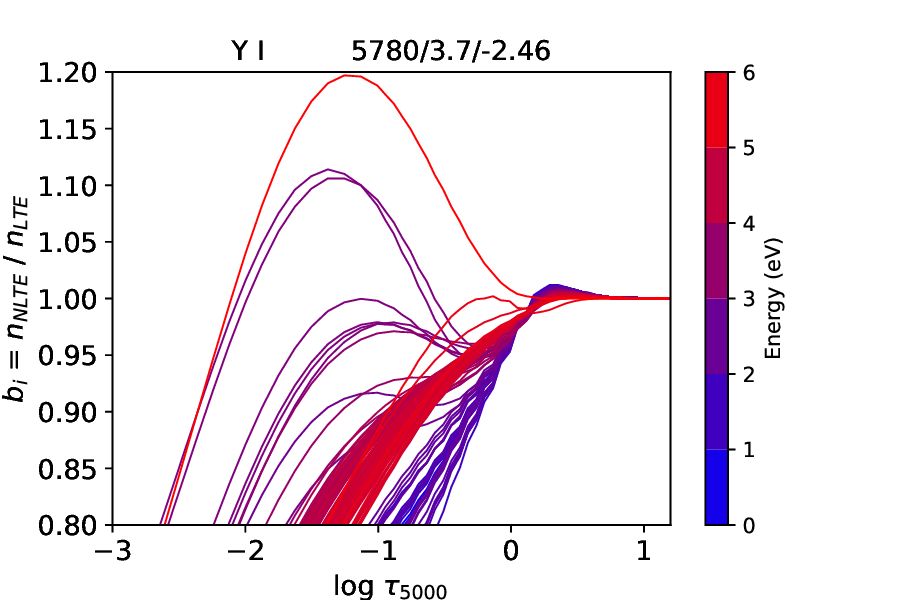}\\
  \centering}
  \parbox{0.45\linewidth}{\includegraphics[scale=0.5]{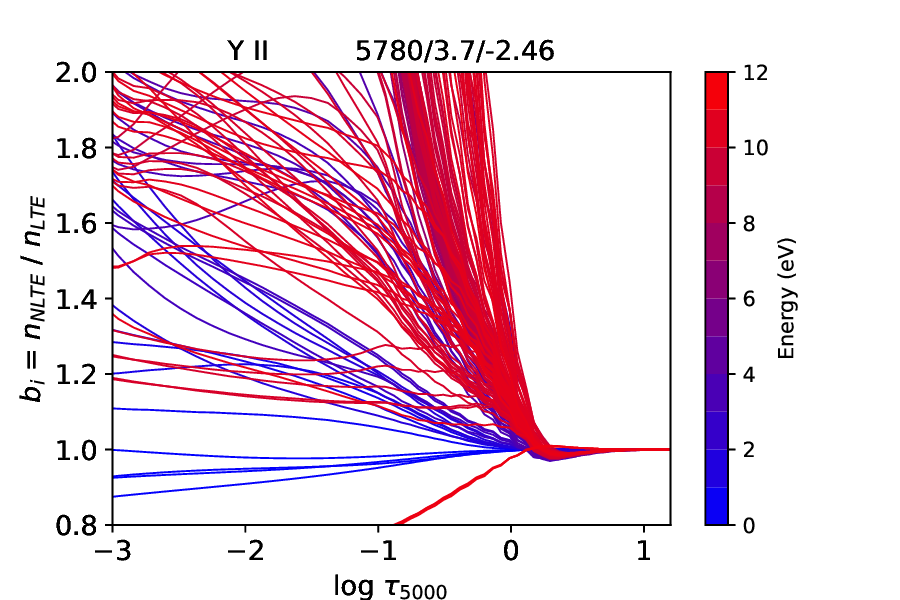}\\
  \centering}
  \hspace{1\linewidth}
  \hfill
  \\[0ex]
  \caption{Departure coefficients for the Y\ione\ levels (left column) and the Y\ii\ levels (right column) as a function of log~$\tau_{5000}$ in the model atmospheres with 5777/4.44/0 (top row) and  5780/3.7/-2.46 (bottom row). }
  \label{bfactors}
   \end{center}
  \end{minipage}
  \end{figure*}

{Figure}\,\ref{bfactors} shows the departure coefﬁcients for all investigated levels of Y\ione\ and Y\ii\ in the models 5777/4.44/0 and 5780/3.7/-2.46, which represent the solar atmosphere and metal-poor HD~140283, respectively. The energy levels are shown by color bars. In deep atmospheric layers, log~$\tau_{5000}>$1, where the gas density is large and collisional processes dominate, the departure coefﬁcients are equal to unity indicating no deviation from LTE. 
The Y\ione\ ground state, a$^2$D (\Eexc =~0~eV; $\lambda_{\rm thr}$~=~2006~\AA\,), and low-excitation levels, z$^2$P$^{\circ}$ (\Eexc ~=~1.305~eV; $\lambda_{\rm thr}$~=~2553~\AA\,) and a$^4$F (\Eexc ~=~1.356~eV; $\lambda_{\rm thr}$~=~2570~\AA\,), are subject to UV overionization, which lead to severe depopulation of these levels in both atmospheres. Population depletion is extended to the remaining Y\ione\ levels due to their coupling to the low-excitation levels. Since Y\ii\ dominates the element abundance over all atmospheric depths, the Y\ii\ ground state keeps its LTE populations. 
The deviations from LTE are stronger in the metal-poor atmosphere with 5780/3.7/-2.46 compared to solar atmosphere due to physical conditions, namely, smaller electron number density and stronger ionization field that lead to stronger overionization of neutral yttrium.

The excited levels of Y\ii\ are overpopulated relative to their LTE populations in both models, outward log~$\tau_{5000}\approx$~0.3 due to the radiative pumping of ultraviolet transitions from the ground state.
In contrast, the highly-excited levels of Y\ii\ with energy more than 11 eV and the ground state of Y\iii\ appear to be strongly depopulated.

To understand the NLTE effects for a given spectral line we analyze the departure coefﬁcients of the lower ($b_l$) and upper ($b_u$) levels at the line-formation depths. 
 In NLTE a given line is stronger compared to LTE, if $b_l > 1$ and (or) the line source function is smaller than the Planck function, that is, $b_l > b_u$.

\begin{figure*}
   \begin{minipage}{170mm}
 \begin{center}
 \parbox{0.3\linewidth}{\includegraphics[scale=0.2]{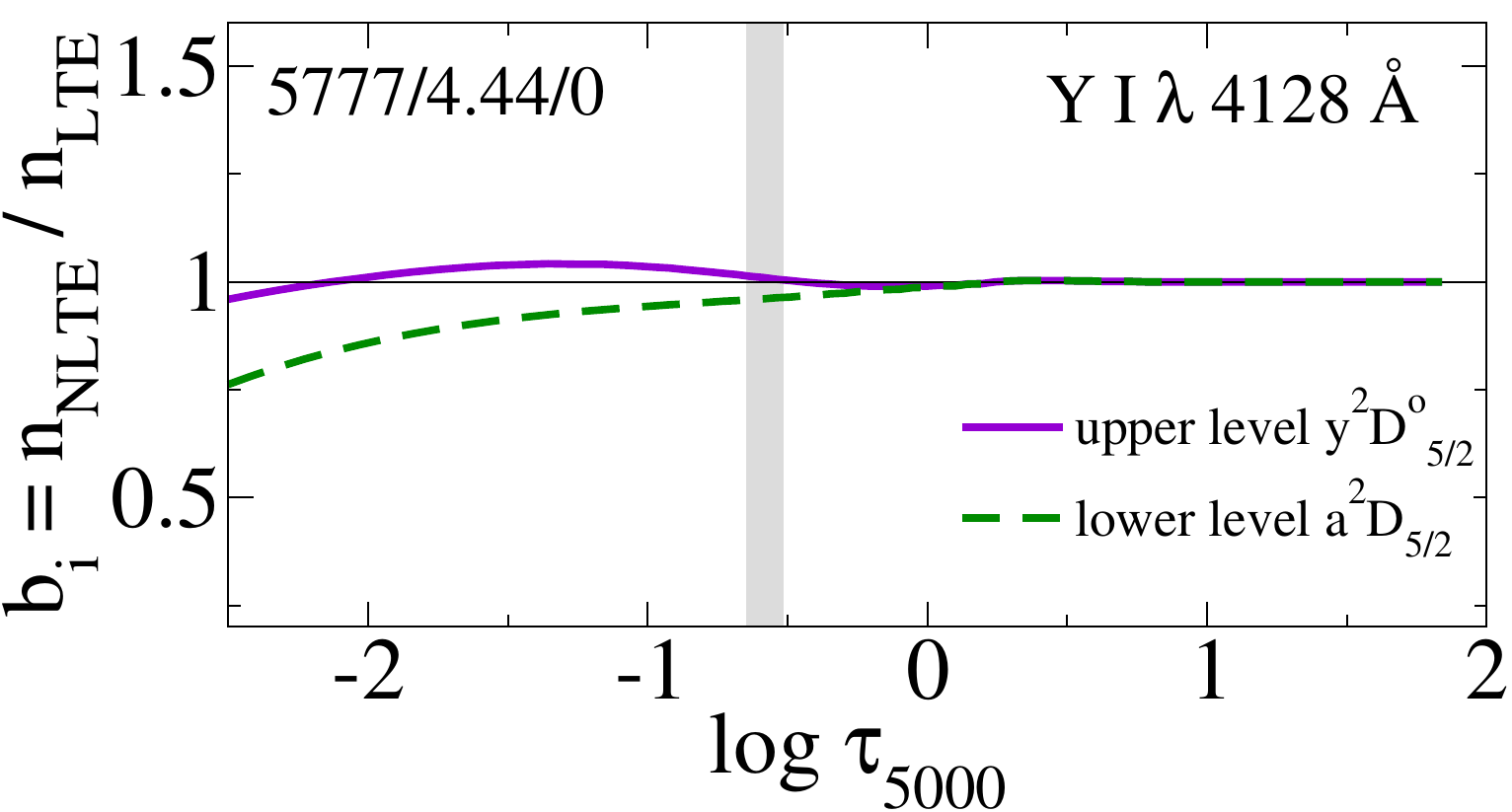}\\
  \centering}
   \parbox{0.22\linewidth}{\includegraphics[scale=0.2]{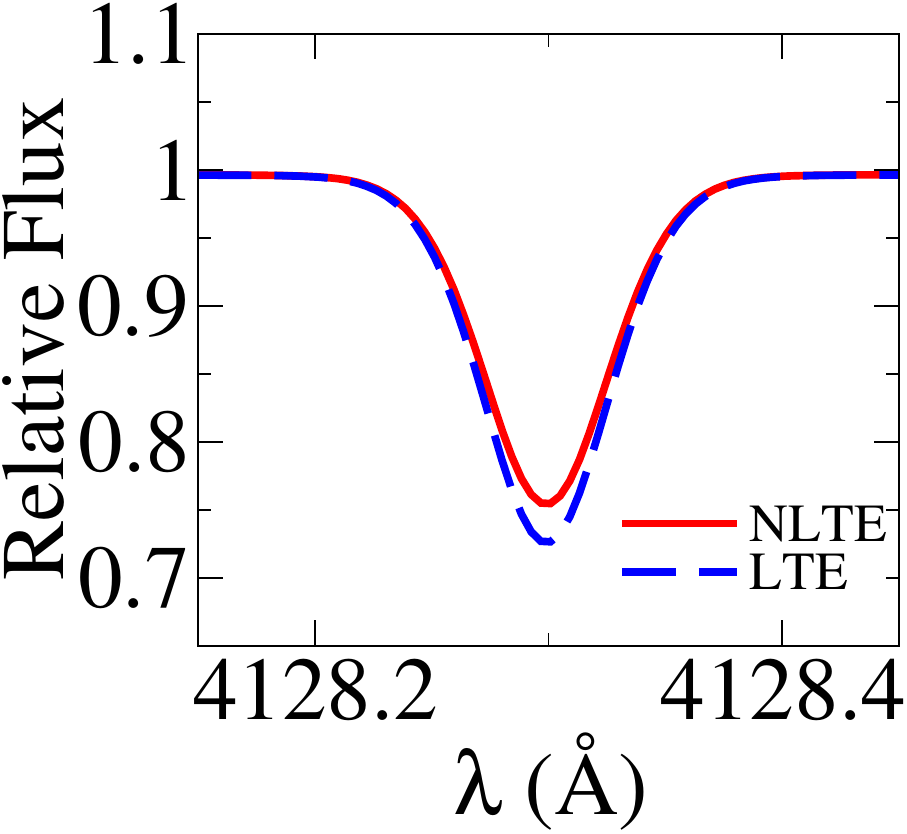}\\
  \centering}
  \hspace{1\linewidth}
  \hfill
  \\[0ex]   
 \parbox{0.3\linewidth}{\includegraphics[scale=0.2]{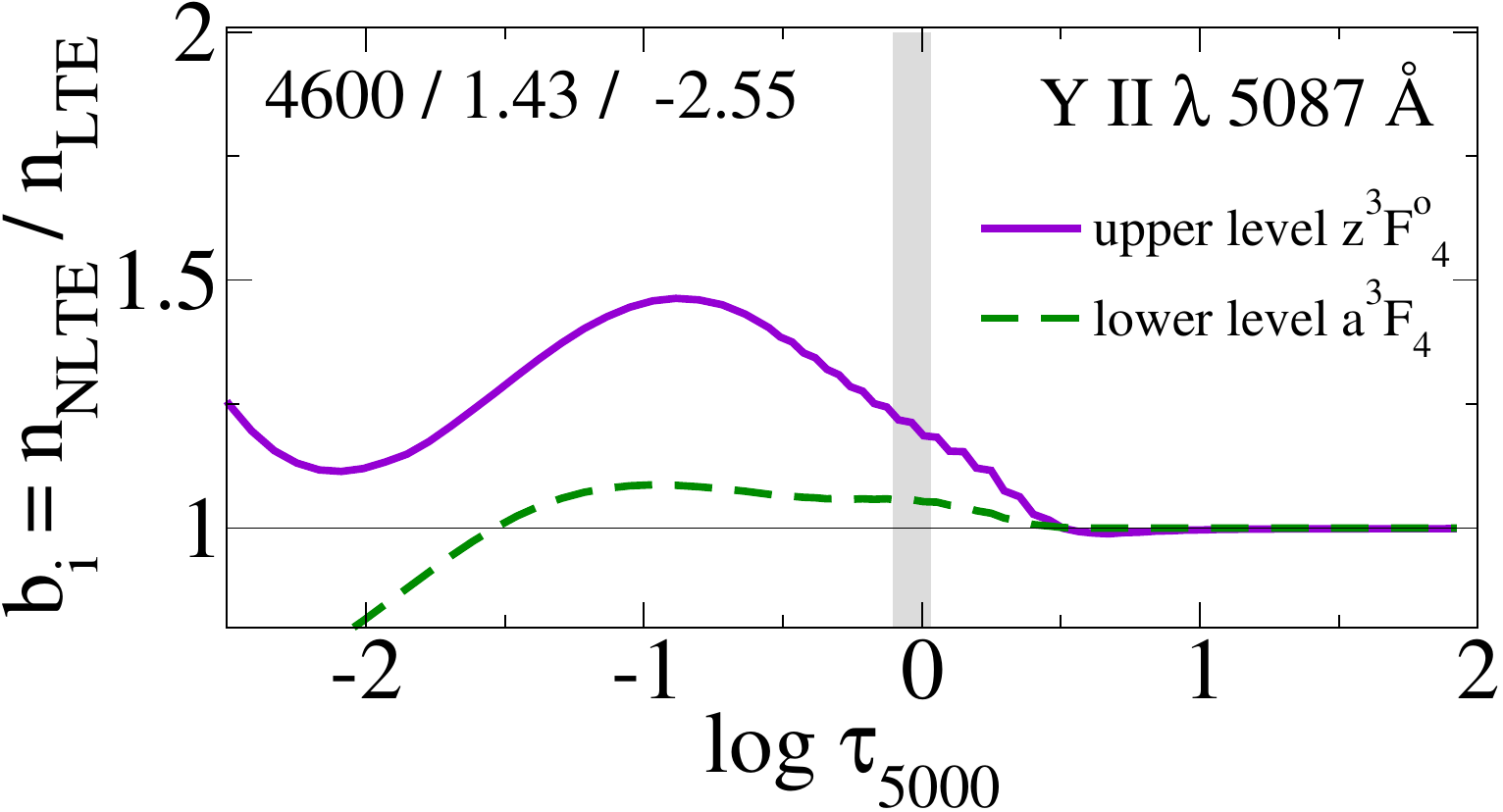}\\
 \centering}
  \parbox{0.22\linewidth}{\includegraphics[scale=0.2]{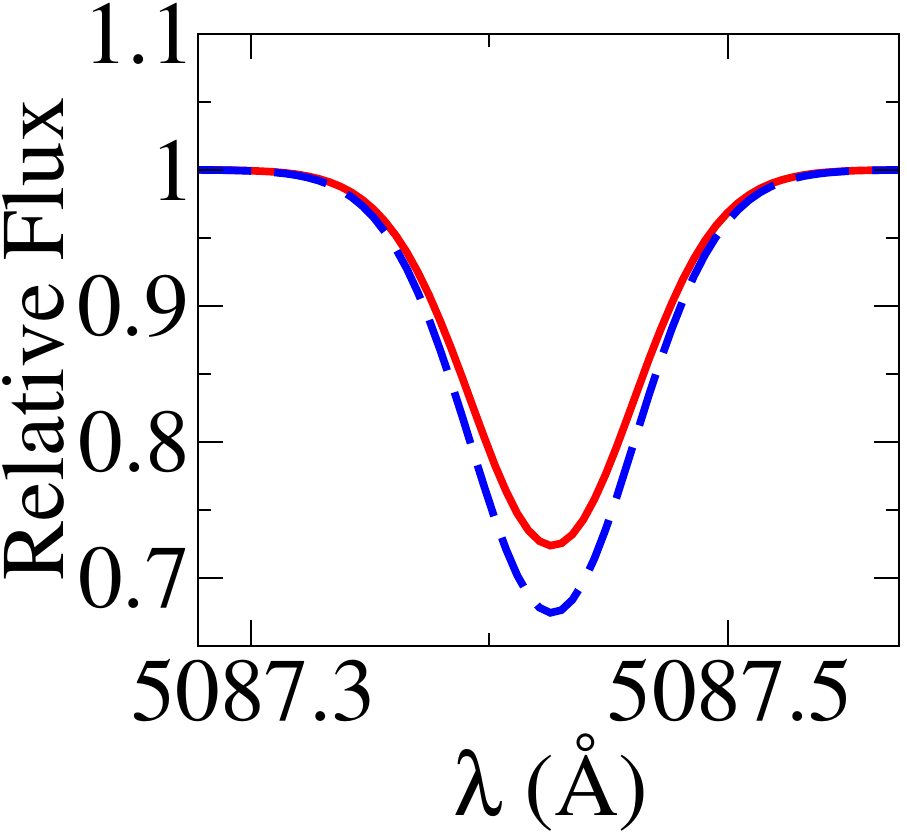}\\
 \centering}
 \hspace{1\linewidth}
 \hfill
 \\[0ex]
 \caption{Departure coefficients as a function of log~$\tau_{5000}$ for the lower and upper levels of Y\ione\ line at 4128~\AA\ in the model 5777/4.44/0 (upper panel) and for the lower and upper levels of Y\ii\ line at 5087~\AA\ in the model 4600/1.43/-2.55 (bottom panel). The line-formation regions are shown by grey shadow vertical lines. Right panels: theoretical line proﬁles of the investigated lines calculated with LTE and NLTE for the particular atmospheric models.  }
 \label{bfactors2}
 \end{center}
 \end{minipage}
 \end{figure*}

First, we consider line of Y\ione\ at 4128~\AA\ in the 5777/4.44/0 model. The departure coefficients for the lower and upper levels for this line are presented in {Figure}\,\ref{bfactors2} (upper panel). Its core forms around log~$\tau_{5000}\approx$~-0.6,
where the departure coefﬁcient of the lower level is less than unity, $b_l < 1$, and $b_l < b_u$ that lead to the weakening line and positive abundance correction of $\Delta_{\rm NLTE}$ = log$\epsilon_{\rm NLTE}$ - log$\epsilon_{\rm LTE}$ = +0.07~dex. The LTE and NLTE profiles are also presented.

Second, we consider line of Y\ii\ at 5087~\AA\ in the 4600/1.43/-2.55 model. The departure coefﬁcients for the lower and upper levels are presented in {Figure}\,\ref{bfactors2} (bottom panel). The core of the line forms around log~$\tau_{5000}\approx$~-0.04, where the departure coefﬁcient of the lower level is less than departure coefﬁcient of the upper level, $b_l < b_u$, that leads to larger source function compared to the Planck function. As a result, the NLTE line is weaker than LTE, and NLTE correction to the yttrium abundance will be positive, $\Delta_{\rm NLTE}$ = +0.10~dex.

\begin{figure*}
   \begin{minipage}{170mm}
 \begin{center}
 \parbox{0.215\linewidth}{\includegraphics[scale=0.2]{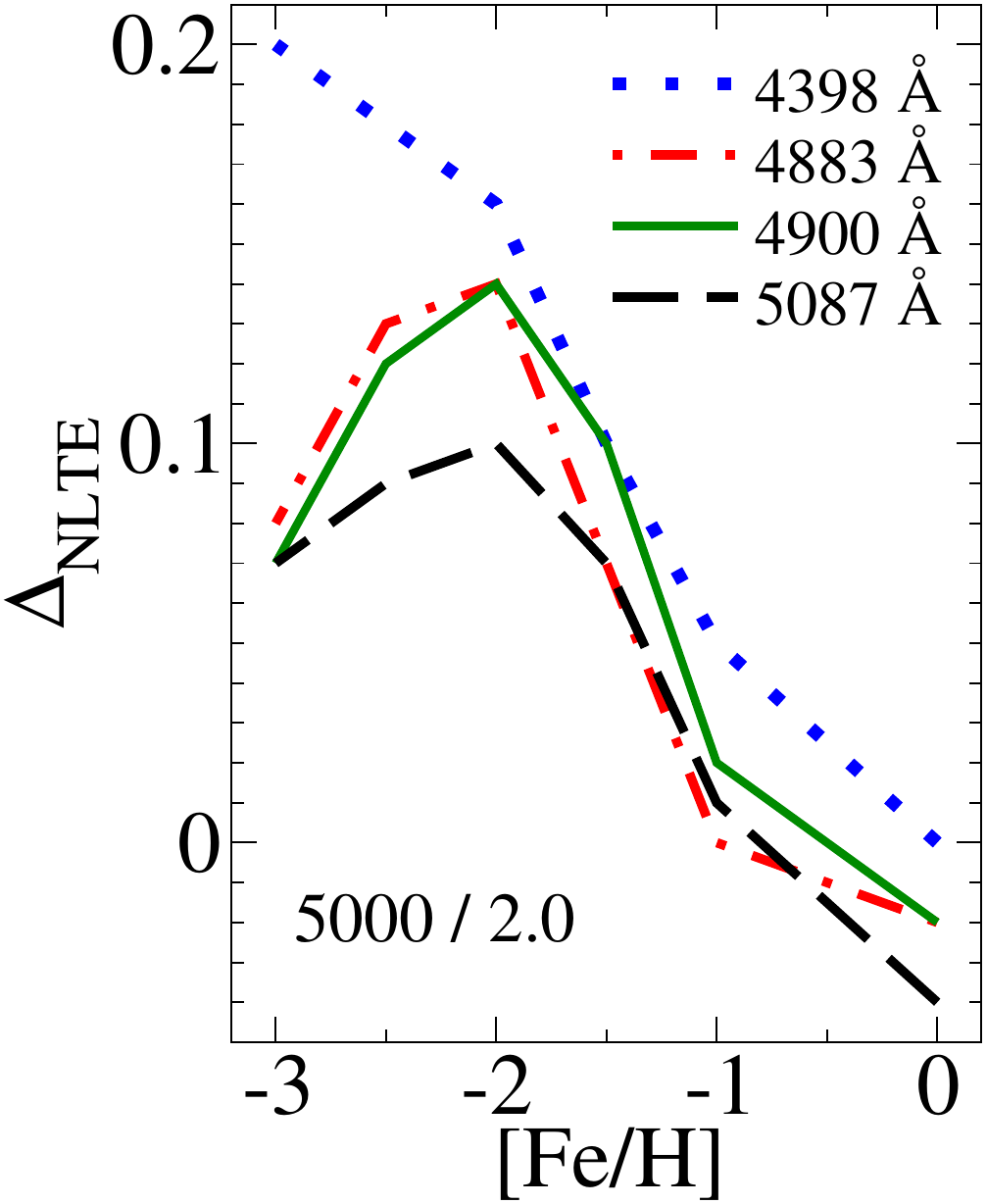}\\
 \centering}
 \parbox{0.18\linewidth}{\includegraphics[scale=0.2]{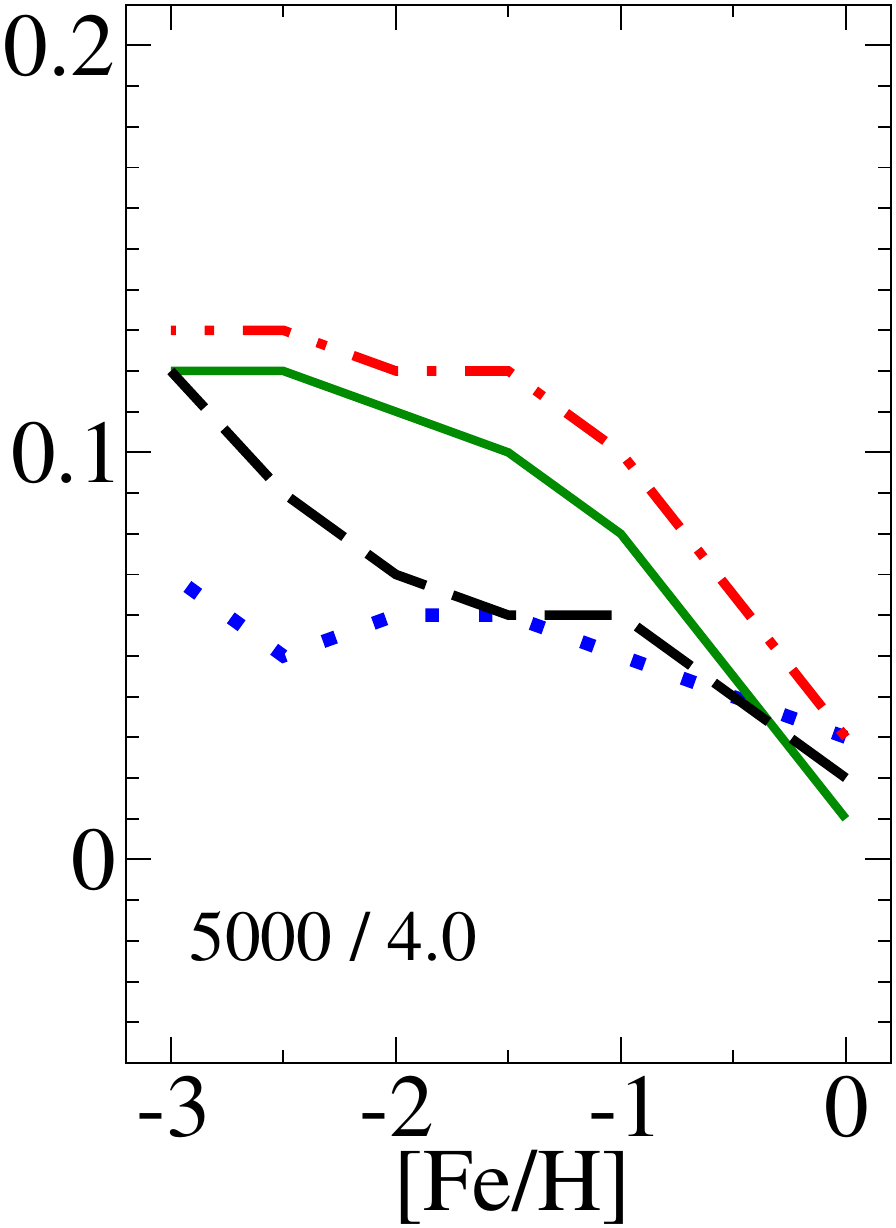}\\
 \centering}
 \parbox{0.2\linewidth}{\includegraphics[scale=0.2]{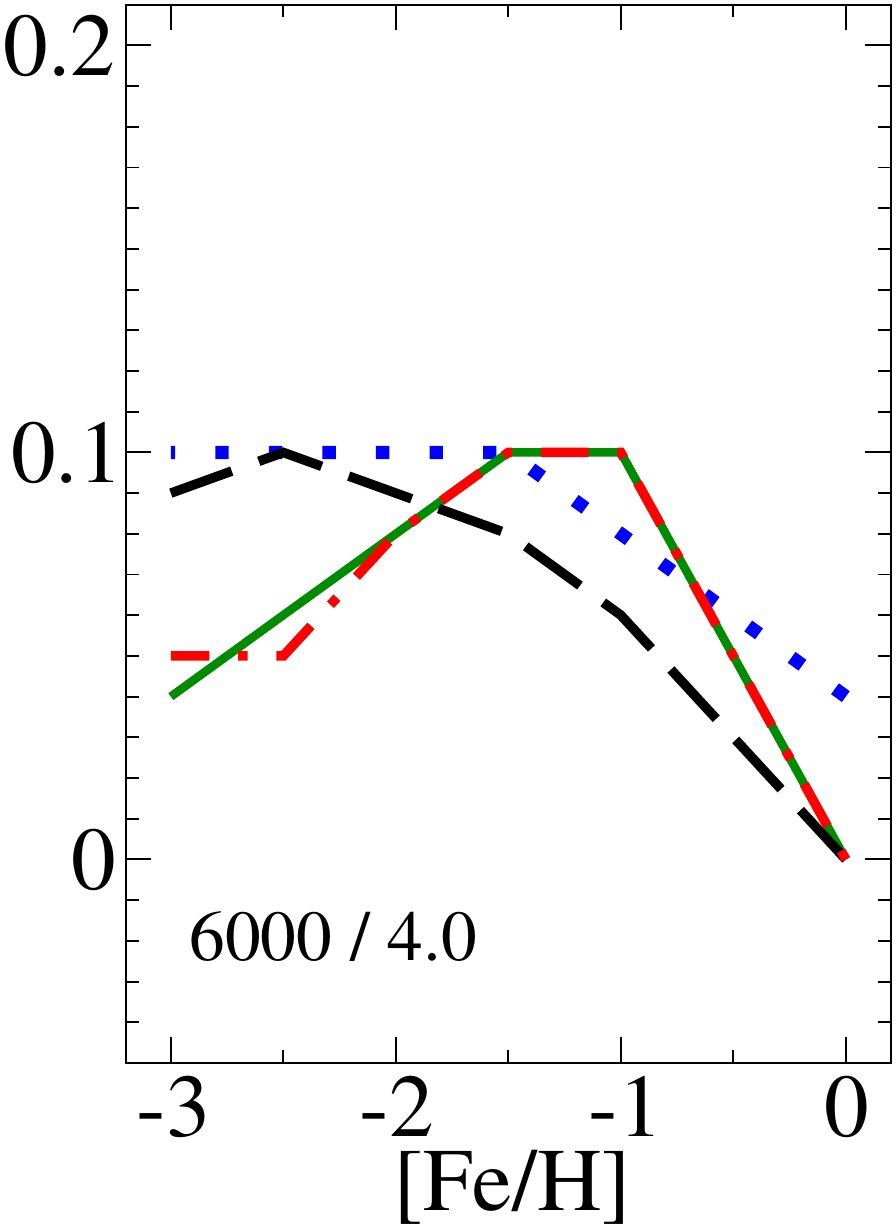}\\
 \centering}
 \hspace{1\linewidth}
 \hfill
 \\[0ex]
 \caption{NLTE abundance corrections as functions of [Fe/H] for selected Y\ii\ lines at 4398, 4883, 4900, and 5087~\AA\ for three sets of stellar parameters \Teff / log~$g$ = 5000 / 2.0, 5000 / 4.0, and 6000 / 4.0. Everywhere, V$_{mic}$ = 2~\kms and [Y/Fe] = 0.}
 \label{corr}
 \end{center}
 \end{minipage}
 \end{figure*}

{Figure}\,\ref{corr} shows the NLTE abundance corrections for selected Y\ii\ lines at 4398, 4883, 4900, and 5087~\AA\ in three representing model atmospheres with \Teff / log~$g$ = 5000/2.0, 5000/4.0, and 6000/4.0, and V$_{mic}$ = 2~\kms and [Y/Fe] = 0 for all of them.

\section{Solar lines of Y\ione\ and Y\ii} \label{subsec:sun}

\subsection{Selection of Y\ione\ and Y\ii\ lines}

\citet{1982ApJ...261..736H} derived the yttrium solar abundance from eight Y\ione\ and 41 Y\ii\ solar lines. However, it should be noticed that almost all Y\ione\ and some Y\ii\ lines are blended with lines of other spices. Nonetheless, the effects of blended lines can be taken into account through the use of spectrum fitting method. Below, we briefly characterize all lines used in this study.

The Y\ione\ line at 4128.304~\AA\ is quite strong in solar spectrum, however, it is located at the wing of strong Mn\ii\ $\lambda$4128.129~\AA, which may affect the abundance from the Y\ione\ line, and
it is also blended with three lines Mn\ione\ $\lambda$4128.330, Ce\ii\ 
$\lambda$4128.36, and Cr\ione\ $\lambda$4128.387~\AA\ (Fig.~\ref{pics}). 

The Y\ione\ lines at 4643.698 and 5630.138~\AA\ are weak and strongly influenced by the lines of other atoms in solar spectrum. However,  in the atmospheres of cool giants, the strength of the lines increases, while the blending lines decrease in strength. The Y\ione\ line at 4674.848~\AA\ is very weak in solar spectrum and can be used with caution. It is blended with Ni\ione\  $\lambda$4674.755, Cu\ione\ $\lambda$4674.78, C$_2$ $\lambda$4674.831, C$_2$ $\lambda$4674.879, Ca\ione\ $\lambda$4674.939, and K\ione\ $\lambda$4674.942~\AA\ (Fig.~\ref{pics}). The intensity of the line becomes stronger once more within the atmospheres of K giants.

The Y\ione\  line at 6023.406~\AA\ is very weak in solar-type spectra and it lies in the wing of the Fe\ione\ $\lambda$6024.057~\AA\ line. In spectra of K-type stars, the line becomes more pronounced and can be utilized for determining abundance by spectrum fitting, when considering the absorption in the wing of Fe\ione\ line. 
The Y\ione\ line at 6222.578~\AA\ is weak and it is strongly blended by the lines of Ca\ione\ $\lambda$6222.497, Fe\ii\ $\lambda$6222.570, and Si\ione\ $\lambda$6222.682 in solar spectrum. 
In cooler atmospheres, for example, HD~220009, the Y\ione\ 6222.578~\AA\ line  becomes more pronounced and can be used for abundance determination. 
The Y\ione\ line at 6402.007~\AA\ is too weak in solar spectrum and it lies in the wing of strong Fe\ione\ $\lambda$6400.316~\AA\ line. In cooler atmospheres, the line can be measured reliably. 
 The Y\ione\ line at 6435.022~\AA\ is quite strong in solar spectrum, however, it is blended with two lines CN $\lambda$6434.894~\AA\ and V\ione\   $\lambda$6435.158~\AA\ from both sides (Fig.~\ref{pics}). In spectra of K-type stars, the line becomes more pronounced, while blending lines remain weak. It is one of the most reliable Y\ione\ lines for abundance determination.
The Y\ione\ line at 6437.168~\AA\ is weak blend free line, which can be found only in the atmospheres of cool giants. 

The lines of Y\ii\ at 4883.682, 5087.418, and 5289.815~\AA\ are strong and free from blends in the spectra of solar-type stars. These lines can be utilized for determining abundance in metal-poor stars. The line Y\ii\ at 3788.693~\AA\ is blended with H$_{10}$ $\lambda$3797~\AA, and Ce\ii\ $\lambda$3788.746~\AA\ in solar spectrum. The line can be measured reliably in metal-poor stars and F-type stars like Procyon. 
The line Y\ii\ at 4124.904~\AA\ has blends Cr\ione\ $\lambda$4124.863~\AA, and Ce\ii\ $\lambda$4124.870~\AA. The Y\ii\ line at 4177.528~\AA\  is blended with two lines Fe\ione\ $\lambda$4177.593\AA, and Fe\ii\ $\lambda$4177.686~\AA. This line was not used for solar analysis. However, the blends become less pronounced in HD~84937 due to its low metallicity and hot temperature. The line Y\ii\ at 4374.933~\AA\ is blended with strong Ti\ii\ $\lambda$4374.816~\AA\ in solar spectrum (Fig.~\ref{pics}) and weak metallic lines  
Co\ione\ $\lambda$4374.914, Nd\ii\ $\lambda$4374.920, Mn\ione\ $\lambda$4374.947, and Sm\ii\ $\lambda$4374.975~\AA. In VMP stars, weak metallic lines typically have a negligible impact. 
The line Y\ii\ at 4398.008~\AA\ is blended with
Fe\ione\ $\lambda$4397.971, Nd\ii\ $\lambda$4398.013, Ti\ione\ $\lambda$4398.073, and CH $\lambda$4398.076~\AA\ in solar spectrum (Fig.~\ref{pics}). The line Y\ii\ at 4900.118~\AA\ is blended with a nearby Ti\ione\ line at 4899.91~\AA, however, it is insignificant when rotational velocity is less then 5 \kms.  Small blend of Er\ii\ at 4900.08~\AA\ is less then 2$\%$ relative to Y\ii\ line. The Y\ii\ line at 4982.128~\AA\ lies between two strong lines Ti\ione\ 4981.730~\AA\ and 4982.499~\AA\ and has very weak blending line Mn\ione\ $\lambda$4982.085~\AA, which is negligible in metal-poor stars.     
The line Y\ii\ at 5205.722~\AA\ lies in the wing of strong line of Cr\ione\ at 5206.041~\AA\ and 
has weak blending lines MgH at 5205.755~\AA\ ($\sim$3$\%$).

\subsection{Solar Yttrium abundance}

\begin{figure*}
  \begin{minipage}{175mm}
  \parbox{0.24\linewidth}{\includegraphics[scale=0.16]{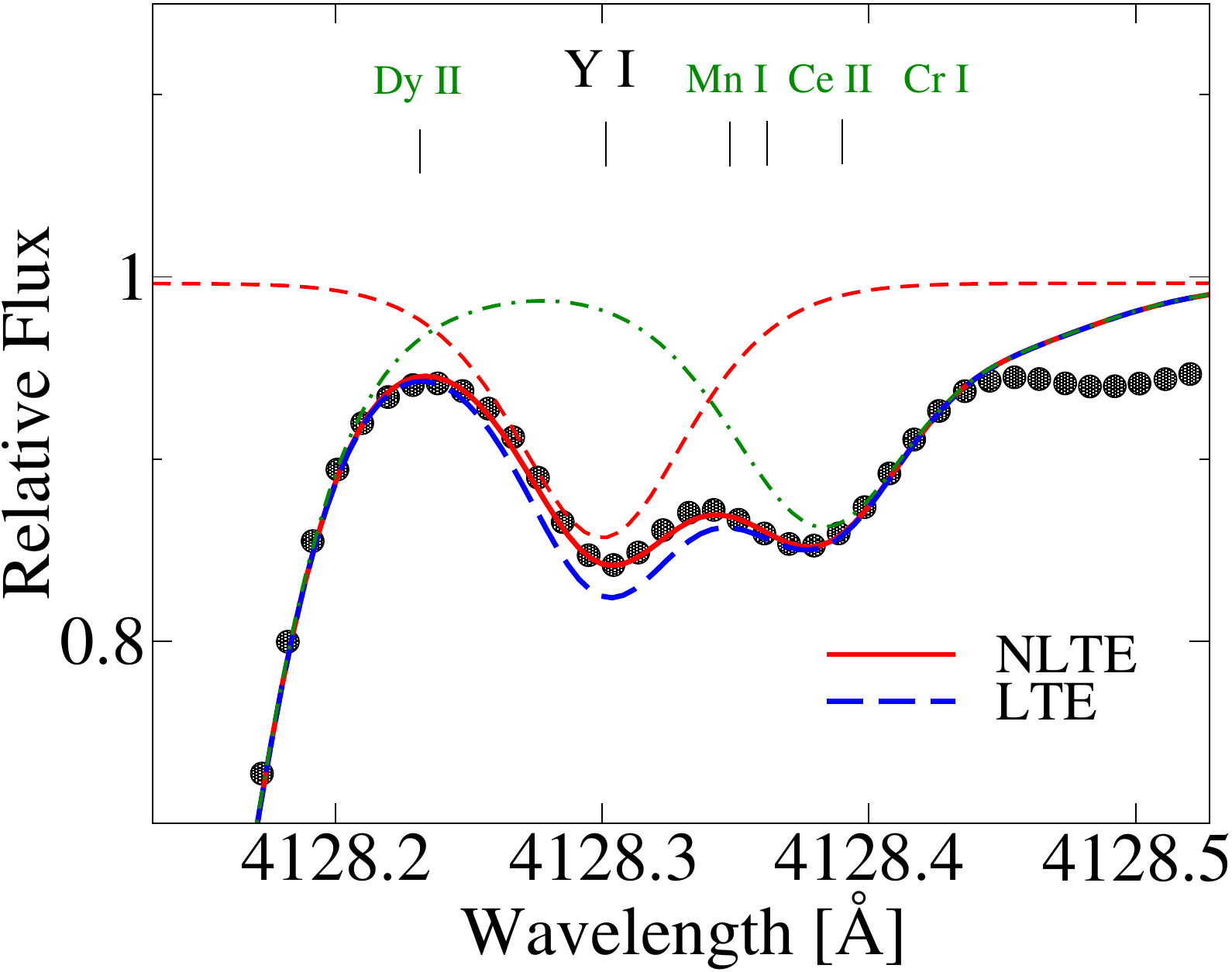}\\
  \centering}
  \parbox{0.24\linewidth}{\includegraphics[scale=0.16]{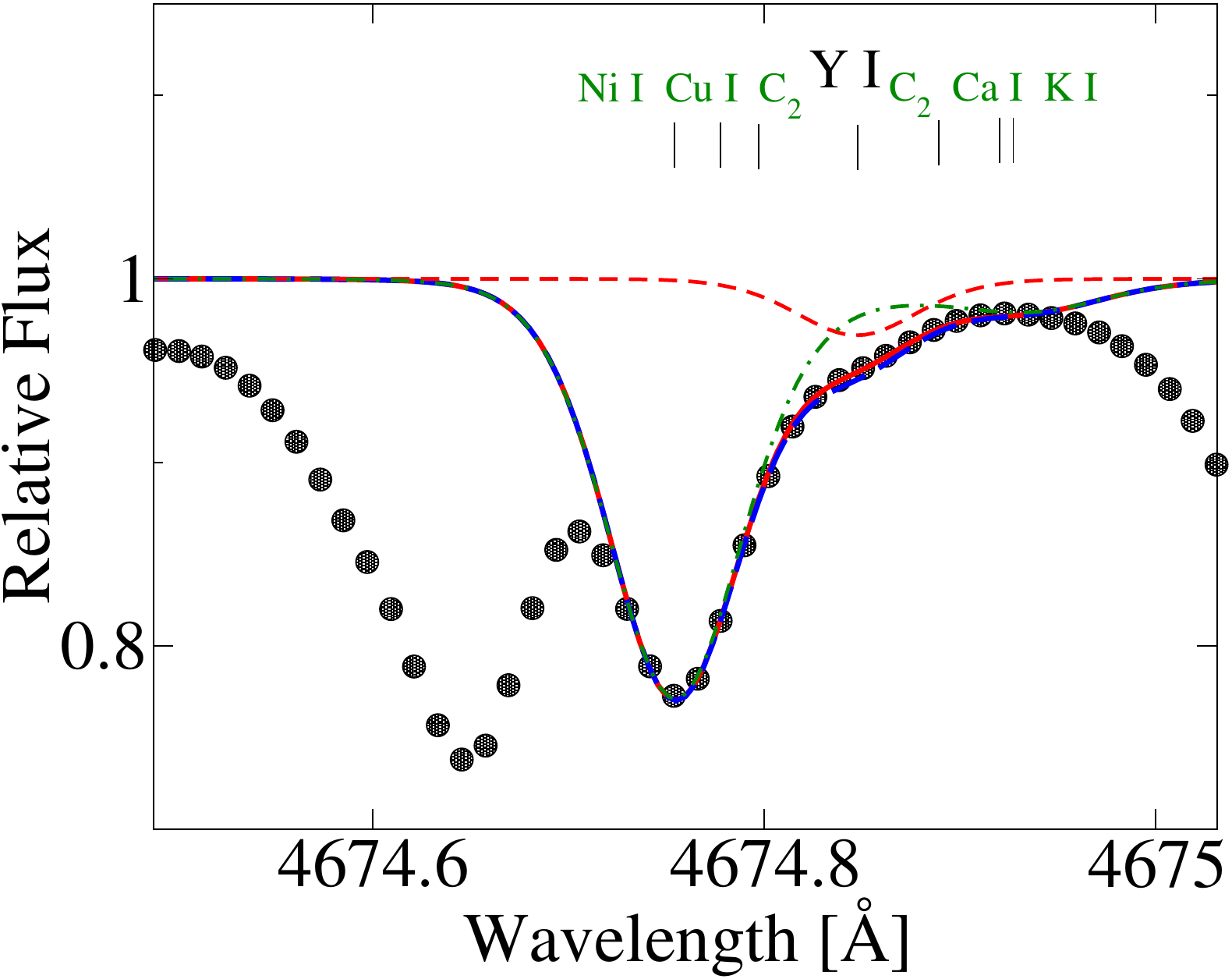}\\
  \centering}
  \parbox{0.24\linewidth}{\includegraphics[scale=0.16]{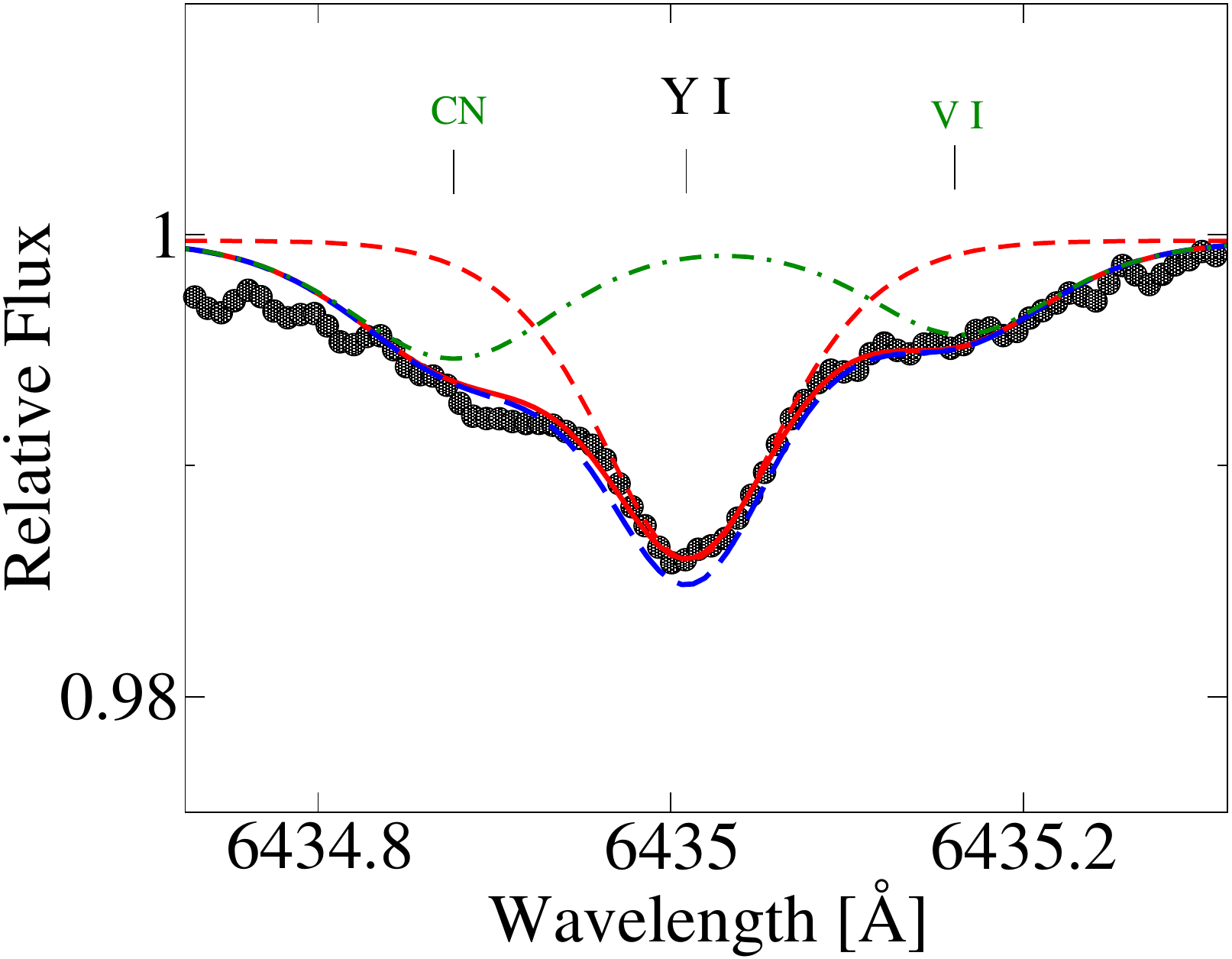}\\
  \centering}
  \parbox{0.24\linewidth}{\includegraphics[scale=0.16]{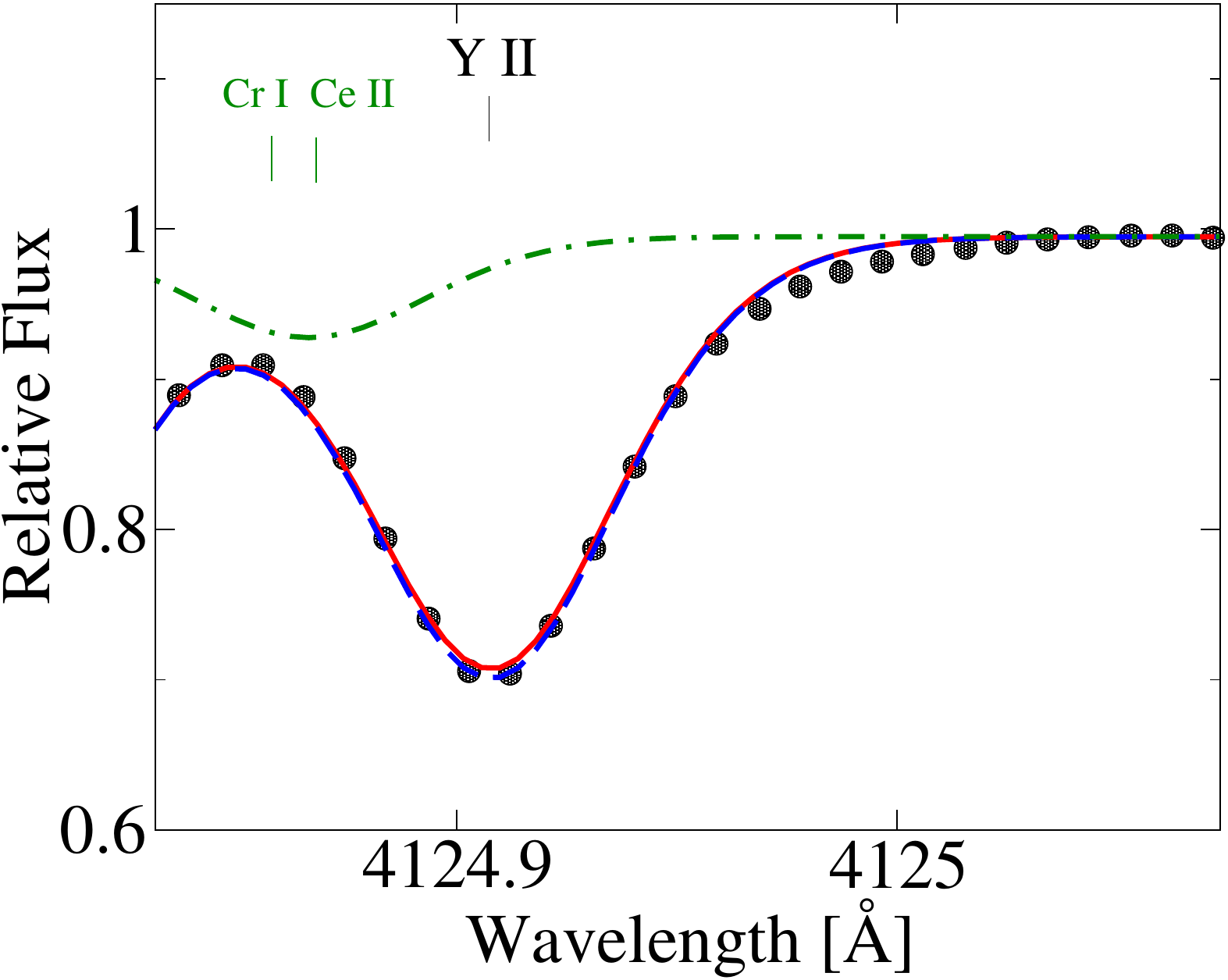}\\
  \centering}
  \hspace{1\linewidth}
  \hfill
  \\[0ex]
  \parbox{0.24\linewidth}{\includegraphics[scale=0.16]{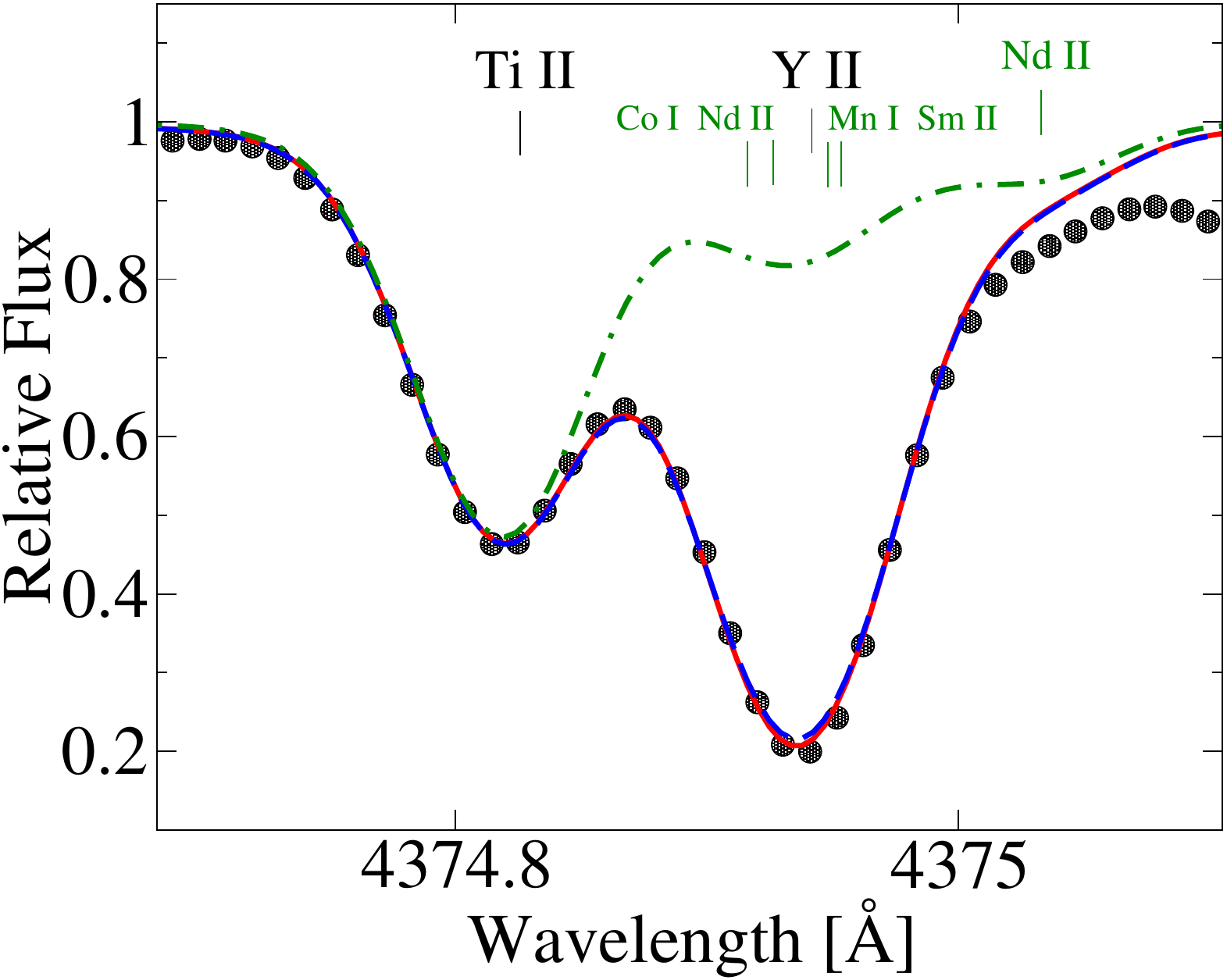}\\
  \centering}
  \parbox{0.24\linewidth}{\includegraphics[scale=0.16]{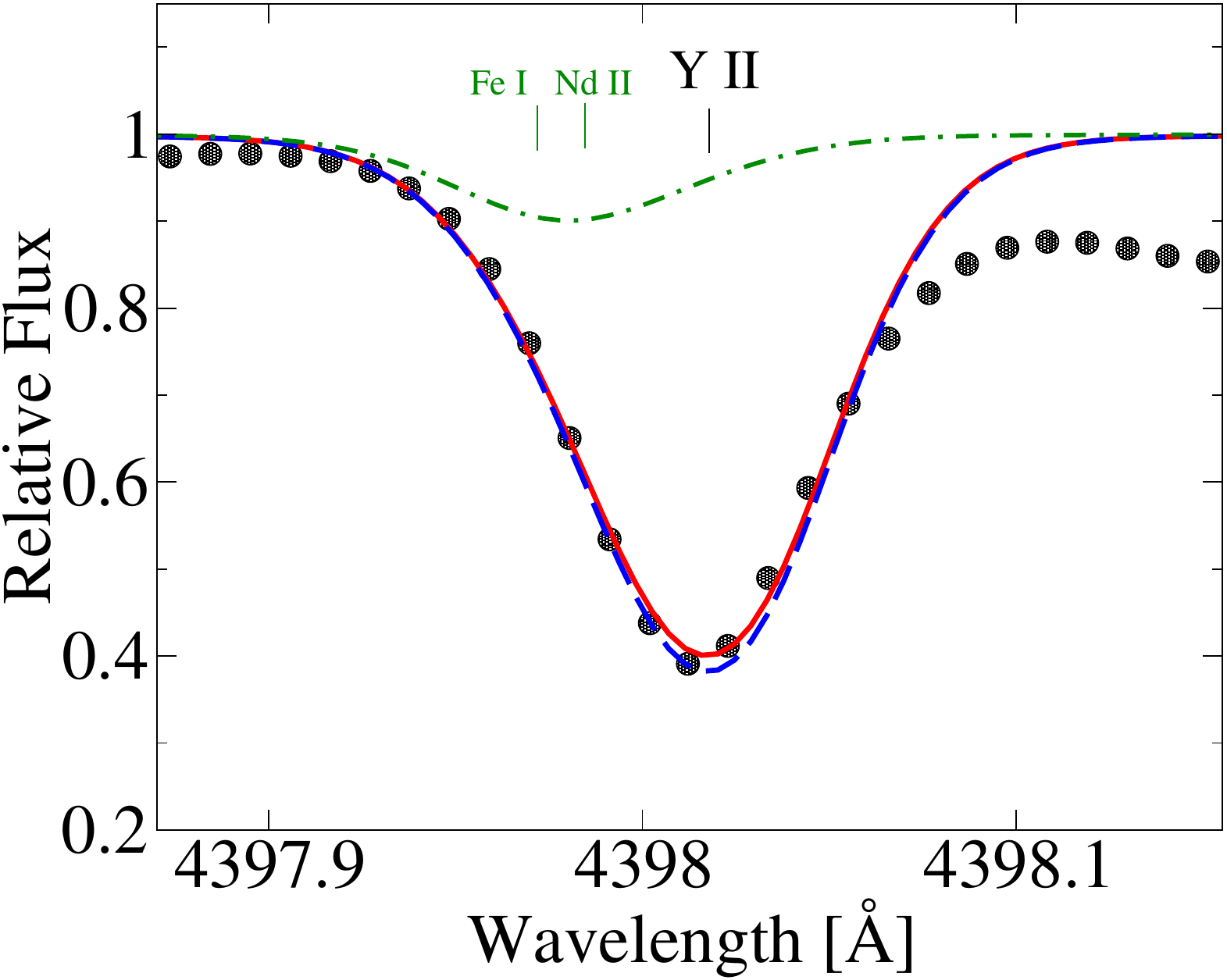}\\
  \centering}
  \parbox{0.24\linewidth}{\includegraphics[scale=0.16]{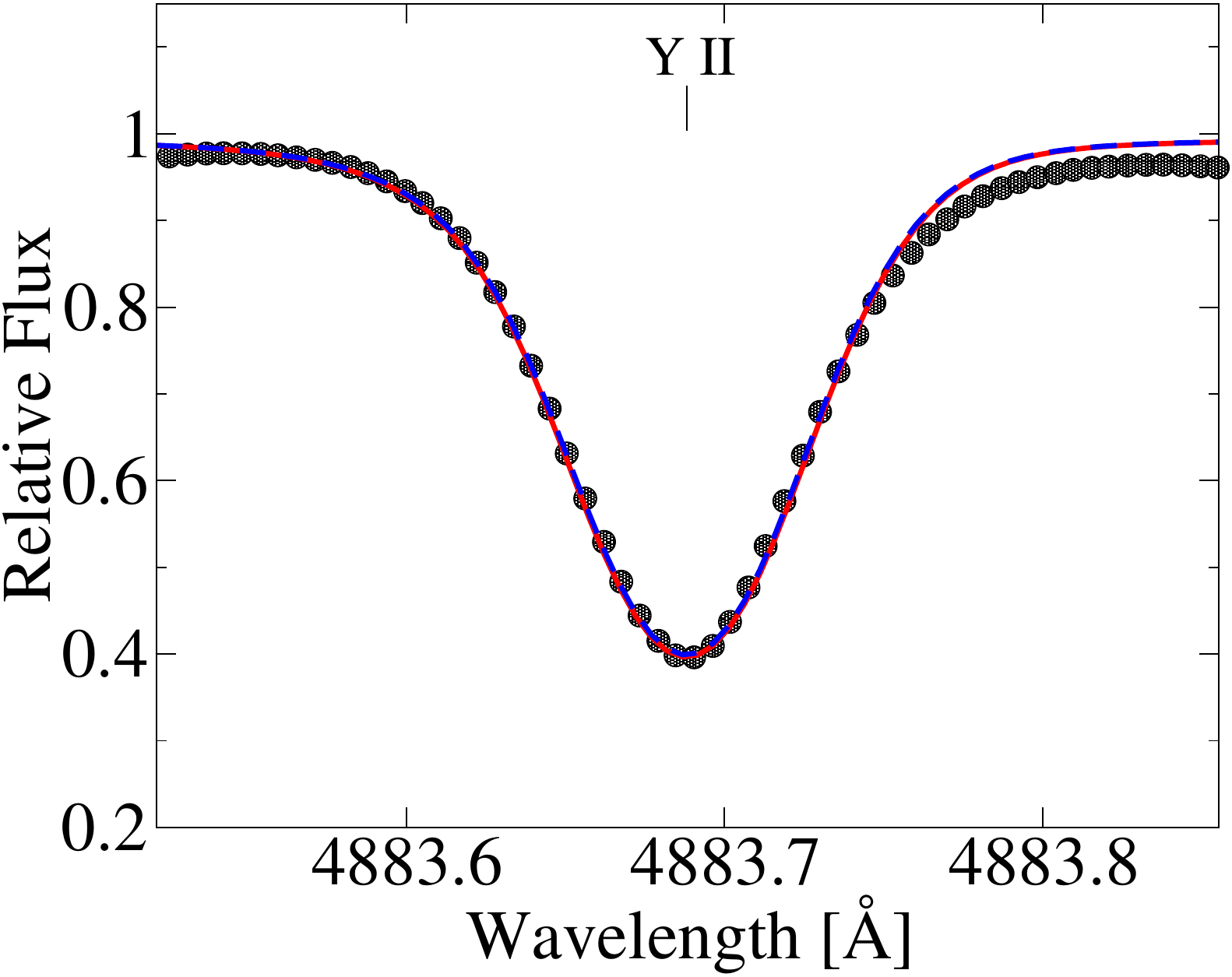}\\
  \centering}
  \parbox{0.24\linewidth}{\includegraphics[scale=0.16]{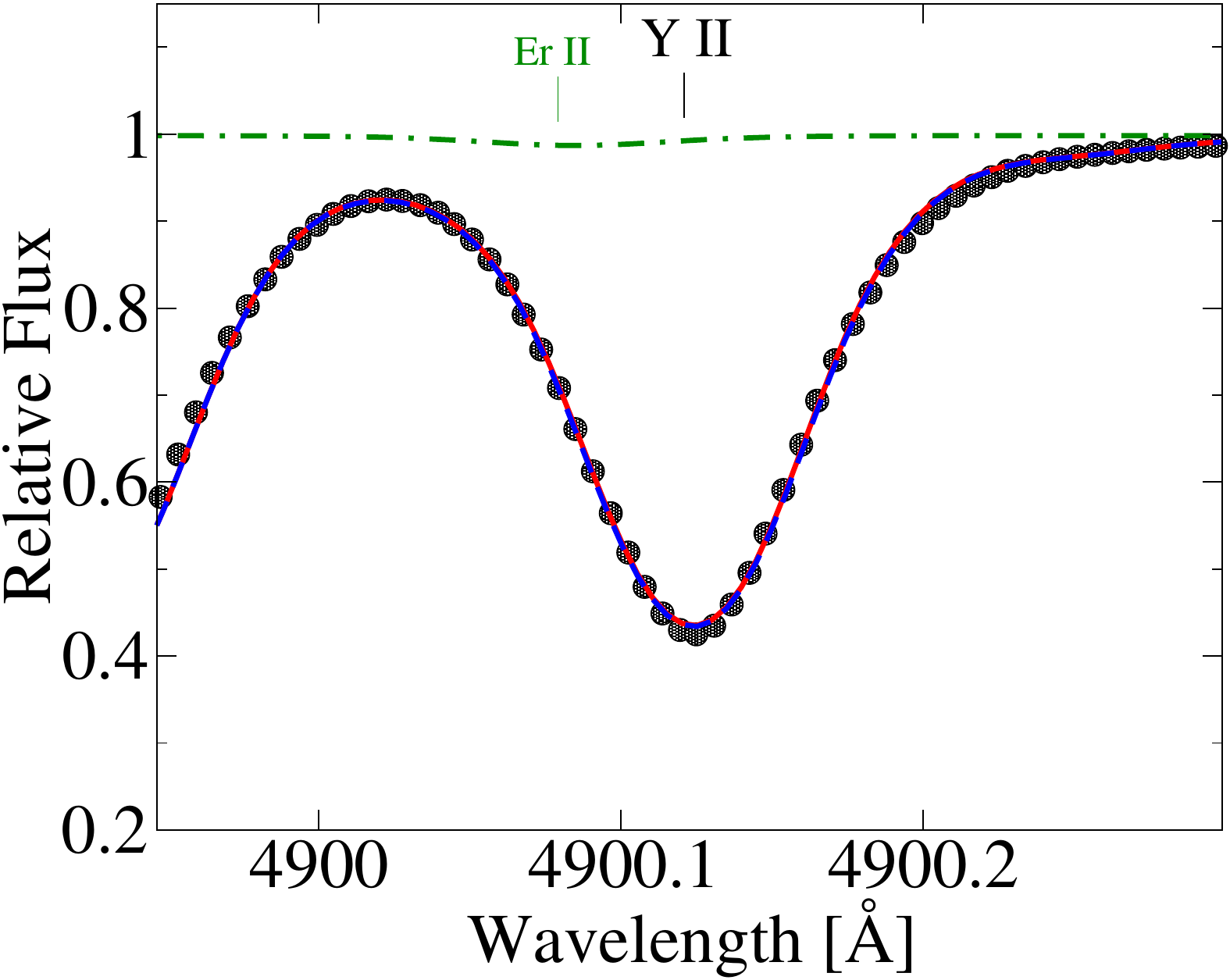}\\
  \centering}
  \hspace{1\linewidth}
  \hfill
  \\[0ex]
  \parbox{0.24\linewidth}{\includegraphics[scale=0.16]{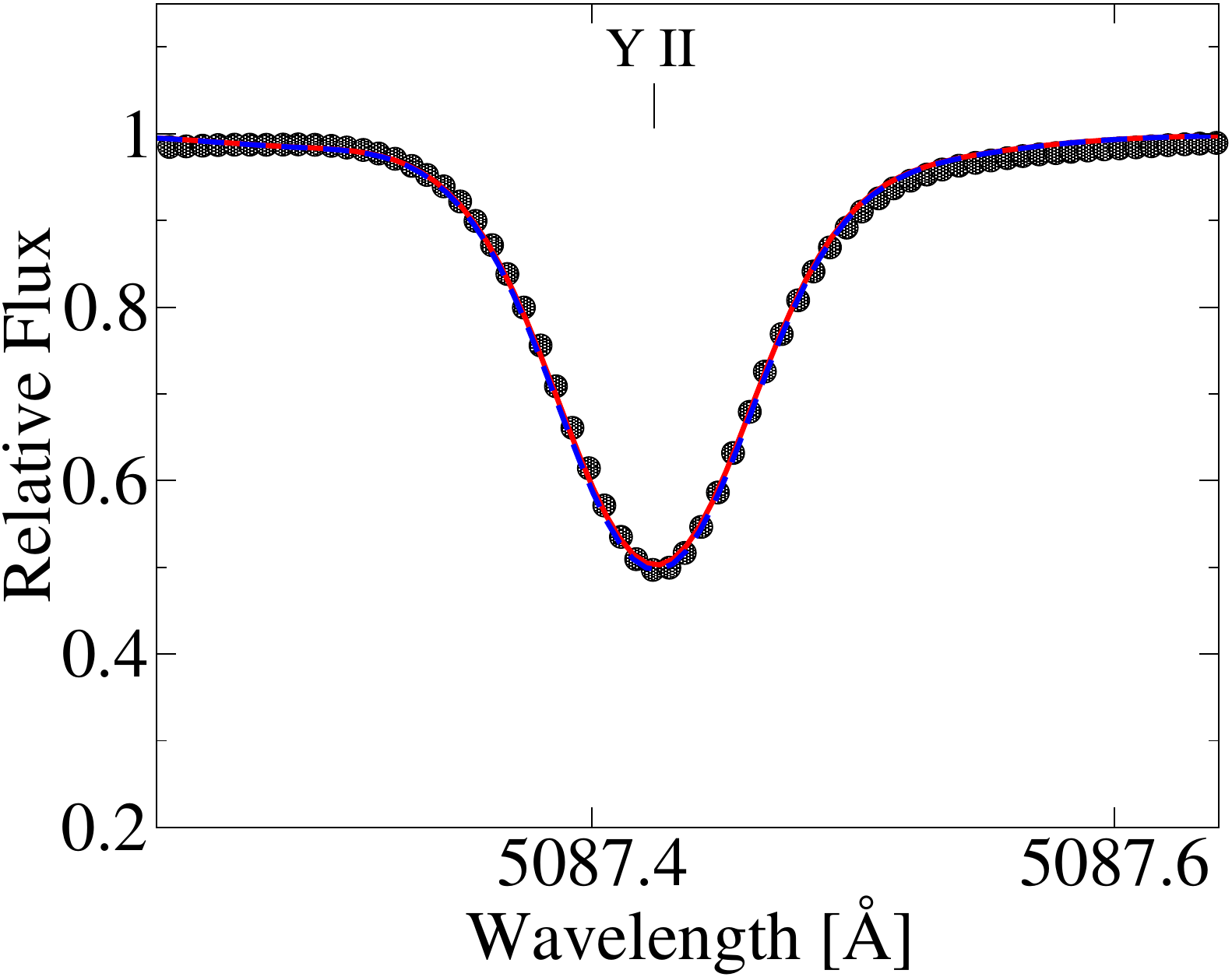}\\
  \centering}
  \parbox{0.24\linewidth}{\includegraphics[scale=0.16]{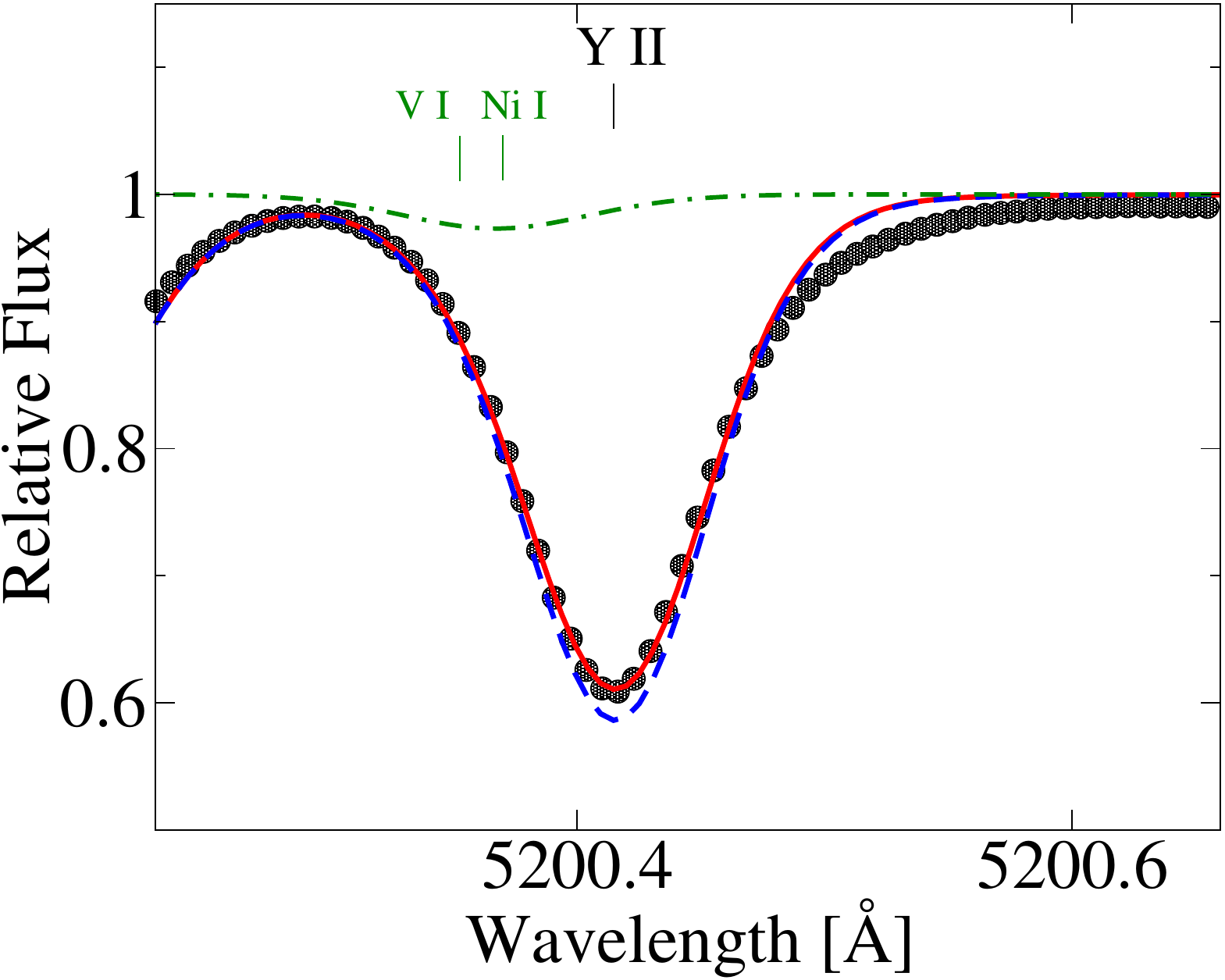}\\
  \centering}
  \parbox{0.24\linewidth}{\includegraphics[scale=0.16]{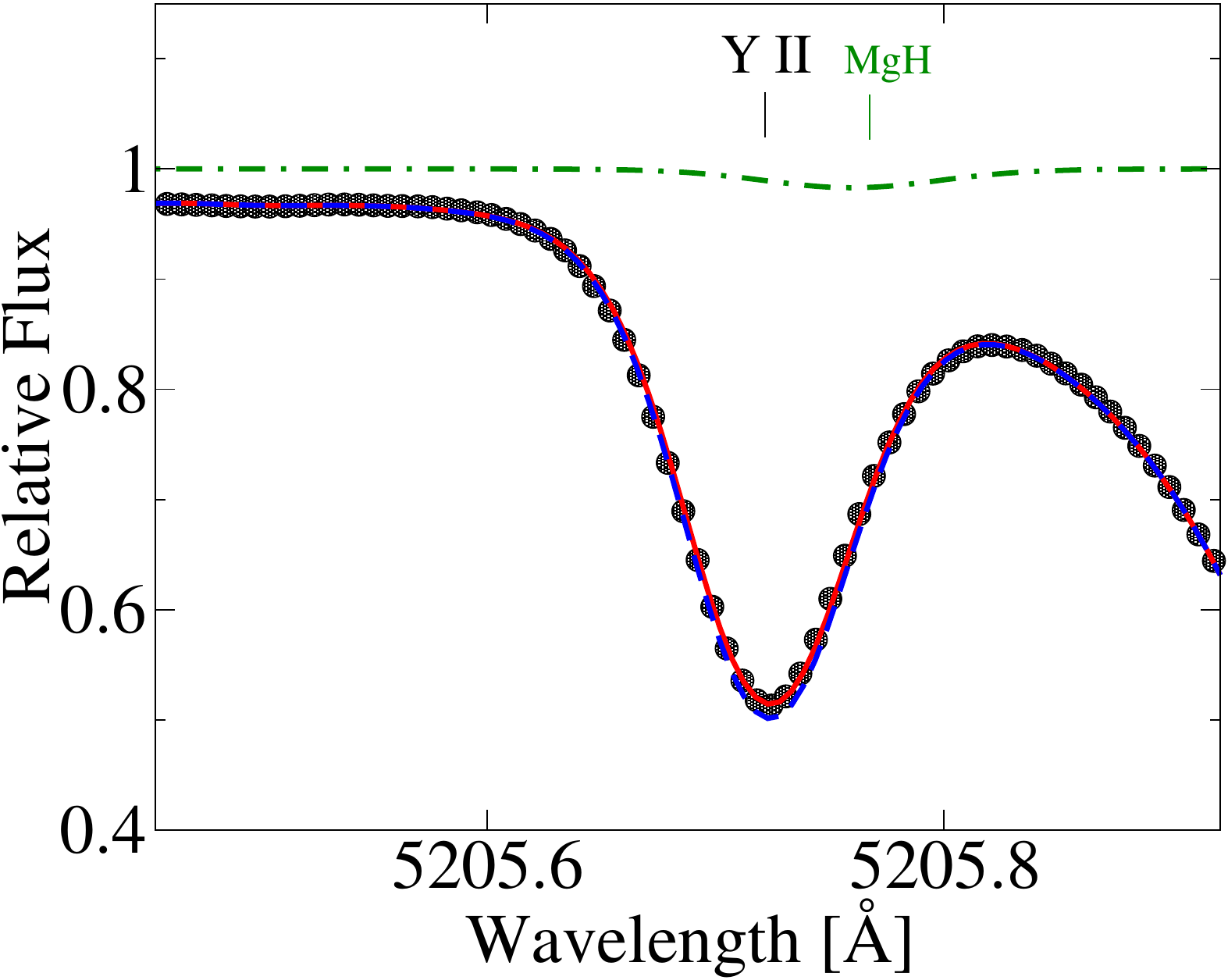}\\
  \centering}
  \parbox{0.24\linewidth}{\includegraphics[scale=0.16]{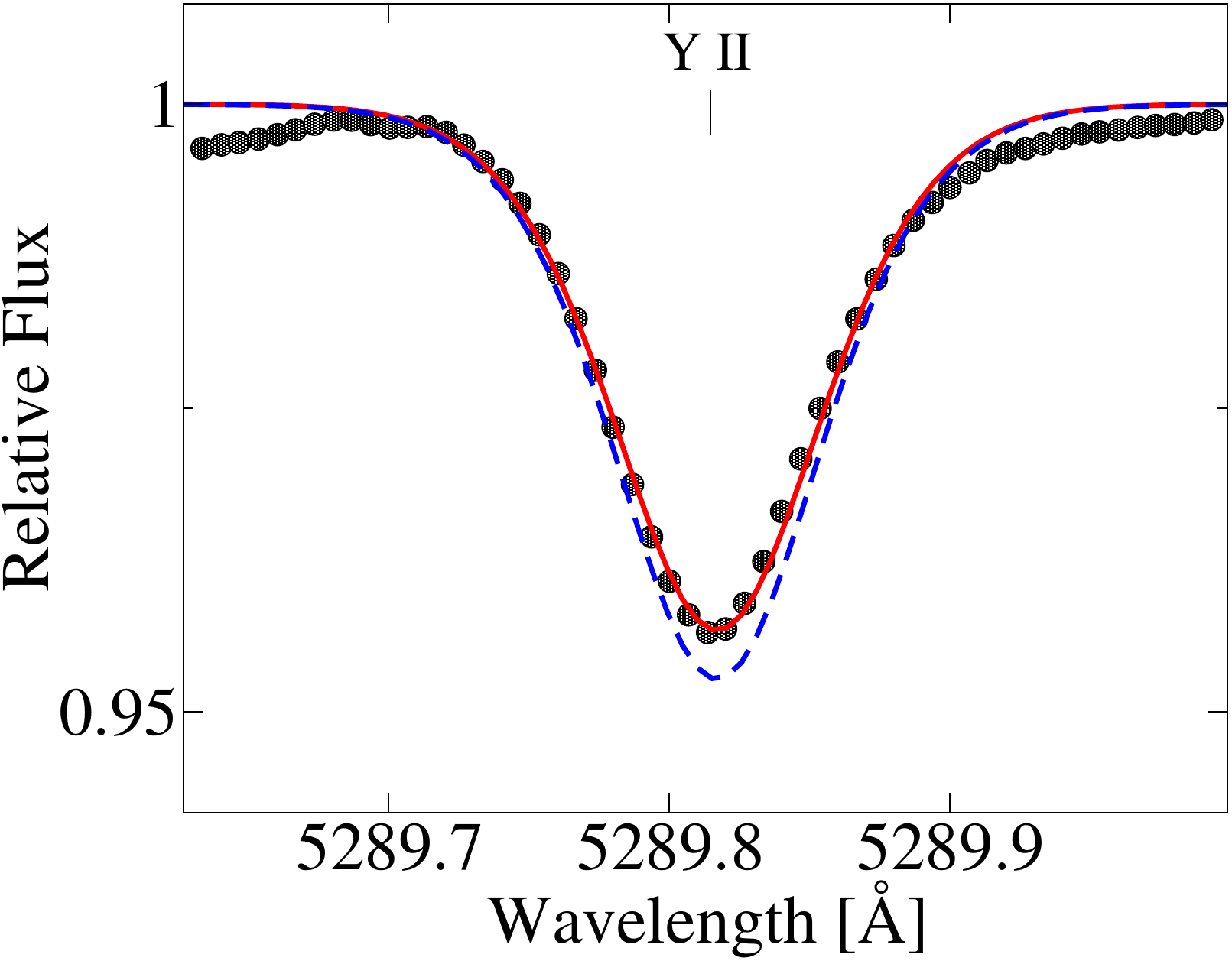}\\
  \centering}
  \hspace{1\linewidth}
  \hfill
  \\[0ex]
  \caption{Best NLTE fits (continuous red curve) to the solar flux line profiles (black circles) of Y\ione\ 4128, 4674, 6435~\AA\ and Y\ii\ 4124, 4374, 4398, 4883, 4900, 5087, 5200, 5205, 5289~\AA. For each line, the LTE profile (dashed blue curve) was computed with the yttrium abundance obtained from the NLTE analysis.
   }
  \label{pics}
  \end{minipage}
  \end{figure*}

As a first application of the presented model atom, we derive the solar yttrium abundance from lines of Y\ione\ and Y\ii. We used the MARCS model atmosphere 5777/4.44/0 and a depth-independent microturbulence of 0.9~\kms. The solar flux observations were taken from the Kitt Peak Solar Atlas \citep{1984sfat.book.....K}. The element abundance was determined from line-profile fitting. The theoretical flux profiles were convolved with a profile that combines a rotational broadening of 1.8~\kms\ and broadening by macroturbulence with a radial-tangential profile. For different lines of Y\ione\ and Y\ii, the most probable macroturbulence velocity, V$_{mac}$, was varied between 2 and 4~\kms. As a rule, the uncertainty in fitting the observed profile is less than 0.02~dex for weak lines and 0.03~dex for strong lines. 
The quality of the fits is illustrated in Fig.~\ref{pics} for the selected lines of Y\ione\ and Y\ii.
Individual line abundances are presented in Table~\ref{tab1}. 
We analyze the Y\ione\ 4128.304 and 4674.848~\AA\ lines but do not include them in the calculations of the mean abundances.  

Hereafter, the element abundances are given in the scale, where for hydrogen log~$\epsilon_{\rm H}$ = 12. In LTE, the abundance difference between Y\ione\ and Y\ii\ lines, $\Delta{\rm log~\epsilon}$(Y\ione\ - Y\ii), amounts to -0.12 dex. NLTE abundances from the Y\ione\ and Y\ii\ lines consistent within  0.07 dex.

Our 1D NLTE solar yttrium abundance obtained from Y\ii\ lines is 

\begin{equation}
{\rm log}~\epsilon_{\rm Y} = 2.21\pm0.05.
\end{equation}

The obtained result is in line with the 3D LTE result of \citet{2015A&A...573A..27G}, who obtained log~$\epsilon_{\rm Y}$ = 2.21$\pm$0.05. Our result is also consistent within the error bars with the meteoritic yttrium abundance, log~$\epsilon_{\rm Y}$ = 2.15$\pm$0.02 \citep{2021SSRv..217...44L}.

 \begin{deluxetable*}{lccccccccccc}
\tablecaption{Lines of Y\ione\ and Y\ii\ used in abundance analysis and solar yttrium abundances, log~$\epsilon$ (cols. 9 -- 10). \label{tab1}}
\tabletypesize{\scriptsize}
\tablehead{
\colhead{$\lambda$} & \colhead{Transition} & \colhead{g$_l$ - g$_u$} & \colhead{\Eexc} & \colhead{log~$gf$} & \colhead{Ref.} & \colhead{log $\gamma_{\rm r}$}   & \colhead{log $\gamma_{\rm S}/N_e$}  & \colhead{log $\gamma_{\rm vW}/N_H$}
  & \colhead{LTE} & \colhead{NLTE} & \colhead{$\Delta_{\rm NLTE}$} \\
\colhead{\AA\,}     & \colhead{}   & \colhead{}       & \colhead{eV}    & \colhead{}         & \colhead{}     &\colhead{rad s$^{-1}$cm$^3$ } & \colhead{rad s$^{-1}$cm$^3$}  & \colhead{rad s$^{-1}$cm$^3$} &  \colhead{} & \colhead{} & \colhead{}
}
\colnumbers
\startdata
 Y\ione\  &     &      &    &       &   &           &           &       &          &         \\
 4128.304$^*$ & a$^2$D  -- y$^2$D$^{\circ}$  & 6  -- 6 & 0.066  & 0.378  & 1   & 8.36  & -5.96 & -7.68 &  1.94 & 2.01 & 0.07  \\
 4643.698 & a$^2$D  -- z$^2$F$^{\circ}$  & 4  -- 6 & 0.000  & -0.46  & 1   & 7.55  & -6.02 & -7.72 &\nodata &\nodata &\nodata \\
 4674.848$^*$ & a$^2$D  -- z$^2$F$^{\circ}$  & 6  -- 8 & 0.066  & -0.46  & 1   & 7.38  & -6.01 & -7.72 & 2.02 & 2.10 & 0.08  \\
 5630.138 & a$^4$F  -- z$^4$G$^{\circ}$  & 4  -- 6 & 1.356  &  0.11  & 2   & 7.83  & -5.85 & -7.76 &\nodata &\nodata &\nodata\\
 6023.406 & a$^2$D  -- z$^4$D$^{\circ}$  & 4  -- 4 & 0.000  & -0.99  & 3   & 6.40  & -6.02 & -7.73 &\nodata &\nodata &\nodata\\
 6222.578 & a$^2$D  -- z$^2$D$^{\circ}$  & 4  -- 6 & 0.000  & -1.74  & 3   & 6.96  & -5.97 & -7.74 &\nodata &\nodata &\nodata\\
 6402.007 & a$^2$D  -- z$^2$D$^{\circ}$  & 6  -- 4 & 0.066  & -1.9   & 3   & 6.99  & -5.27 & -7.74 &\nodata &\nodata &\nodata\\
 6435.022 & a$^2$D  -- z$^2$D$^{\circ}$  & 6  -- 6 & 0.066  & -0.82  & 1   & 6.96  & -5.97 & -7.74 & 2.08 & 2.14 & 0.06  \\
 6437.168 & a$^2$G  -- x$^2$F$^{\circ}$  & 10 -- 8 & 2.294  & -0.62  & 1   & 8.24  & -5.85 & -7.63 &\nodata &\nodata & \nodata\\
 Mean     &                                                            &          &        &        &     &       &       &       &  2.08 &  2.14 & \\
$\sigma$  &                                                               &          &        &        &     &       &       &       &  0.03 & 0.03 & \\
   Y\ii\  &                                                            &          &        &        &     &       &       &       &      &      &       \\
 3788.693 & a$^3$D  -- z$^3$F$^{\circ}$ & 3 -- 5  & 0.104  & -0.11  & 4   & 8.30  & -6.23 & -7.80 & 2.13  & 2.13 &  0.00  \\
 3950.349 & a$^3$D  -- z$^1$D$^{\circ}$ & 3 -- 5  & 0.104  & -0.45  & 4   & 8.29  & -6.23 & -7.80 & \nodata &\nodata &\nodata \\
 4124.904 & a$^1$D  -- z$^3$F$^{\circ}$ & 5 -- 7  & 0.409  & -1.57  & 4   & 8.29  & -6.26 & -7.81 & 2.16  & 2.17 &  0.01  \\
 4177.528 & a$^1$D  -- z$^3$F$^{\circ}$ & 5 -- 5  & 0.409  & -0.07  & 4   & 8.30  & -6.26 & -7.80 & \nodata &\nodata &\nodata \\
 4235.727 & a$^3$D  -- z$^3$P$^{\circ}$ & 5 -- 5  & 0.130  & -1.59  & 4   & 7.36  & -6.23 & -7.84 & \nodata &\nodata &\nodata \\
 4374.933 & a$^1$D  -- z$^1$D$^{\circ}$ & 5 -- 5  & 0.409  & 0.09   & 4   & 8.29  & -6.26 & -7.80 & 2.26  & 2.25 &  -0.01 \\
 4398.008 & a$^3$D  -- z$^3$P$^{\circ}$ & 5 -- 3  & 0.130  & -1.1   & 4   & 7.44  & -6.23 & -7.83 & 2.22  & 2.25 &  0.03  \\
 4682.321 & a$^1$D  -- z$^3$P$^{\circ}$ & 5 -- 5  & 0.409  & -1.5   & 4   & 7.36  & -6.26 & -7.84 & \nodata &\nodata &\nodata \\
 4823.304 & a$^3$F  -- z$^3$D$^{\circ}$ & 5 -- 5  & 0.992  & -1.09  & 4   & 8.46  & -6.40 & -7.76 & \nodata &\nodata &\nodata \\
 4854.861 & a$^3$F  -- z$^3$D$^{\circ}$ & 5 -- 3  & 0.992  & -0.29  & 4   & 8.46  & -6.40 & -7.76 & \nodata &\nodata &\nodata \\
 4883.682 & a$^3$F  -- z$^3$D$^{\circ}$ & 9 -- 7  & 1.084  & 0.1    & 4   & 8.47  & -6.40 & -7.76 & 2.18  & 2.18 &  0.00  \\
 4900.118 & a$^3$F  -- z$^3$D$^{\circ}$ & 7 -- 5  & 1.033  & -0.07  & 4   & 8.46  & -6.40 & -7.76 & 2.27  & 2.27 &  0.00  \\
 4982.128 & a$^3$F  -- z$^3$F$^{\circ}$ & 7 -- 9  & 1.033  & -1.36  & 4   & 8.31  & -6.40 & -7.81 & 2.18  & 2.19 &  0.01  \\
 5087.418 & a$^3$F  -- z$^3$F$^{\circ}$ & 9 -- 9  & 1.084  & -0.19  & 4   & 8.31  & -6.40 & -7.81 & 2.19  & 2.20 &  0.01  \\
 5119.110 & a$^3$F  -- z$^3$F$^{\circ}$ & 5 -- 7  & 0.992  & -1.39  & 4   & 8.29  & -6.40 & -7.81 & 2.21  & 2.23 &  0.02  \\
 5123.209 & a$^3$F  -- z$^1$P$^{\circ}$ & 5 -- 3  & 0.992  & -1.2   & 4   & 8.30  & -6.40 & -7.78 & \nodata &\nodata &\nodata \\
 5200.409 & a$^3$F  -- z$^3$F$^{\circ}$ & 5 -- 5  & 0.992  & -0.67  & 4   & 8.30  & -6.40 & -7.80 & 2.23  & 2.26 &  0.03  \\
 5205.722 & a$^3$F  -- z$^3$F$^{\circ}$ & 7 -- 7  & 1.033  & -0.34  & 4   & 8.29  & -6.40 & -7.81 & 2.15  & 2.17 &  0.02  \\
 5289.815 & a$^3$F  -- z$^3$F$^{\circ}$ & 7 -- 5  & 1.033  & -1.78  & 4   & 8.30  & -6.40 & -7.80 & 2.13  & 2.16 &  0.03  \\
 5402.773 & b$^1$D  -- z$^1$F$^{\circ}$ & 5 -- 7  & 1.839  & -0.35  & 4   & 8.33  & -6.34 & -7.74 & \nodata &\nodata &\nodata \\
 5473.384 & a$^3$P  -- y$^3$P$^{\circ}$ & 3 -- 5  & 1.738  & -0.93  & 4   & 8.64  & -6.36 & -7.75 & 2.30  & 2.33 &  0.03  \\
 5521.562 & a$^3$P  -- y$^3$P$^{\circ}$ & 3 -- 3  & 1.738  & -1.1   & 4   & 8.64  & -6.36 & -7.75 & 2.24  & 2.25 &  0.01  \\
 5544.610 & a$^3$P  -- y$^3$P$^{\circ}$ & 3 -- 1  & 1.738  & -0.99  & 4   & 8.64  & -6.36 & -7.75 & \nodata &\nodata &\nodata \\
 5546.008 & a$^3$P  -- y$^3$P$^{\circ}$ & 5 -- 3  & 1.748  & -0.94  & 4   & 8.64  & -6.36 & -7.75 & \nodata &\nodata &\nodata \\
 6613.731 & a$^3$P  -- z$^3$D$^{\circ}$ & 5 -- 7  & 1.748  & -0.99  & 4   & 8.47  & -6.36 & -7.76 & \nodata &\nodata &\nodata \\
 6795.415 & a$^3$P  -- z$^3$D$^{\circ}$ & 3 -- 5  & 1.738  & -1.22  & 4   & 8.46  & -6.36 & -7.76 & 2.15  & 2.17 &  0.02  \\
 7881.878 & b$^1$D  -- z$^1$P$^{\circ}$ & 5 -- 3  & 1.839  & -0.49  & 4   & 8.30  & -6.34 & -7.78 & \nodata &\nodata &\nodata  \\
 Mean     &                                                                &      &        &        &     &       &       &       & 2.20 & 2.21 &        \\
$\sigma$  &                                                                &      &        &        &     &       &       &       & 0.05 & 0.05 &        \\ \hline
\multicolumn2l{ Y\ione\ - Y\ii\ }                                          &       &        &        &     &       &       &       &  -0.12 &  -0.07 &        \\ \hline
\enddata
\tablecomments{ {\bf References:} (1) \citet{NIST_ASD}; (2) \citet{2015JPhB...48h5001S}; (3) \citet{2018JOSAB..35.2244L}; (4) \citet{2017MNRAS.471..532P}. Lines and abundances, which were not used in mean calculations are marked by ($^*$). The fitting error (0.03~dex) for Y\ione\ 6435.022~\AA\ line was adopted as $\sigma$. For Y\ii\ lines, $\sigma$ means standard deviation.}
\end{deluxetable*}

\section{Yttrium abundances of the selected stars}\label{Sect:Stars}

\subsection{Stellar Sample, Observations, and Stellar Parameters}

Our sample includes 11 bright stars with well-known atmospheric parameters  (Table~\ref{tab_param}). All of them were established as $Gaia$ FGK benchmark stars with a range of different metallicities and have been used as reference or test objects in previous spectroscopic investigations and as targets of asteroseismological campaigns. The sample includes three FGK dwarfs: HD~22049, Procyon, and HD~49933, three K giants: Aldebaran (HD~29139), Arcturus (HD~124897), and Pollux (HD~62509), three FG dwarfs with low metallicity: HD~22879, HD~84937, and HD~140283, and two metal-poor giants: HD~122563, HD~220009. Their V magnitudes range from -0.05 to 8.3. Most of the stars are close to the Sun, and their parameters should be fairly accurate. Below we comment brieﬂy on some
individual stars.

HD~49933 is a solar-type star, with V~sin$i\sim6$~\kms. The physical parameters of HD~49933 were obtained previously by different authors with different methods. However, they display some scatter both in \Teff\ and log~$g$. Among the recent determinations, one of the lowest values of \Teff(K)/log~$g$ = 6476/4.18 were determined by \citet{2020AJ....160..120J}, while the highest ones, 6727/4.50 were presented in \citet{2020A&A...636A..85S}. We adopted \Teff/log~$g$ = 6635/4.21, obtained from angular diameter measurements and bolometric fluxes and from a homogeneous mass determination based on stellar evolution models \citep{2015AA...582A..49H}. 

Procyon is a known spectroscopic binary system. The proximity of Procyon to Earth makes it possible to measure directly its angular diameter. We adopted \Teff=6590 K and log~$g$=3.95 
calculated as average values from eleven sets of parameters in \citet{2016MNRAS.456.1221R}.

HD~22879 is a metal poor high-velocity star, with V~sin$i\sim4.2$~\kms\ \citep{2019A&A...624A..19B}, classified by \citet{2001AJ....121.2148G} as G0 V mF2, because the metallic line spectrum is approximately the strength of the F2 V standard. We adopted \Teff(K)/log~$g$/[Fe/H]=5893/4.45/-0.86 from spectroscopic analysis of \citet{2020AA...639A.127C}.

HD~220009 is a K subgiant with high proper motion and rotation velocity V~sin$i\sim10$~\kms\ \citep{1970CoAsi.239....1B}.  We adopted \Teff/log~$g$/[Fe/H] = 4266/1.43/-0.74 from \citet{2015AA...582A..49H, 2014AA...564A.133J}, whose parameters are in agreement with other studies, where stellar atmospheric parameters were derived purely spectroscopically (e.g. \citet{2020AJ....160...83S}). 

The physical parameters for three K giants: Aldebaran (HD~29139), Arcturus (HD~124897), and Pollux (HD~62509) were taken from the sample of {\it Gaia} FGK benchmark stars \citep{2015AA...582A..49H}, and the NLTE [Fe/H] values for these stars were taken from \citet{2014AA...564A.133J}.

Three metal-poor stars in the -2.55$\leq$[Fe/H]$\leq$-2.12 metallicity range, HD~84937, HD~122563, and HD~140283, were used for our analysis. 
Their stellar parameters are based on numerous photometric measurements and accurate distances, that is why they are traditionally featured in many NLTE studies, e.g. \citet{2015MNRAS.453.1619A, 2016MNRAS.461.1000S, 2017A&A...605A..53M, 2018ApJ...866..153A, 2019A&A...631A..43M, 2022MNRAS.515.1510S}.

Being a subject of scientific interest and also popular science movies for the last sixty years, HD~22049 ($\epsilon$ Eri) is the closest solar-type star to the Sun at a distance of 3.220$\pm$0.001~pc \citep{2021A&A...649A...1G} with a mass of 0.82$\pm$0.05 M$\odot$ \citep{2012ApJ...748...72B}. $\epsilon$ Eri is a young (360--720 Myr; \citet{2015A&A...574A.120J}) K2~V star, which displays relatively high stellar variability and stellar activity, see e.g. \citet{1995ApJ...441..436G, 2018ApJ...857..133B, 2020A&A...636A..49C, 2021AJ....161..272H}. This single star hosts at least one exoplanet with an estimated mass 1.55$\pm$0.24 times that of Jupiter discovered by \citet{2000ApJ...544L.145H}.

The spectrum of Arcturus in the visible spectral range (3772$-$7900~\AA) was obtained in 2019/06/05 (under programm ID 0103.D-0118(A)) with the Echelle SPectrograph for Rocky Exoplanets and Stable Spectroscopic Observations (ESPRESSO) spectrograph installed at the Combined-Coude Laboratory of the VLT and linked to the four 8.2 m Unit Telescopes \citep{2021A&A...645A..96P}. The spectra of the remaining stars in the visible spectral range (3781$-$6912~\AA) were obtained with the High-Accuracy Radial velocity Planetary Searcher (HARPS) echelle spectrograph installed on ESO’s 3.6-m Telescope at La Silla \citep{2003Msngr.114...20M}. The extracted and wavelength calibrated spectra were taken from ESO Archive\footnote{http://archive.eso.org/eso/eso$\_$archive$\_$main.html}. All spectra were obtained with a high spectral resolving power, $R$, and high signal-to-noise ratio (S/N; from 186 to 424). The characteristics of the observed spectra for individual stars are given in Table~\ref{tab_obs}. The normalisation of stellar spectra to their continuum was performed with the \textsc{RASSINE} $Python$ code \citep{2020A&A...640A..42C}.

 \begin{deluxetable*}{lccccccccc}
\tablecaption{Atmospheric parameters of the reference stars and sources of the data. \label{tab_param}}
\tablewidth{0pt}
\tablehead{
\colhead{Star} & \colhead{Name} & \colhead{Sp. T.} & \colhead{ \Teff } & \colhead{log$g$} & \colhead{[Fe/H]}  & \colhead{ $\xi_t$ }  & \colhead{Ref.} & \colhead{[Y/Fe]$_{\rm LTE}$} & \colhead{[Y/Fe]$_{\rm NLTE}$} \\
\colhead{} & \colhead{} & \colhead{} & \colhead{ K} & \colhead{CGS} & \colhead{dex} & \colhead{ \kms }    & \colhead{}  & \colhead{} & \colhead{}
}
\decimalcolnumbers
\startdata                                                              
 HD~22049   &$\epsilon$ Eri&  K2 V       & 5076 & 4.61  &  -0.09  & 1.1  & 1, 2  & -0.08$\pm$0.06  & -0.05$\pm$0.03   \\
 HD~22879   &             &   G0 V mF2   & 5893 & 4.45  &  -0.86  & 1.8  & 3     & 0.03$\pm$0.07   & 0.10$\pm$0.06    \\
 HD~29139   &  Aldebaran  &   K5+III     & 3930 & 1.11  &  -0.37  & 1.7  & 1     & -0.20$\pm$0.16  & -0.10$\pm$0.07   \\
 HD~49933   &             &   F3 V       & 6635 & 4.21  &  -0.41  & 1.9  & 1, 2  & 0.04$\pm$0.09   & 0.11$\pm$0.09    \\
 HD~61421   &  Procyon    &   F5 IV-V    & 6590 & 3.95  &  -0.02  & 1.8  & 4     & 0.04$\pm$0.04   & 0.07$\pm$0.04    \\
 HD~62509   &  Pollux     &   K0 IIIb    & 4860 & 2.9   &  0.13   & 1.5  & 1     & -0.18$\pm$0.13  & -0.17$\pm$0.13   \\
 HD~84937   &             &   F8 V       & 6350 & 4.09  &  -2.12  & 1.7  & 5     & -0.01$\pm$0.03  & 0.07$\pm$0.03    \\
 HD~122563  &             &   G8 III     & 4600 & 1.43  &  -2.55  & 1.6  & 6     & -0.40$\pm$0.08  & -0.28$\pm$0.07   \\
 HD~124897  &  Arcturus   &   K1.5 III   & 4290 & 1.6   &  -0.52  & 1.7  & 1     & -0.21$\pm$0.09  & -0.17$\pm$0.08   \\
 HD~140283  &             &   F9 V       & 5780 & 3.7   &  -2.46  & 1.6  & 5     & -0.42$\pm$0.04  & -0.32$\pm$0.04   \\
 HD~220009  &             &   K1 IV      & 4266 & 1.43  &  -0.74  & 1.3  & 1, 2  & -0.06$\pm$0.06  & -0.03$\pm$0.07    \\
\enddata                                                
\tablecomments{{\bf Note.} References: 
  (1) \citet{2015AA...582A..49H}; (2) \citet{2014AA...564A.133J}; (3) \citet{2020AA...639A.127C};
 (4) \citet{2016MNRAS.456.1221R}; (5) \citet{2015ApJ...808..148S}; (6) \citet{mash_fe}.
 Spectral types are extracted from the SIMBAD database. }
\end{deluxetable*}

\begin{deluxetable*}{lccccccccc}
\tablecaption{Characteristics of observed spectra of the reference stars. \label{tab_obs}}
\tablewidth{0pt}
\tablehead{
\colhead{Star}& \colhead{Name}& \colhead{V$^1$}&\colhead{Telescope/} & \colhead{Spectral range}& \colhead{$t_{exp}$}& \colhead{Observing run} &\colhead{$R$}& \colhead{$S/N$}  &\colhead{Run/Program ID}  \\
\colhead{}    & \colhead{}    & \colhead{mag}  &\colhead{Spectrograph }& \colhead{\AA\,}     & \colhead{s}        & \colhead{year/month/day}  &\colhead{} & \colhead{} & \colhead{}
}
\decimalcolnumbers
\startdata
  HD~22049    &$\epsilon$ Eri& 3.7    &  1 & 3781-6913  & 30   &  2019/12/01 &  115\,000  &  186.3 & 0104.C-0863(A)\\
  HD~22879    &              & 6.7    &  1 & 3781-6913  & 900  &  2005/10/30 &  115\,000  &  306.5 & 072.C-0488(E)\\
  HD~29139    &  Aldebaran   & 0.9    &  1 & 3781-6913  & 15   &  2007/10/22 &  115\,000  &  372.0 & 080.D-0347(A)\\
  HD~49933    &              & 5.8    &  1 & 3781-6913  & 279  &  2006/02/11 &  115\,000  &  234.8 & 076.C-0279(A)\\
  HD~61421    &  Procyon     & 0.4    &  1 & 3781-6913  & 10   &  2007/01/11 &  115\,000  &  301.8 & 078.D-0492(A) \\
  HD~62509    &  Pollux      & 1.1    &  1 & 3781-6913  & 8    &  2007/11/06 &  115\,000  &  257.6 & 080.D-0347(A) \\
  HD~84937    &              & 8.3    &  1 & 3781-6913  & 2700 &  2007/12/31 &  115\,000  &  216.5 & 080.D-0347(A) \\
  HD~122563   &              & 6.2    &  1 & 3781-6913  & 1500 &  2008/02/24 &  115\,000  &  318.5 & 080.D-0347(A) \\
  HD~124897   &  Arcturus    & $-$0.05&  2 & 3772-7900  & 15   &  2019/06/05 &  190\,000  &  424.5 & 0103.D-0118(A) \\
  HD~140283   &              & 7.2    &  1 & 3781-6913  & 2700 &  2008/03/06 &  115\,000  &  267.4 & 080.D-0347(A) \\
  HD~220009   &              & 5.1    &  1 & 3781-6913  & 300  &  2007/09/29 &  115\,000  &  332.2 & 080.D-0347(A) \\
\enddata
\tablecomments{{\bf Notes:} $^1$V is a visual magnitude from the SIMBAD data base.
Telescope/spectrograph: 1 = ESO-3.6/HARPS Echelle; 2 = ESO-VLT-U3/VIS.}
\end{deluxetable*}

\subsection{Analysis of Y\ione\ and Y\ii\ lines in the reference stars} 

\begin{figure*}
   \begin{minipage}{170mm}
 \begin{center}
 \parbox{0.45\linewidth}{\includegraphics[scale=0.2]{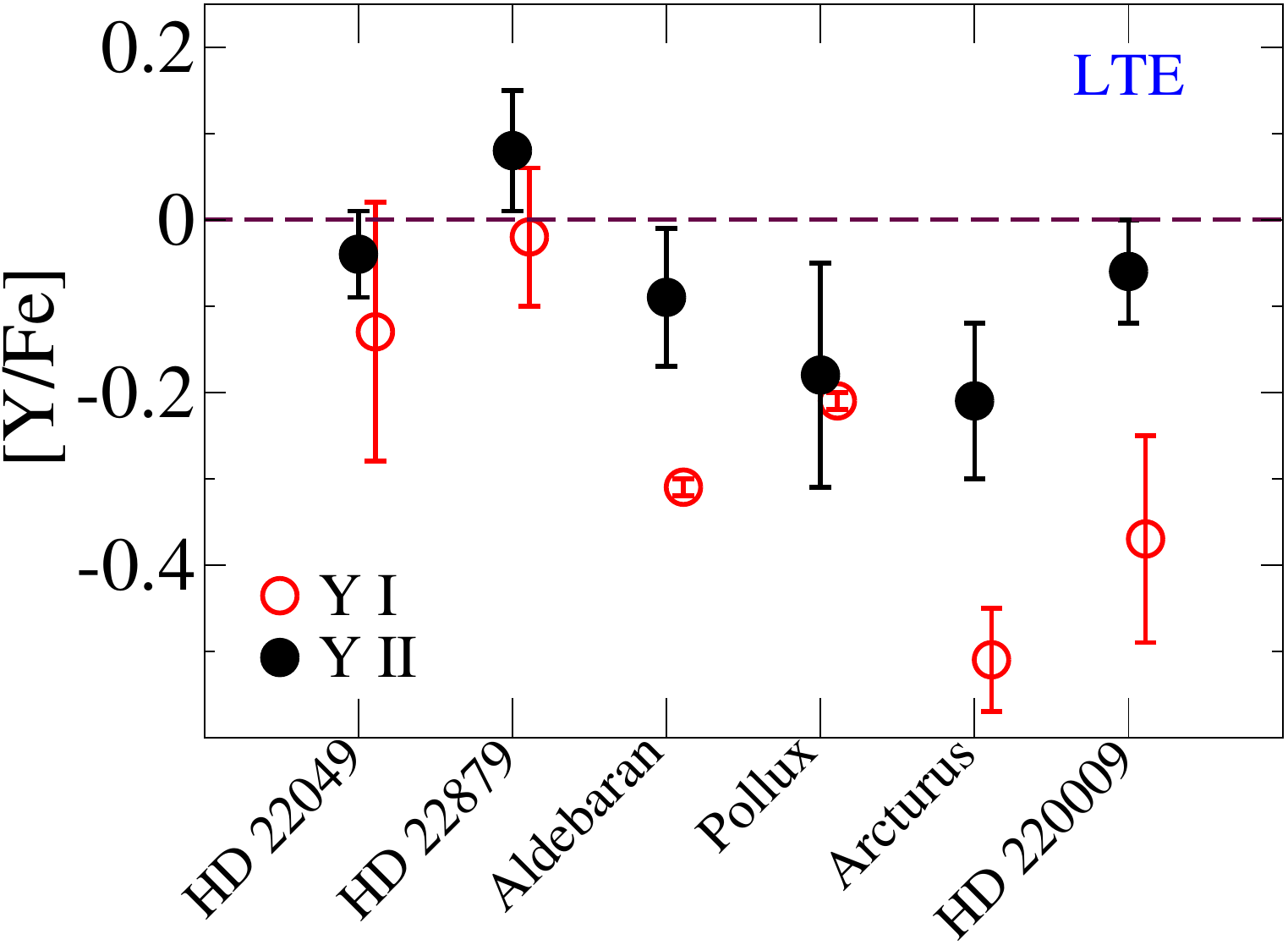}\\
 \centering}
 \parbox{0.45\linewidth}{\includegraphics[scale=0.2]{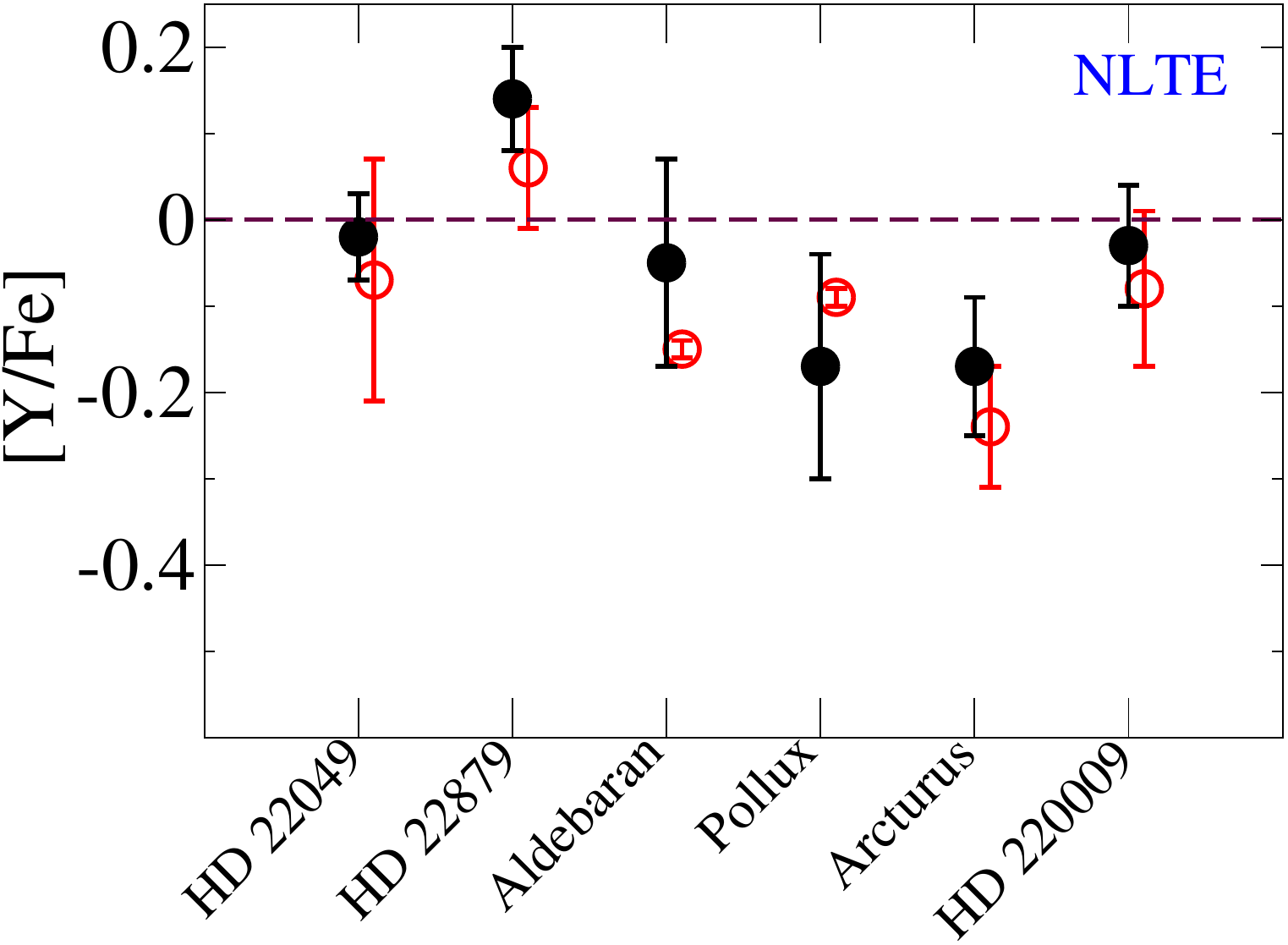}\\
 \centering}
 \hspace{1\linewidth}
 \hfill
 \\[0ex]
 \caption{[Y/Fe] LTE (left panel) and NLTE (right panel) ratios obtained from Y\ione\ and Y\ii\ lines. The error bars correspond to the dispersion in the single line measurements about the mean.}
 \label{Y_Fe}
 \end{center}
 \end{minipage}
 \end{figure*}

In this section, we derive the yttrium abundances of 11
reference stars using lines of Y\ione\ and Y\ii\ in the visible spectral range. The abundance results on the individual lines are presented in Table~\ref{tab_abn}. The derived
LTE and NLTE [Y/Fe] ratios are given in Table~\ref{tab_param}. 
For six stars where lines of both ionization stages are available, the NLTE abundances derived from the Y\ione\ and Y\ii\ lines agree within the error bars, while in LTE, the abundance differences can reach up to -0.31~dex. That is well illustrated by Figure~\ref{Y_Fe}.
For the most of the stars, NLTE for Y\ione\ gives a smaller line-to-line scatter compared to LTE (see Table~\ref{tab_abn}).

For the most of the stars, the obtained yttrium NLTE abundance is close to the solar value. The exceptions are two metal poor stars HD~12256 and HD~140283 with [Y/Fe] = -0.28~dex and -0.32~dex, respectively. Below we give detail information on individual stars.

{\it $\epsilon$ Eri.}  This is a solar-type star; therefore the yttrium  NLTE abundance, log$\epsilon$ = 2.05$\pm$0.14,  was derived from four Y\ione\ lines, and log$\epsilon$ = 2.10$\pm$0.05 from nine Y\ii\ lines. The NLTE abundance corrections are significant and positive for Y\ione\ lines at 4128, 4674, 6023, 6435~\AA, ($\Delta_{\rm NLTE}$ $\sim$0.06~dex), while for  Y\ii\ lines, they are slightly positive ($\Delta_{\rm NLTE}$ $\leq$ 0.03~dex). 

\begin{figure*}
  \begin{minipage}{170mm}
  \begin{center}
  \parbox{0.32\linewidth}{\includegraphics[scale=0.2]{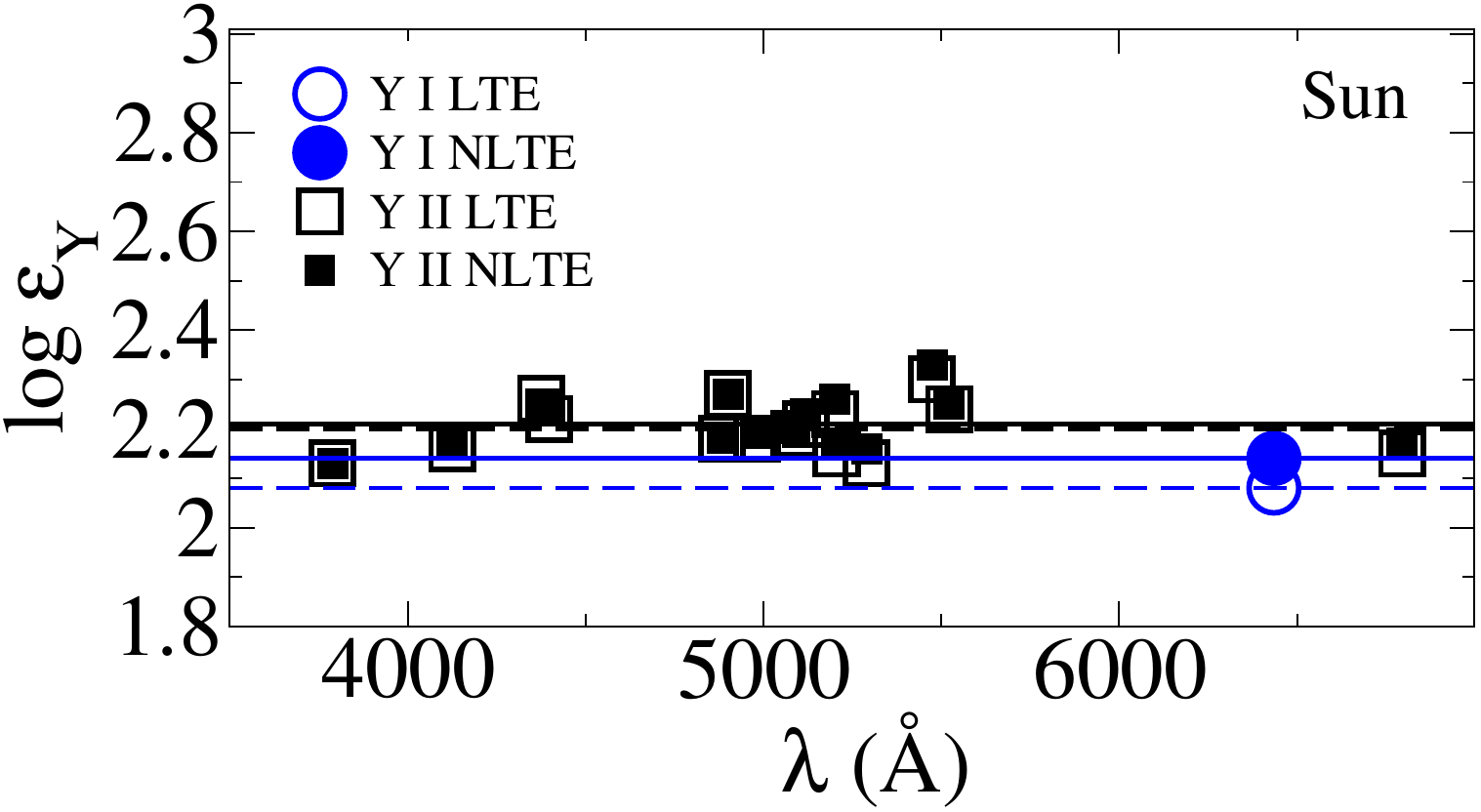}\\
  \centering}
  \parbox{0.32\linewidth}{\includegraphics[scale=0.2]{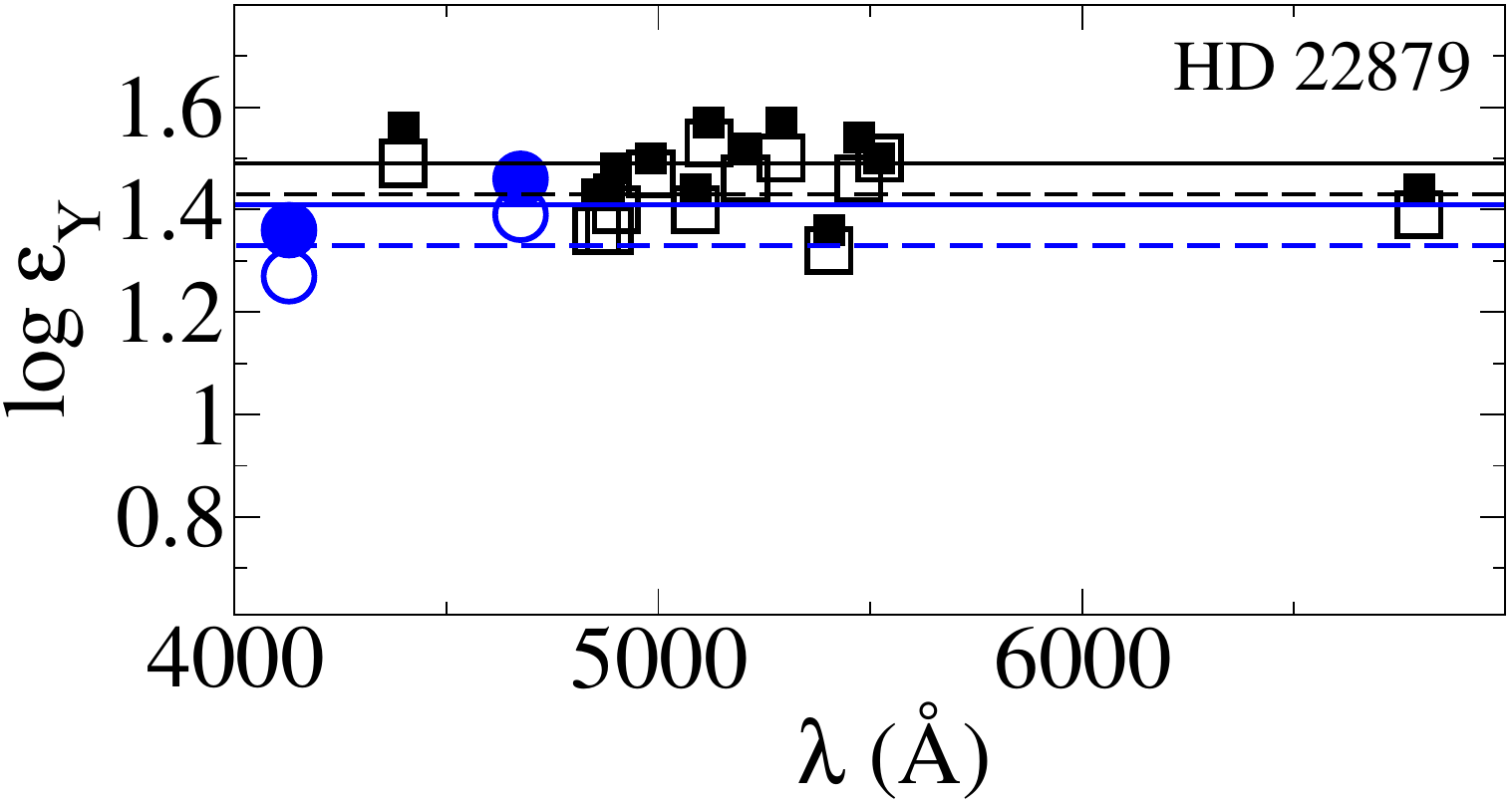}\\
  \centering}
  \parbox{0.32\linewidth}{\includegraphics[scale=0.2]{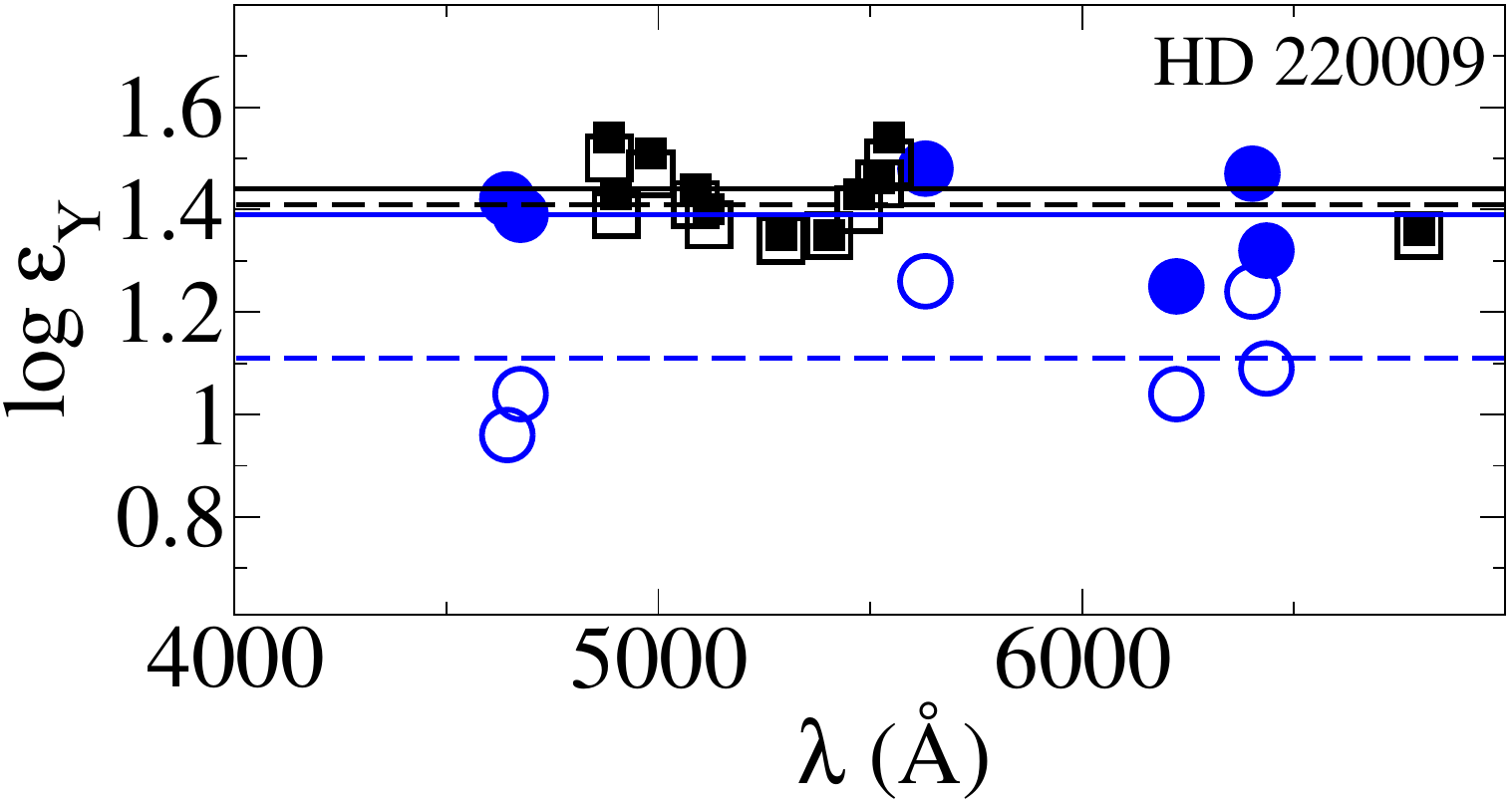}\\
  \centering}
  \hspace{1\linewidth}
  \hfill
  \\[0ex]
  \caption{Yttrium LTE (open symbols) and NLTE (filled symbols) abundances of few program stars derived from lines of Y\ione\ (circles) and Y\ii\ (squares) in wide spectral region. }
  \label{pics_equil}
    \end{center}
  \end{minipage}
  \end{figure*}

{\it HD~22879.} NLTE for both Y\ione\ and Y\ii\ gives a smaller line-to-line scatter compared to LTE (Figure~\ref{pics_equil}). 
NLTE leads to weakened both lines of Y\ione\ and Y\ii, and the obtained NLTE abundances from the two ionization stages are consistent within 0.08~dex. 

{\it Aldebaran.} Most of the Y\ione\ and Y\ii\ lines in the investigated spectral range are strongly contaminated with blending lines. 
Even though our line profile analysis allows to separate pure yttrium lines in the spectra, we limited with two lines of Y\ione\ and three lines of Y\ii\ only. Compared to solar-type atmosphere, the NLTE effects are stronger in Y\ione\ with $\Delta_{\rm NLTE}$ $\leq$0.16~dex, and weaker in Y\ii. The NLTE abundances from lines of the two ionization
stages, Y\ione\ and Y\ii, are consistent, however, we notice quite big scatter in both LTE and NLTE abundances for Y\ii\ lines.  

{\it HD~49933.} This is the hottest star in our sample; therefore the NLTE yttrium abundance, log$\epsilon$ = 1.91$\pm$0.09, was derived from the Y\ii\ lines only. Totally, ten lines were employed and all of them have slightly positive NLTE abundance corrections, from 0.00~dex to 0.12~dex. 

{\it Procyon.} The NLTE effects for the Y\ii\ lines are similar to those found for HD~49933, although they are slightly
smaller. We obtained [Y/Fe]$_{\rm NLTE}$ = 0.04$\pm$0.04 and [Y/Fe]$_{\rm LTE}$ = 0.07$\pm$0.04 from eighteen Y\ii\ line.

{\it Pollux.} The NLTE abundance corrections are from 0.0 to 0.02 for all investigated Y\ii\ lines. 
While, for Y\ione\ lines at 6402, and 6435~\AA, the $\Delta_{\rm NLTE}$ $=$ 0.11, and 0.12~dex, respectively. The obtained NLTE abundances from the two ionization stages are consistent within 0.08~dex (Figure~\ref{pics_equil}).

{\it HD~84937.} This is turn-off VMP star.
This star has similar physical parameters (\Teff\ and log$g$) as Procyon, but because of its low metallicity ([Fe/H] = -2.12), the NLTE effects for Y\ii\ lines are stronger compared to Procyon's atmosphere and less amount of lines are available for measurement. 
The NLTE yttrium abundance, averaged over 6 Y\ii\ lines, is log$\epsilon$ = 0.16$\pm$0.03 and [Y/Fe] = 0.07$\pm$0.03.

{\it HD~122563.} In the spectrum of this cool metal-poor giant, we found 12 Y\ii\ lines only, which can be used for yttrium abundance determination. The NLTE corrections are positive for all investigated lines and do not exceed 0.21~dex. The Y NLTE abundance, averaged over 12 Y\ii\ lines, is log$\epsilon$ = -0.62$\pm$0.07 and [Y/Fe] = -0.28$\pm$0.07 that is by 0.12~dex higher compared to LTE abundance.

{\it Arcturus.} The NLTE effects for both Y\ione\ and Y\ii\ lines are similar to those found for Aldebaran. The obtained
NLTE abundances from the two ionization stages are consistent within 0.07~dex. While the LTE abundance based on the Y\ione\ lines is 0.31~dex lower compared with that for Y\ii.

{\it HD~140283.} In the atmosphere of this VMP subgiant, only few Y\ii\ lines are avaliable.
From six Y\ii\ lines, we derived NLTE yttrium abundance -0.57$\pm$0.04. The NLTE corrections are positive for all investigated lines and do not exceed 0.14~dex. The mean NLTE yttrium abundance is higher by 0.10~dex compared to LTE value.

{\it HD~220009.} Six lines of Y\ione\ and eleven lines of Y\ii\ were examined in the spectrum of this star. The obtained NLTE abundances from the two ionization stages are consistent within 0.05~dex (Figure~\ref{pics_equil}), while the LTE abundance based on the Y\ione\ lines is 0.30~dex lower compared with that for Y\ii.

 \begin{deluxetable*}{ccccccccccccccccccc}
 \tabletypesize{\scriptsize}
 \tablecaption{NLTE and LTE abundances of the program stars. \label{tab_abn}}
\def\arraystretch{0.7}
\setlength{\tabcolsep}{2pt}
\tabletypesize{\scriptsize }
\tablewidth{20pt}
\tablehead{
\colhead{$\lambda$ (\AA\,)} & \colhead{ \tiny LTE} & \colhead{ \tiny NLTE} & \colhead{$\Delta$} & \colhead{ \tiny LTE} & \colhead{ \tiny NLTE} & \colhead{$\Delta$} & \colhead{\tiny LTE} & \colhead{ \tiny NLTE} & \colhead{$\Delta$} & \colhead{ \tiny LTE} & \colhead{ \tiny NLTE} & \colhead{$\Delta$} & \colhead{ \tiny LTE} & \colhead{ \tiny NLTE} & \colhead{$\Delta$} & \colhead{ \tiny LTE} & \colhead{ \tiny NLTE} & \colhead{$\Delta$} }
\decimalcolnumbers
\startdata
 \multicolumn{19}c{ }      \\
&\multicolumn3c{HD22049}   &\multicolumn3c{HD22879} & \multicolumn3c{Aldebaran} &\multicolumn3c{Pollux}  & \multicolumn3c{Arcturus} &\multicolumn3c{HD220009}  \\ \hline
 Y\ione\ & \multicolumn{18}c{ }      \\
 4128.304       & 1.82    &  1.88  & 0.06   & 1.27    & 1.36   & 0.09   & \nodata &\nodata &\nodata & \nodata &\nodata &\nodata & \nodata &\nodata &\nodata & \nodata &\nodata &\nodata \\
 4643.698       & \nodata &\nodata &\nodata & \nodata &\nodata &\nodata & \nodata &\nodata &\nodata & \nodata &\nodata &\nodata & \nodata &\nodata &\nodata & 0.96    & 1.42   & 0.46   \\
 4674.848       & 1.93    &  2.00  & 0.07   & 1.39    & 1.46   & 0.07   & \nodata &\nodata &\nodata & \nodata &\nodata &\nodata & 1.17    & 1.48   & 0.31   & 1.04    & 1.39   & 0.35   \\
 5630.138       & \nodata &\nodata &\nodata & \nodata &\nodata &\nodata & \nodata &\nodata &\nodata & \nodata &\nodata &\nodata & \nodata &\nodata &\nodata & 1.26    & 1.48   & 0.22   \\
 6023.406       & \nodata &\nodata &\nodata & \nodata &\nodata &\nodata & \nodata &\nodata &\nodata & \nodata &\nodata &\nodata & \nodata &\nodata &\nodata & \nodata &\nodata &\nodata \\
 6222.578       & 2.15    &  2.20  & 0.05   & \nodata &\nodata &\nodata & \nodata &\nodata &\nodata & \nodata &\nodata &\nodata & \nodata &\nodata &\nodata & 1.04    & 1.25   & 0.21  \\
 6402.007       & \nodata &\nodata &\nodata & \nodata &\nodata &\nodata & 1.52    & 1.68   & 0.16   & 2.13    & 2.24   & 0.11   & 1.25    & 1.49   & 0.24   & 1.24    & 1.47   & 0.23  \\
 6435.022       & 2.08    &  2.13  & 0.05   & \nodata &\nodata &\nodata & 1.54    & 1.70   & 0.16   & 2.14    & 2.26   & 0.12   & 1.13    & 1.37   & 0.24   & 1.09    & 1.32   & 0.23  \\
  Mean          & 2.00    &  2.05  &\nodata & 1.33    & 1.41   &\nodata & 1.53    & 1.69   &\nodata & 2.14    & 2.25   &\nodata & 1.18    & 1.45   &\nodata & 1.11    & 1.39   &\nodata \\
  $\sigma$      & 0.15    &  0.14  &\nodata & 0.08    & 0.07   &\nodata & 0.01    & 0.01   &\nodata & 0.01    & 0.01   &\nodata & 0.06    & 0.07   &\nodata & 0.12    & 0.09   &\nodata \\\hline
  Y\ii\         & \multicolumn{18}c{ }      \\
 4398.008       & 2.00    & 2.03   &  0.03  & 1.49    & 1.56   & 0.07   & \nodata &\nodata &\nodata & 2.16    & 2.16   &  0.00  & \nodata &\nodata &\nodata & \nodata &\nodata &\nodata \\
 4854.861       & \nodata &\nodata &\nodata & 1.36    & 1.43   & 0.07   & \nodata &\nodata &\nodata & 2.07    & 2.07   &  0.00  & \nodata &\nodata &\nodata & \nodata &\nodata &\nodata \\
 4883.682       & 2.17    & 2.20   & 0.03   & 1.36    & 1.44   & 0.08   & \nodata &\nodata &\nodata & 2.08    & 2.08   &  0.00  & 1.63    & 1.67   & 0.04   & 1.50    & 1.54   & 0.04   \\
 4900.118       & 2.11    & 2.13   & 0.02   & 1.40    & 1.48   & 0.08   & \nodata &\nodata &\nodata & 2.12    & 2.12   &  0.00  & 1.39    & 1.45   & 0.06   & 1.39    & 1.43   & 0.04   \\
 4982.128       & \nodata &\nodata &\nodata & 1.47    & 1.50   & 0.03   & \nodata &\nodata &\nodata & 2.31    & 2.32   &  0.01  & 1.44    & 1.48   & 0.04   & 1.47    & 1.51   & 0.04   \\
 5087.418       & 2.01    & 2.03   &  0.02  & 1.40    & 1.44   & 0.04   & 1.73    & 1.75   & 0.02   & 1.96    & 1.97   &  0.01  & 1.43    & 1.46   & 0.03   & 1.41    & 1.44   & 0.03   \\
 5119.110       & \nodata &\nodata &\nodata & 1.53    & 1.57   & 0.04   & \nodata &\nodata &\nodata & \nodata &\nodata &\nodata & \nodata &\nodata &\nodata & 1.37    & 1.40   & 0.03   \\
 5205.722       & \nodata &\nodata &\nodata & 1.46    & 1.52   & 0.06   & \nodata &\nodata &\nodata & \nodata &\nodata &\nodata & \nodata &\nodata &\nodata & \nodata &\nodata &\nodata \\
 5289.815       & 2.07    & 2.08   & 0.01   & 1.50    & 1.57   & 0.07   & 1.69    & 1.70   & 0.01   & 2.31    & 2.32   &  0.01  & 1.52    & 1.54   & 0.02   & 1.34    & 1.35   & 0.01   \\
 5402.773       & 2.08    & 2.08   & 0.00   & 1.32    & 1.36   & 0.04   & \nodata &\nodata &\nodata & 2.06    & 2.06   &  0.00  & \nodata &\nodata &\nodata & 1.35    & 1.35   & 0.00   \\
 5473.384       & \nodata &\nodata &\nodata & 1.46    & 1.54   & 0.08   & \nodata &\nodata &\nodata & \nodata &\nodata &\nodata & 1.50    & 1.54   & 0.04   & 1.40    & 1.43   & 0.03  \\
 5521.562       & \nodata &\nodata &\nodata & 1.50    & 1.50   & 0.00   & \nodata &\nodata &\nodata & 2.17    & 2.17   &  0.00  & \nodata &\nodata &\nodata & 1.45    & 1.46   & 0.01  \\
 5544.610       & \nodata &\nodata &\nodata & \nodata &\nodata &\nodata & 1.84    & 1.92   & 0.08   & 2.37    & 2.39   &  0.02  & \nodata &\nodata &\nodata & 1.49    & 1.54   & 0.05 \\
 5546.008       & 2.08    & 2.08   &  0.00  & \nodata &\nodata &\nodata & \nodata &\nodata &\nodata & \nodata &\nodata &\nodata & \nodata &\nodata &\nodata & \nodata &\nodata &\nodata \\
 6613.731       & 2.09    & 2.12   &  0.03  & \nodata &\nodata &\nodata & \nodata &\nodata &\nodata & \nodata &\nodata &\nodata & \nodata &\nodata &\nodata & \nodata &\nodata &\nodata \\
 6795.415       & 2.10    & 2.11   &  0.01  & 1.39    & 1.44   & 0.05   & \nodata &\nodata &\nodata & 2.20    & 2.20   &  0.00  & \nodata &\nodata &\nodata & 1.35    & 1.36   & 0.01  \\\hline
  Mean          & 2.08    & 2.10   &\nodata & 1.43    & 1.49   &\nodata & 1.75    & 1.79   &\nodata & 2.16    & 2.17   &\nodata & 1.49    & 1.52   &\nodata & 1.41    & 1.44   &\nodata \\
  $\sigma$      & 0.05    & 0.05   &\nodata & 0.07    & 0.06   &\nodata & 0.08    & 0.12   &\nodata & 0.13    & 0.13   &\nodata & 0.09    & 0.08   &\nodata & 0.06    & 0.07   &\nodata \\\hline
              \multicolumn{19}c{ }      \\
 Y\ione$-$Y\ii\ & -0.08   & -0.05  &\nodata & -0.10   & -0.08  &\nodata & -0.22   & -0.10  &\nodata & -0.02   & 0.08   &\nodata & -0.31   & 0.07   &\nodata & -0.30   & -0.05   &\nodata \\
 \multicolumn{19}c{ }      \\\hline\hline
            \multicolumn{19}c{ }      \\
           & \multicolumn3c{HD49933} & \multicolumn3c{Procyon} & \multicolumn3c{HD84937} & \multicolumn3c{HD122563} & \multicolumn3c{HD140283} & \multicolumn3c{} \\\hline
  Y\ii\    & \multicolumn{18}c{ }      \\
 3788.693 &\nodata  & \nodata & \nodata & \nodata &\nodata &\nodata & \nodata &\nodata &\nodata & -0.87   & -0.66  & 0.21   & -0.66   & -0.52  &  0.14  & &  & \\
 4124.904 &\nodata  & \nodata & \nodata & 2.26    & 2.29   & 0.03   & \nodata &\nodata &\nodata & \nodata &\nodata &\nodata & \nodata &\nodata &\nodata & &  & \\
 4177.528 &\nodata  & \nodata & \nodata & \nodata &\nodata &\nodata & 0.03    & 0.16   & 0.13   & \nodata &\nodata &\nodata & \nodata &\nodata &\nodata & &  & \\
 4235.727 &\nodata  & \nodata & \nodata & 2.21    & 2.25   & 0.04   & 0.11    & 0.19   & 0.08   & \nodata &\nodata &\nodata & \nodata &\nodata &\nodata & &  & \\
 4374.933 &\nodata  & \nodata & \nodata & \nodata &\nodata &\nodata & 0.07    & 0.17   & 0.10   & -0.82   & -0.72  & 0.10   & -0.73   & -0.64  & 0.09   & &  & \\
 4398.008 & 1.84    & 1.92    & 0.08    & 2.24    & 2.32   & 0.08   & 0.10    & 0.20   & 0.10   & -0.62   & -0.45  & 0.17   & -0.68   & -0.56  & 0.12   & &  & \\
 4682.321 &\nodata  & \nodata & \nodata & \nodata &\nodata &\nodata & \nodata &\nodata &\nodata & -0.66   & -0.63  & 0.03   & \nodata &\nodata &\nodata & &  & \\
 4823.304 &\nodata  & \nodata & \nodata & \nodata &\nodata &\nodata & \nodata &\nodata &\nodata & -0.80   & -0.67  & 0.13   & \nodata &\nodata &\nodata & &  & \\
 4854.861 & 1.72    & 1.80    & 0.08    & 2.17    & 2.20   & 0.03   & \nodata &\nodata &\nodata & \nodata &\nodata &\nodata & \nodata &\nodata &\nodata & &  & \\
 4883.682 & 1.83    & 1.88    & 0.05    & 2.28    & 2.27   & -0.01  & 0.07    & 0.13   & 0.06   & -0.77   & -0.64  & 0.13   & -0.67   & -0.61  & 0.06   & &  & \\
 4900.118 &\nodata  & \nodata & \nodata & 2.28    & 2.28   & 0.00   & 0.07    & 0.13   & 0.06   & -0.78   & -0.65  & 0.13   & -0.63   & -0.56  & 0.07   & &  & \\
 4982.128 &\nodata  & \nodata & \nodata & 2.23    & 2.25   & 0.02   & \nodata &\nodata &\nodata & -0.60   & -0.54  & 0.06   & \nodata &\nodata &\nodata & &  & \\
 5087.418 & 1.81    & 1.86    & 0.05    & 2.23    & 2.23   & 0.00   & \nodata &\nodata &\nodata & -0.74   & -0.64  & 0.10   & -0.64   & -0.55  & 0.09   & &  & \\
 5119.110 & 1.93    & 2.00    & 0.07    & 2.20    & 2.26   & 0.06   & \nodata &\nodata &\nodata & -0.76   & -0.64  & 0.12   & \nodata &\nodata &\nodata & &  & \\
 5200.409 & 1.88    & 2.00    & 0.12    & 2.26    & 2.34   & 0.08   & \nodata &\nodata &\nodata & -0.73   & -0.57  & 0.16   & \nodata &\nodata &\nodata & &  & \\
 5205.722 & 1.70    & 1.79    & 0.09    & 2.17    & 2.25   & 0.08   & \nodata &\nodata &\nodata & -0.76   & -0.65  & 0.11   & \nodata &\nodata &\nodata & &  & \\
 5289.815 & 1.95    & 2.04    & 0.09    & 2.17    & 2.24   & 0.07   & \nodata &\nodata &\nodata & \nodata &\nodata &\nodata & \nodata &\nodata &\nodata & &  & \\
 5402.773 & 1.77    & 1.79    & 0.02    & \nodata &\nodata &\nodata & \nodata &\nodata &\nodata & \nodata &\nodata &\nodata & \nodata &\nodata &\nodata & &  & \\
 5473.384 &\nodata & \nodata  & \nodata & 2.26    & 2.31   & 0.05   & \nodata &\nodata &\nodata & \nodata &\nodata &\nodata & \nodata &\nodata &\nodata & &  & \\
 5521.562 &\nodata & \nodata  & \nodata & 2.28    & 2.29   & 0.01   & \nodata &\nodata &\nodata & \nodata &\nodata &\nodata & \nodata &\nodata &\nodata & &  & \\
 5544.610 &\nodata & \nodata  & \nodata & 2.26    & 2.30   & 0.04   & \nodata &\nodata &\nodata & \nodata &\nodata &\nodata & \nodata &\nodata &\nodata & &  & \\
 5546.008 & 1.97    & 1.97    & 0.00    & 2.18    & 2.19   & 0.01   & \nodata &\nodata &\nodata & \nodata &\nodata &\nodata & \nodata &\nodata &\nodata & &  & \\
 6613.731 &\nodata  & \nodata & \nodata & 2.24    & 2.26   & 0.02   & \nodata &\nodata &\nodata & \nodata &\nodata &\nodata & \nodata &\nodata &\nodata & &  & \\
 6795.415 &\nodata  & \nodata & \nodata & 2.18    & 2.20   & 0.02   & \nodata &\nodata &\nodata & \nodata &\nodata &\nodata & \nodata &\nodata &\nodata & &  & \\\hline
 Mean     & 1.84    & 1.91    & \nodata & 2.23    & 2.26   &\nodata & 0.08    & 0.16   &\nodata &  -0.74  & -0.62  &\nodata & -0.67   & -0.57  &\nodata & &  & \\
 $\sigma$ & 0.09    & 0.09    & \nodata & 0.04    & 0.04   &\nodata & 0.03    & 0.03   &\nodata &  0.08   & 0.07   &\nodata & 0.04    & 0.04   &\nodata & &  & \\\hline
\enddata
\tablecomments{ Here, $\Delta$ means $\Delta_{NLTE}$.}
\end{deluxetable*}

\section{Yttrium abundance trend}\label{sec:application}

\subsection{Stellar sample}

We present the NLTE yttrium abundances in a sample consisting of 65 FGK stars, which are uniformly distributed across the -2.62$\leq$[Fe/H]$\leq$+0.24 metallicity range. This range is relevant to research on the Galactic chemical evolution. Most of the stars are dwarfs, with a few subgiants added. Stellar atmosphere parameters and yttrium LTE and NLTE abundances for 61 stars are given in Table~\ref{Sample}.
Other four stars of this sample, HD~49933 (Thin disc), HD~22879 (Thick disc), HD~84937 and HD~140283 (Halo) can be found in Table~\ref{tab_param},
where their LTE and NLTE yttrium-to-iron abundance ratios [Y/Fe] are also presented.
For the first 47 stars from Table~\ref{Sample}, we adopted the stellar parameters from \citet{2015ApJ...808..148S}.
For stars $\#$ 48--54, the stellar parameters were adopted from \citet{2003A&A...397..275M}, and, for stars $\#$55--61, from \citet{2019ARep...63..726M}.

The Galactic thin disk stellar population is represented by 28 stars, with [Fe/H] $\geq$ -0.78.
We included 17 thick-disk stars in the -1.49$\leq$[Fe/H]$\leq$-0.17 range, overlapping with that of the thin-disk and 20 halo stars, which are distributed in the -2.62$\leq$[Fe/H]$\leq$-1.08 range.

For the first 47 stars from Table~\ref{Sample}, we employed high-resolution ($R >$ 45,000) and high signal-to-noise ratio S/N $>$ 60
spectra obtained at the 3-m telescope of the Lick
Observatory with the Hamilton spectrograph or taken from ESPaDOnS\footnote
{http://www.cadc-ccda.hia-iha.nrc-cnrc.gc.ca/en/search/} archive. The characteristics of the spectra of the stars $\#$48--61 are presented in Table~\ref{tab_obs2}.
The spectra were downloaded from ESO Archive
and normalized with the \textsc{RASSINE} $Python$ code \citep{2020A&A...640A..42C}.

We used six lines of Y\ii\ at 4374, 4398, 4883, 4900, 5087, 5205~\AA\ for yttrium abundance determination in the sample. When fitting the Y\ii\ 4374~\AA\ line, the Ti\ii\ 4374.813~\AA\ line was taken into account. When fitting the Y\ii\ 5205~\AA\ line, the Cr\ione\ 5206.041~\AA\ line was taken into account. The abundances were determined using a line-by-line differential approach, comparing them to the solar reference abundances.

\subsection{Yttrium Abundance Trend}

\begin{figure}
\begin{center}
 \includegraphics[scale=0.4]{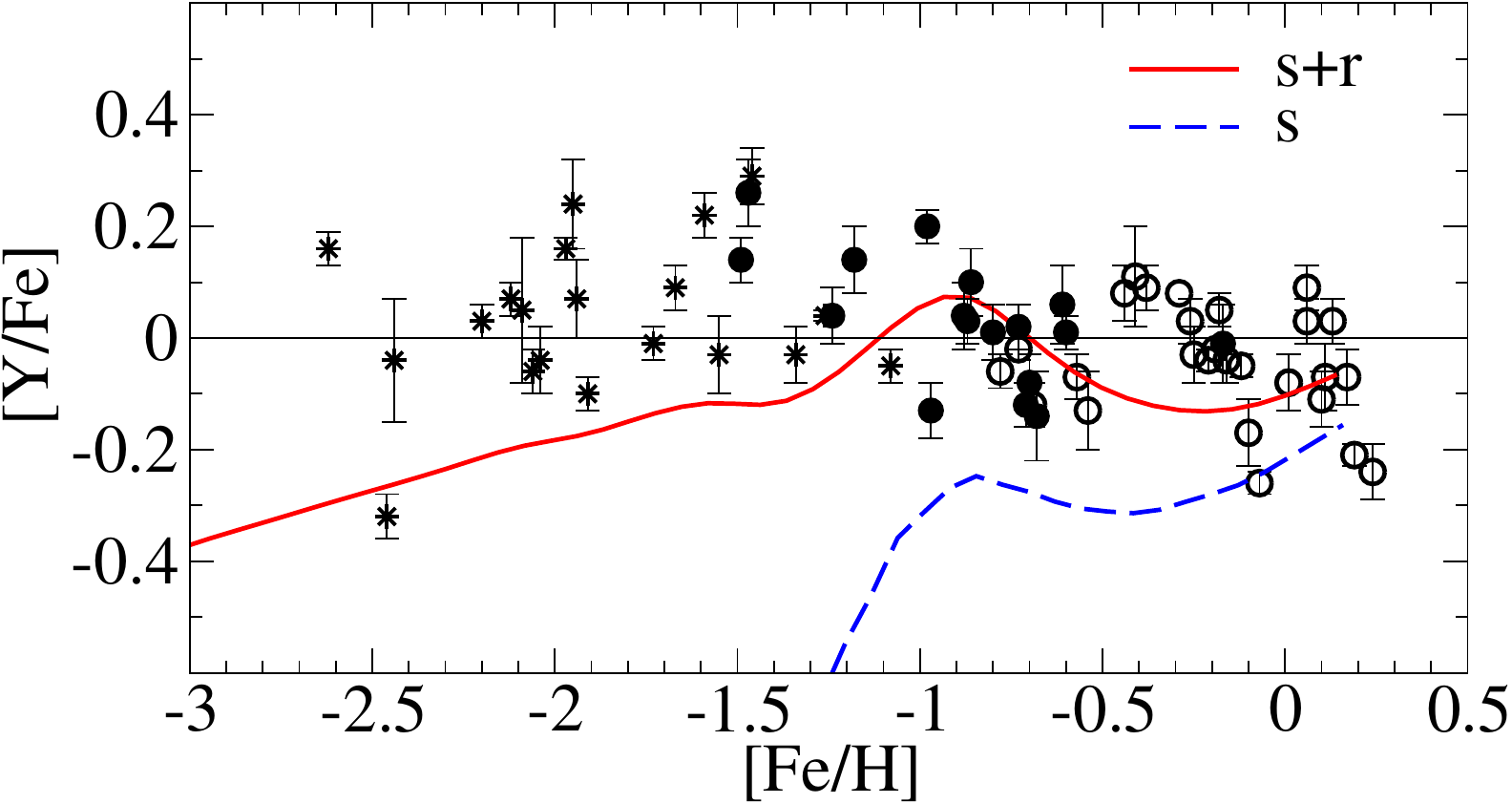}
 \caption{ Stellar NLTE yttrium-to-iron abundance ratios as a function of metallicity for 65 stars; and the [Y/Fe]--[Fe/H] ratio for the solar neighborhood models with $s$-process only (dashed line) and with $s-$ and $r-$processes (solid line) from \citet{2020ApJ...900..179K}. Different symbols correspond
to different stellar populations, namely, the thin disk (open circles), the thick disk (ﬁlled circles), and the halo (asterisks).  }
 \label{Y_Fe_tot}
 \end{center}
 \end{figure}

Halo stars with -2.62$\leq$[Fe/H]$\leq$-1.46 metallicity range show positive trend (Slope: +0.18$\pm$0.11) from [Y/Fe] $\approx$ -0.07 to 0.14~dex (Figure~\ref{Y_Fe_tot}).  However, if we remove HD~140283 from the group of stars with this metallicity range, the positive trend will have slope +0.09$\pm$0.10. Further NLTE [Y/Fe] determination in very metal poor stars with [Fe/H]~$<$~-2.5 would be helpful for extention the trend towards lower metallicity and studies of the earliest stages of Galactic yttrium formation. 
 This behaviour argues that yttrium is mainly produced in supernova type II (massive stars) explosions \citep[SNe,][]{2020ApJ...900..179K}, and their yields depend on metallicity. The more metal-rich SNe, the higher yttrium production. A famous star HD~140283, which has been extensively studied and found to be deficient in $r$-process elements see e.g. \citet{2015ApJ...808..148S, 2015A&A...584A..86S} and revealed a significant underabundance of Sr, Zr, and Ba compared to Fe ratios when compared to other stars in the galactic halo, exposes [Y/Fe] = -0.32~dex, that is lower compared to other stars in our sample.

In the -1.49$\leq$[Fe/H]$\leq$-0.6 metallicity range, the thick disc stars show negative trend (Slope: -0.26$\pm$0.08) in [Y/Fe] from 0.18 to -0.05~dex.
Thin disc stars at -0.44$\leq$[Fe/H]$\leq$0.24 range also show negative trend (Slope: -0.32$\pm$0.09) in [Y/Fe] from 0.08 to -0.14~dex.
 It looks like that at [Fe/H]$\approx$-1.50 the positive trend turns to negative one. The observed knee at [Fe/H]$\approx$-1.50 could indicate the onset of the Fe production by Type Ia SNe (SNe Ia), however, this contradicts with the earlier studies, where the Galactic trends for different $\alpha$-elements (O, Mg Si, Ca) have 'the knee' and onset of iron production in SN Ia at [Fe/H] = -0.9. The reason why we found 'the knee' at [Fe/H]$\approx$-1.50 instead of [Fe/H] = -0.9 is not clear.  

Figure~\ref{Y_Fe_tot} shows the comparison between our observational data and the predictions from the GCE models from \citet{2020ApJ...900..179K}. The $s$ model takes into account the $s$-process only (dashed line), and the $s+r$ model (solid line) takes into account not only the $s$-process, but also high-mass super-AGB stars explode as electron-capture supernovae (ECSNe), NS--NS/NS--BH mergers, and core-collapse supernovae induced by strong magnetic ﬁelds and/or fast rotation (MRSNe). The trend of the $s+r$ model is in excellent agreement with our observational data at -1.60$\leq$[Fe/H]$\leq$+0.24.
However, at metallicity range -2.62$\leq$[Fe/H]$\leq$-1.60, the $s+r$ model predicts $\sim$0.23~dex lower [Y/Fe]. This indicates that the input of Y yields needs to be improved in the GCE model.

\begin{figure*}
\begin{minipage}{170mm}
\begin{center}
\parbox{0.44\linewidth}{\includegraphics[scale=0.27]{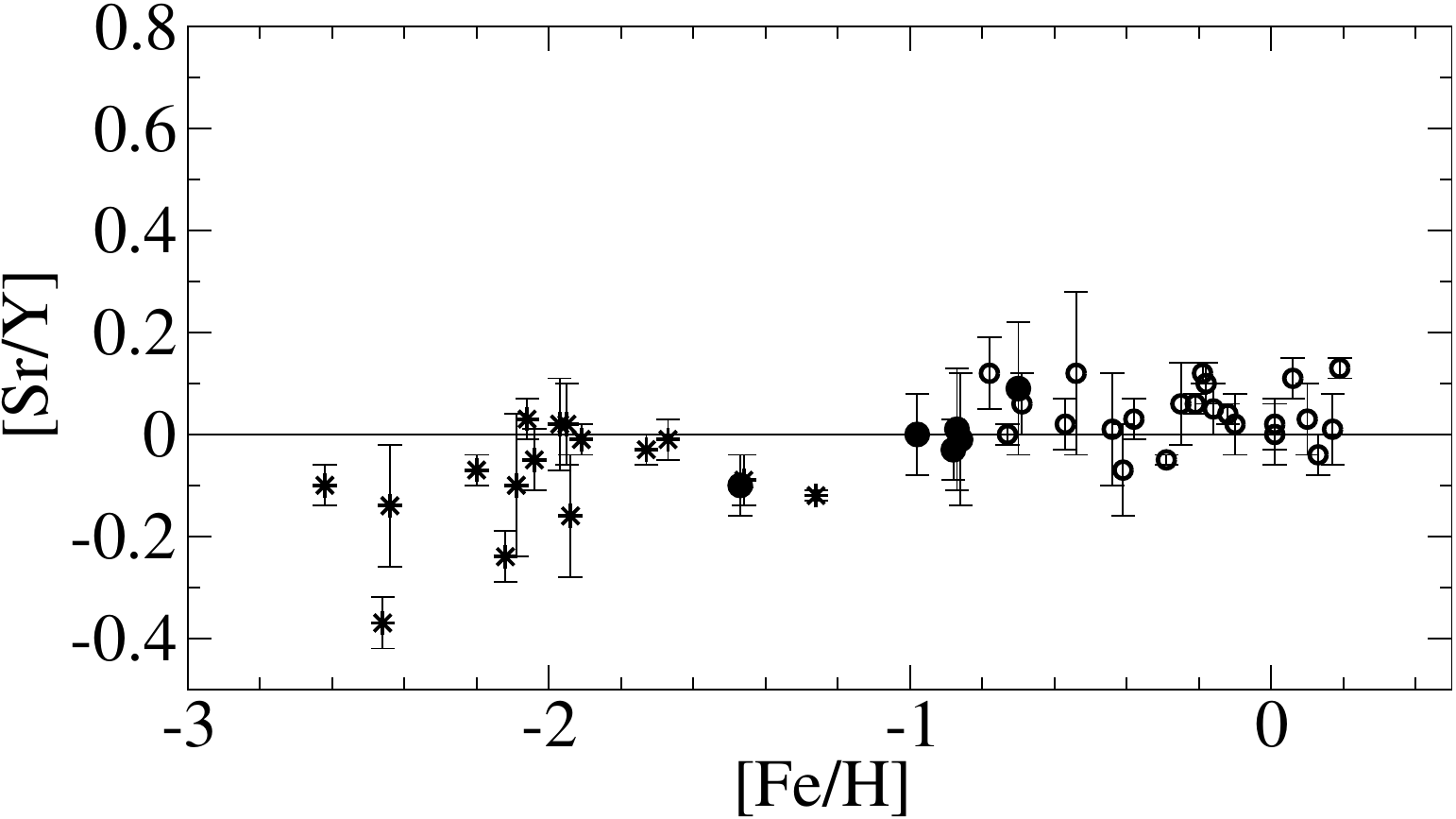}\\
\centering}
\parbox{0.44\linewidth}{\includegraphics[scale=0.27]{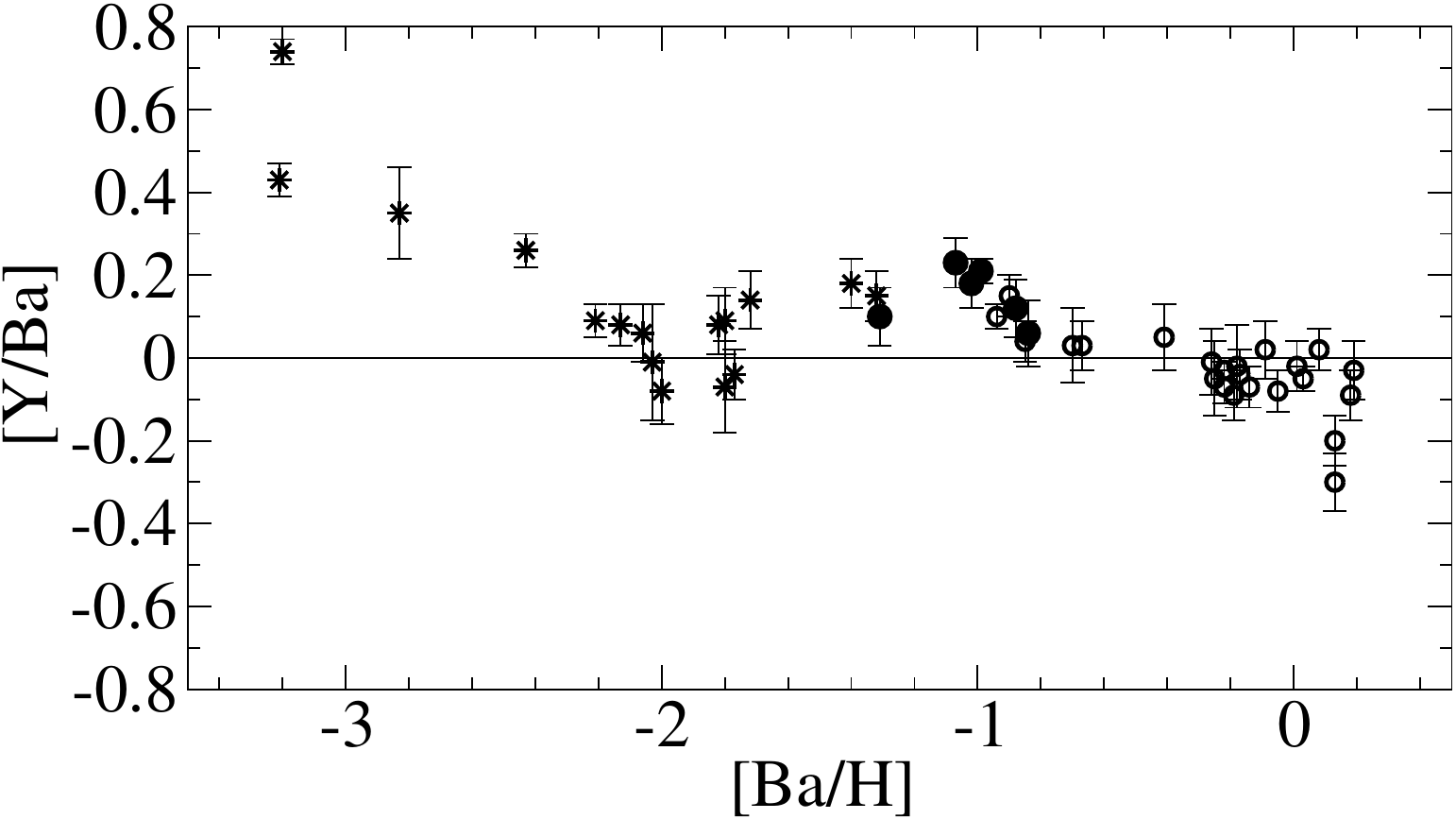}\\
\centering}
\hspace{1\linewidth}
\hfill
\\[0ex]
 \parbox{0.44\linewidth}{\includegraphics[scale=0.27]{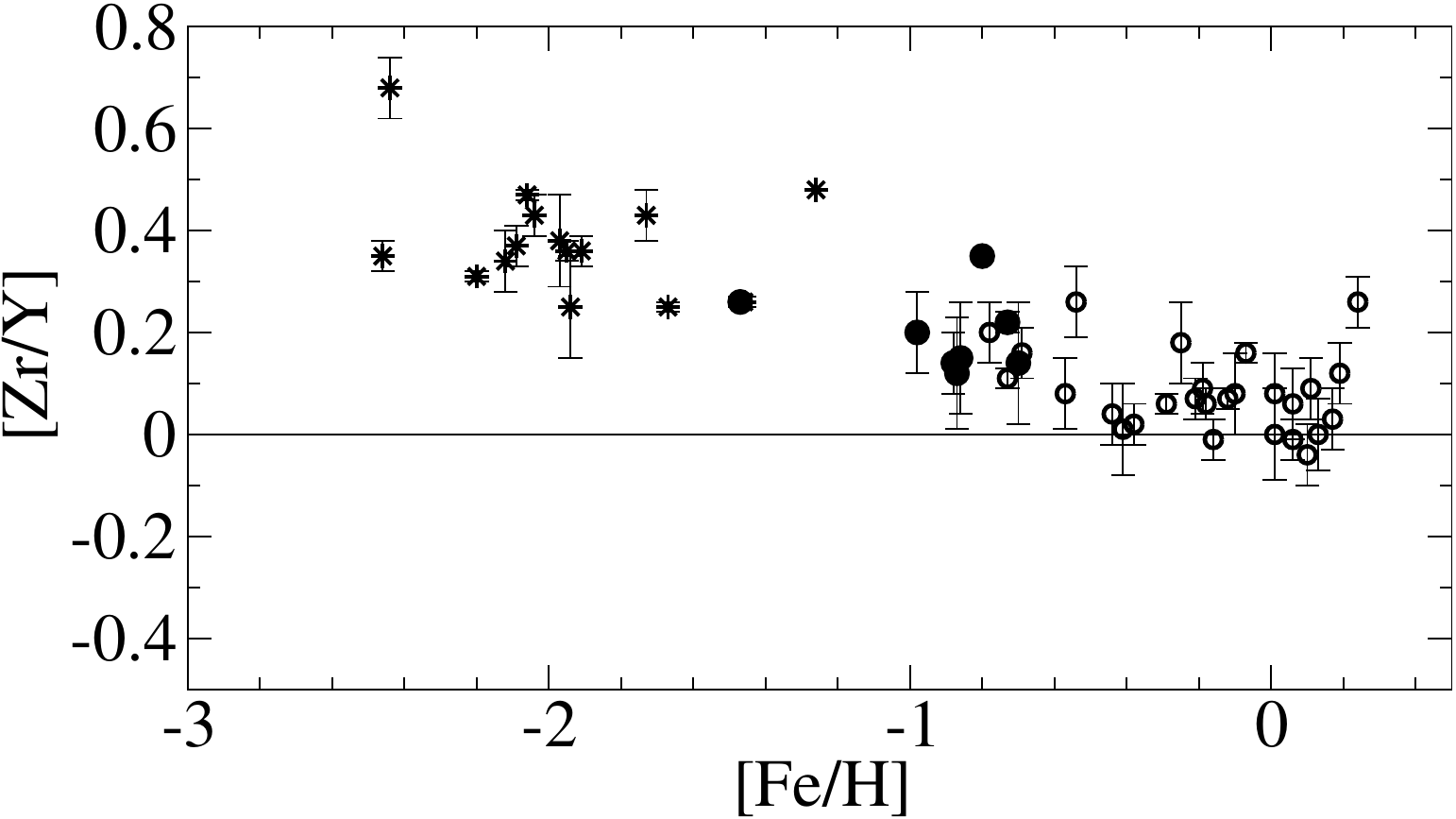}\\
 \centering}
 \parbox{0.44\linewidth}{\includegraphics[scale=0.27]{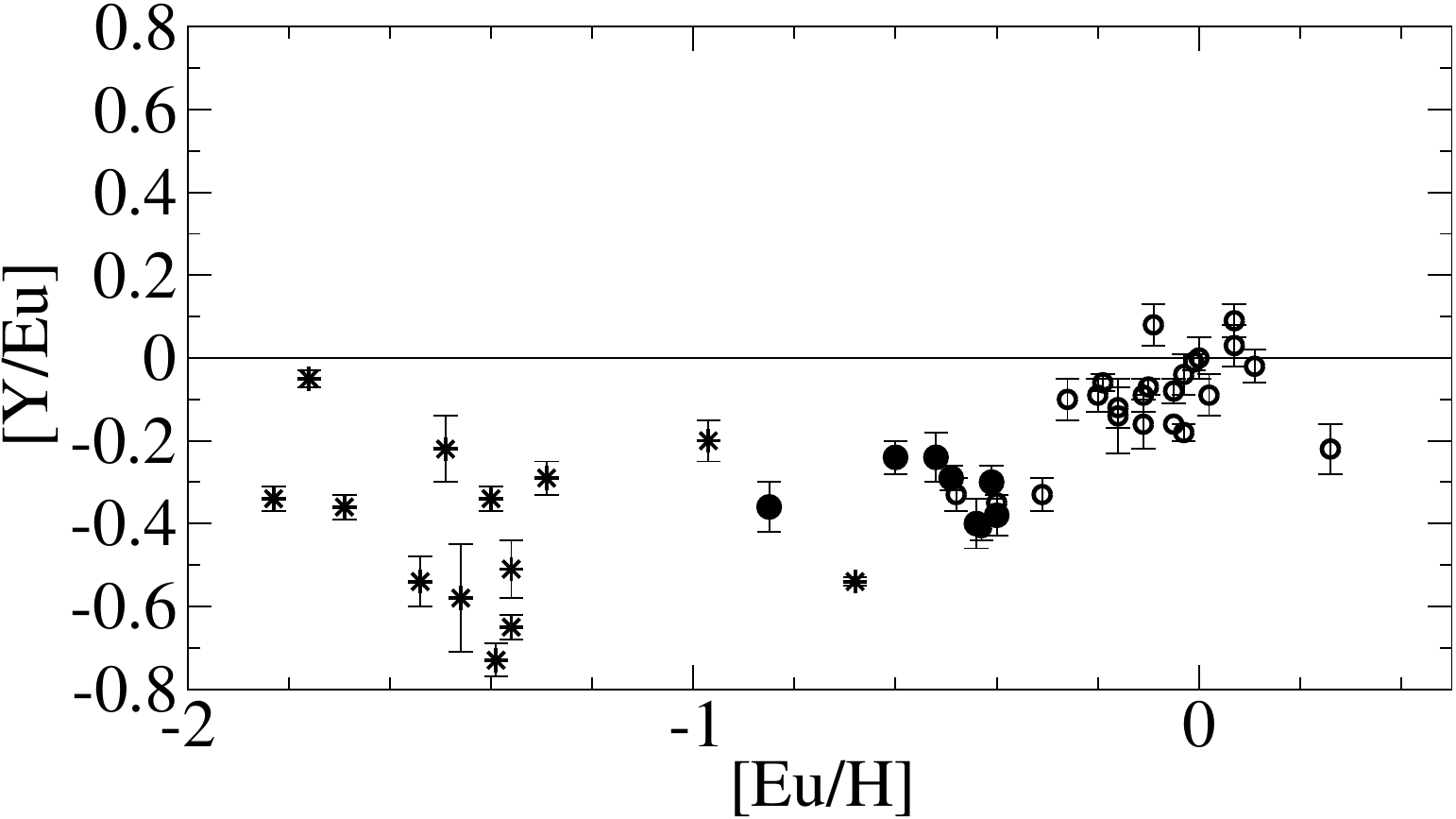}\\
 \centering}
  \hspace{1\linewidth}
  \hfill
  \\[0ex]
 \parbox{0.44\linewidth}{\includegraphics[scale=0.27]{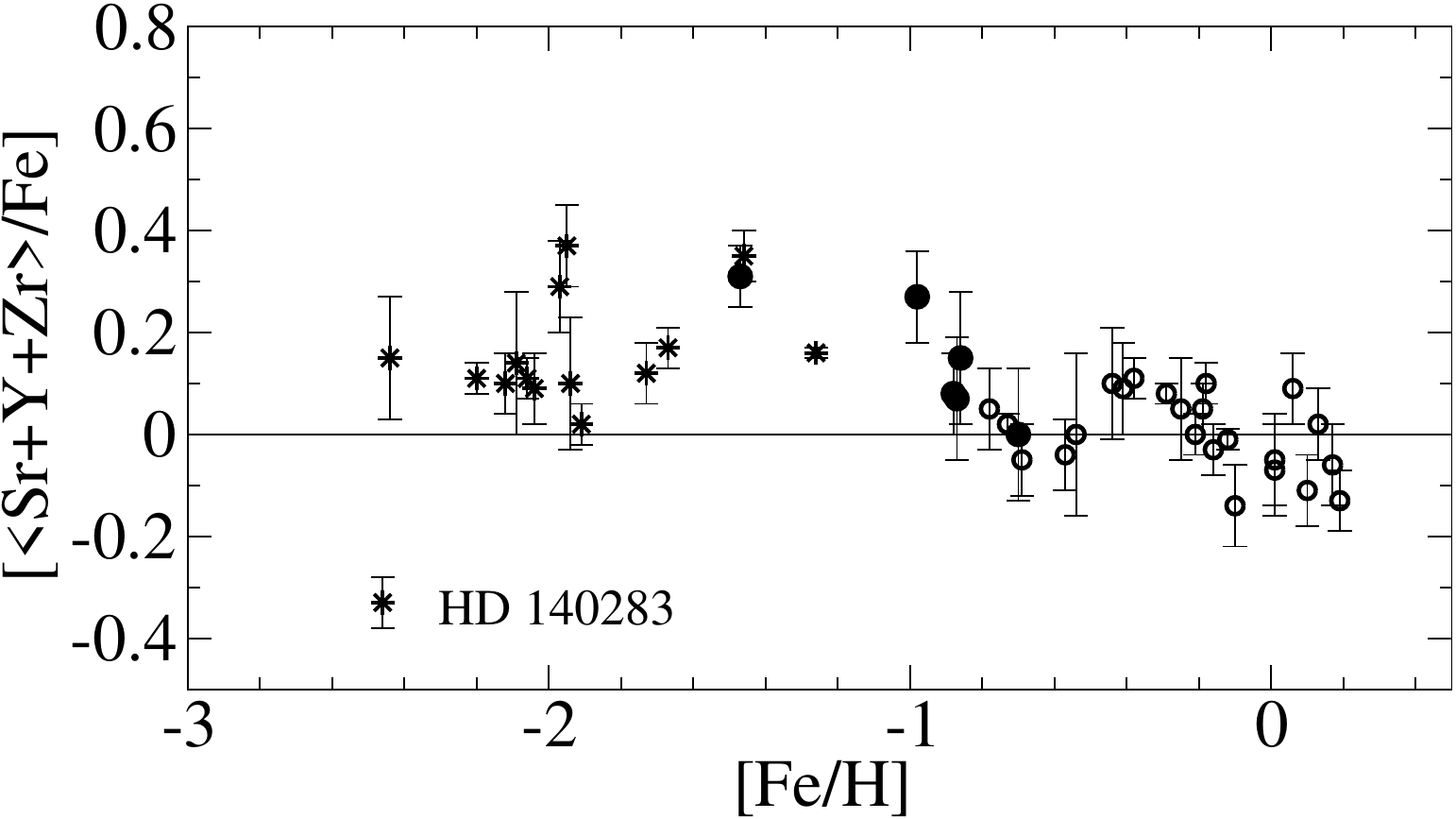}\\
 \centering}
  \parbox{0.44\linewidth}{\includegraphics[scale=0.27]{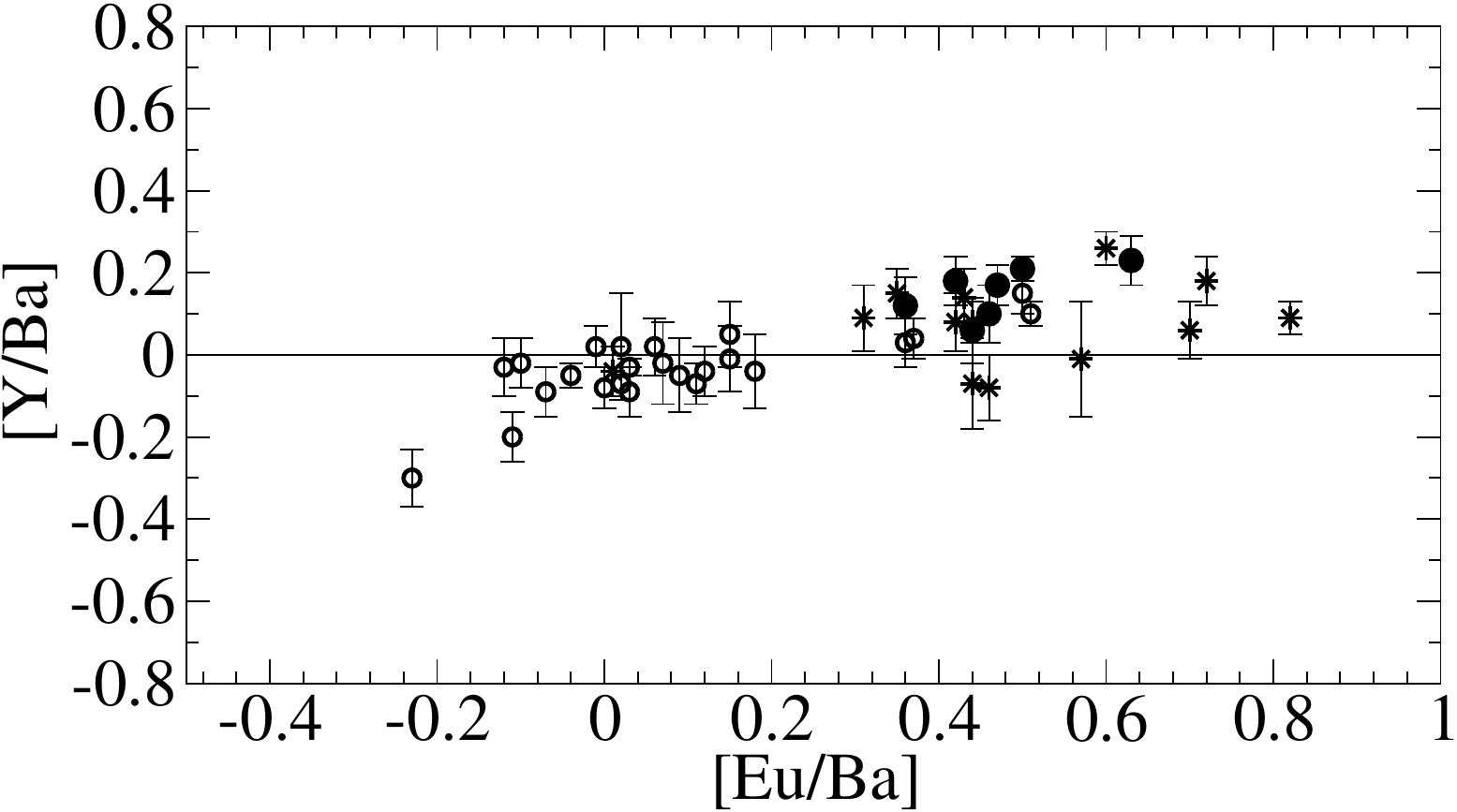}\\
 \centering}
 \hspace{1\linewidth}
 \hfill
 \\[0ex]
 \caption{NLTE abundance ratios between the neutron-capture elements. The same symbols are used as in Figure~\ref{Y_Fe_tot}.}
 \label{Sr_Zr_Y}
 \end{center}
 \end{minipage}
 \end{figure*}

We undertake a comparison between yttrium and two other light neutron-capture elements, namely, strontium (Sr) and zirconium (Zr), which are believed to have a common origin in nucleosynthesis. The NLTE abundances of Sr and Zr for our sample stars were taken from \citet{2016ApJ...833..225Z}.
In Figure~\ref{Sr_Zr_Y}, we study the behavior of [Zr/Y] and [Sr/Zr] with metalicity. The [Zr/Y] ratio is close to the solar value in the thin-disk stars, but it grows steeply in the thick-disk and halo stars, approaching [Zr/Y]~=~0.68 at [Fe/H]~=~-2.4. Similar behavior was found for [Zr/Sr] in \citet{2016ApJ...833..225Z}. The [Sr/Y] ratio is also close to the solar value in the thin-disk stars, however, it is decreasing in the thick-disk and halo stars and reaches [Sr/Y]~=~-0.37 at [Fe/H]~=~-2.5. The difference in behaviour at [Fe/H]$\lesssim$-0.7 between two neighbouring elements, Y and Zr, was noticed in earlier studies, e.g. \cite{1991A&A...244..425Z, 2007A&A...476..935F}.

By observing the trends in [Sr/Fe], [Y/Fe], and [Zr/Fe] (Figures \ref{Y_Fe_tot} and \ref{Sr_Zr_Y}), it becomes evident that the [X/Fe] ratios of these elements increase with higher Z values in metal-poor stars. Specifically, for Sr, Y, and Zr with Z values of 38, 39, and 40 respectively, it can be observed that in metal-poor stars, the order of [Sr/Fe]~$<$~[Y/Fe]~$<$~[Zr/Fe] holds true. This outcome could provide valuable insights into unraveling the source of light neutron capture elements.
If the outlined [X/Fe] ratio sequence is rooted in physical principles, it is reasonable to expect 
[Rb/Fe]~$<$~[Sr/Fe]. This is due to Rb, with an atomic number of 37, serving as another instance of a light neutron capture element found in stars \citep[see e.g.][]{2021A&A...651A..20C}.

Figure~\ref{Sr_Zr_Y} shows the average [$<$Sr+Y+Zr$>$/Fe] ratio for our stars as a function of [Fe/H]. Average [$<$Sr+Y+Zr$>$/Fe] is slightly enhanced in halo stars with [$<$Sr+Y+Zr$>$/Fe] = 0.16$\pm$0.10~dex, and [$<$Sr+Y+Zr$>$/Fe] = 0.01$\pm$0.10~dex in think disk stars. HD~140283 stands out due to its pronounced scarcity of all three elements among metal-poor stars and was not considered in the average value. 

The investigation of [Sr/Fe], [Y/Fe], and [Zr/Fe] trends at lower metallicity down to -4.0 dex will help to better understand the mechanism of the production of these light neutron capture elements at the early stage of the galactic evolution. 

 We compare Y with the heavier neutron-capture elements Barium (Ba) and Europium (Eu) (Fig.~\ref{Sr_Zr_Y}). Barium originates from both the $r$-process and the main components of the $s$-process, whereas Europium is primarily associated with the $r$-process. 
It is evident that both patterns, namely [Y/Ba] vs. [Ba/H] and [Y/Eu] vs. [Eu/H], exhibit significant data dispersion. The dispersion reduces when examining the relationship between [Y/Ba] and [Eu/Ba].
The [Eu/Ba] ratio depends on the relative contributions of the $s$- and $r$-processes to the formation of these elements. The relationship between the [Y/Ba] ratio and the [Eu/Ba] ratio shows a strong correlation (see Fig.~\ref{Sr_Zr_Y}), particular Y and Eu excesses relative to Ba for thick-disk and halo stars. The found correlation is important for understanding the origin of yttrium.

\subsection{Comparison with literature}

\begin{figure}
\begin{center}
 \includegraphics[scale=0.4]{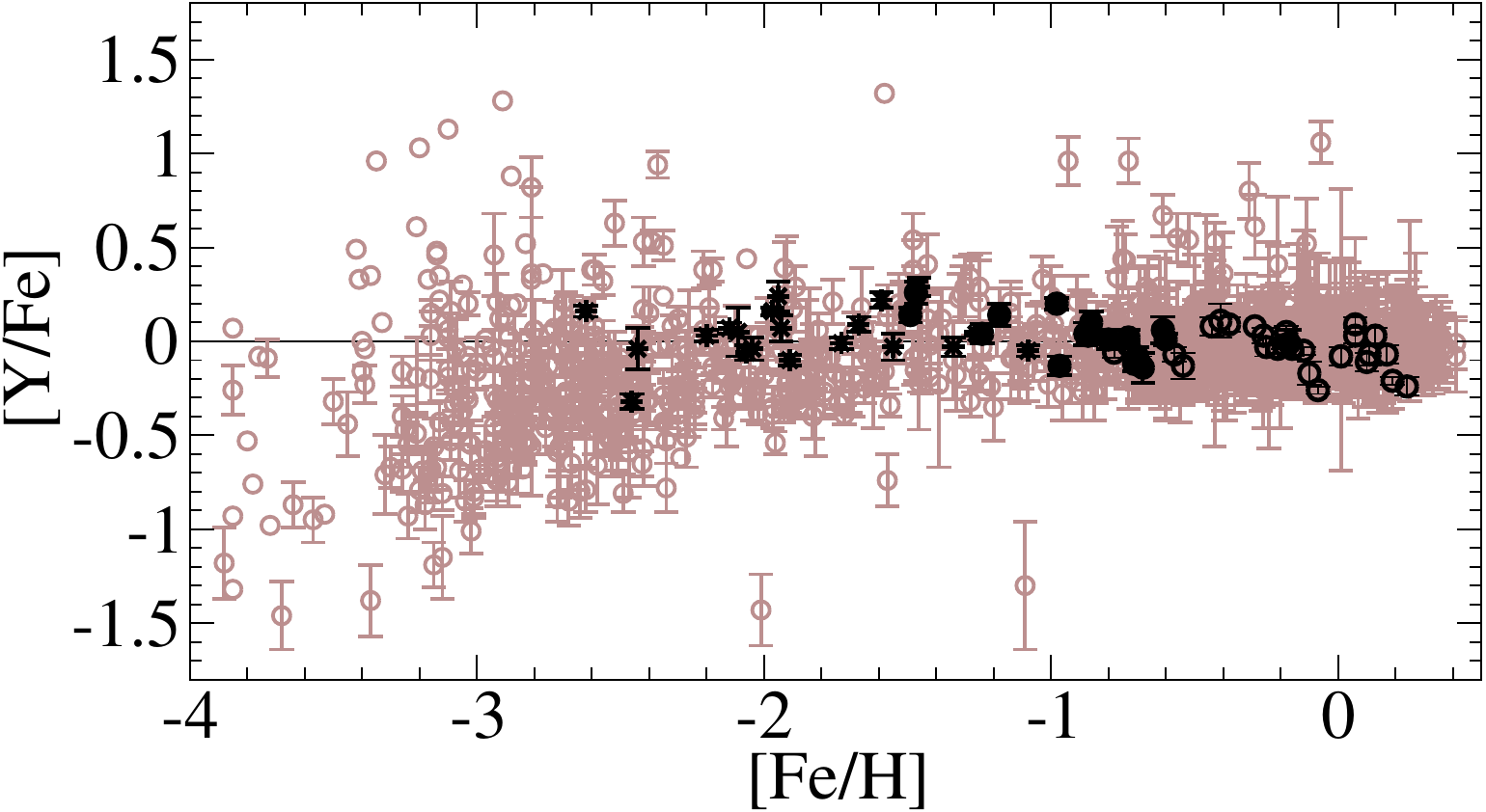}
 \caption{ Our stellar NLTE yttrium-to-iron abundance ratios as a function of metallicity for 65 stars and
 compilation from LTE studies from literature (brown circles). The compilation was made of data from \citet{2004ApJ...607..474H, 2012A&A...545A..31H,2014A&A...562A..71B, 2014AJ....147..136R,  2017A&A...608A..46R}. The same symbols are used as in Figure~\ref{Y_Fe_tot}.}
 \label{Y_Fe_7}
 \end{center}
 \end{figure}

The abundances we obtained align with the existing literature data within the range of metallicities that overlap with our study, despite the fact that most studies were performed under the LTE assumption. The reason for this is that the NLTE abundance corrections for Y\ii\ lines are small within the stellar parameter range that we are focusing on. Fig.~\ref{Y_Fe_7} presents the compilation of literature data from \citet{2004ApJ...607..474H, 2012A&A...545A..31H,2014A&A...562A..71B, 2014AJ....147..136R,  2017A&A...608A..46R} obtained under LTE assumption and our NLTE yttrium-to-iron abundance ratios.

The obtained positive slope for halo stars in metallicity range -2.62$\leq$[Fe/H]$\leq$-1.46 in this work is different to
the result of \citet{2017A&A...608A..46R}, who obtained negative slope -0.06$\pm$0.12 for 21 stars in the overlapping metallicity range.

\subsection{Inﬂuence of NLTE on the Galactic Abundance Trends}

\begin{figure}
\begin{center}
 \includegraphics[scale=0.4]{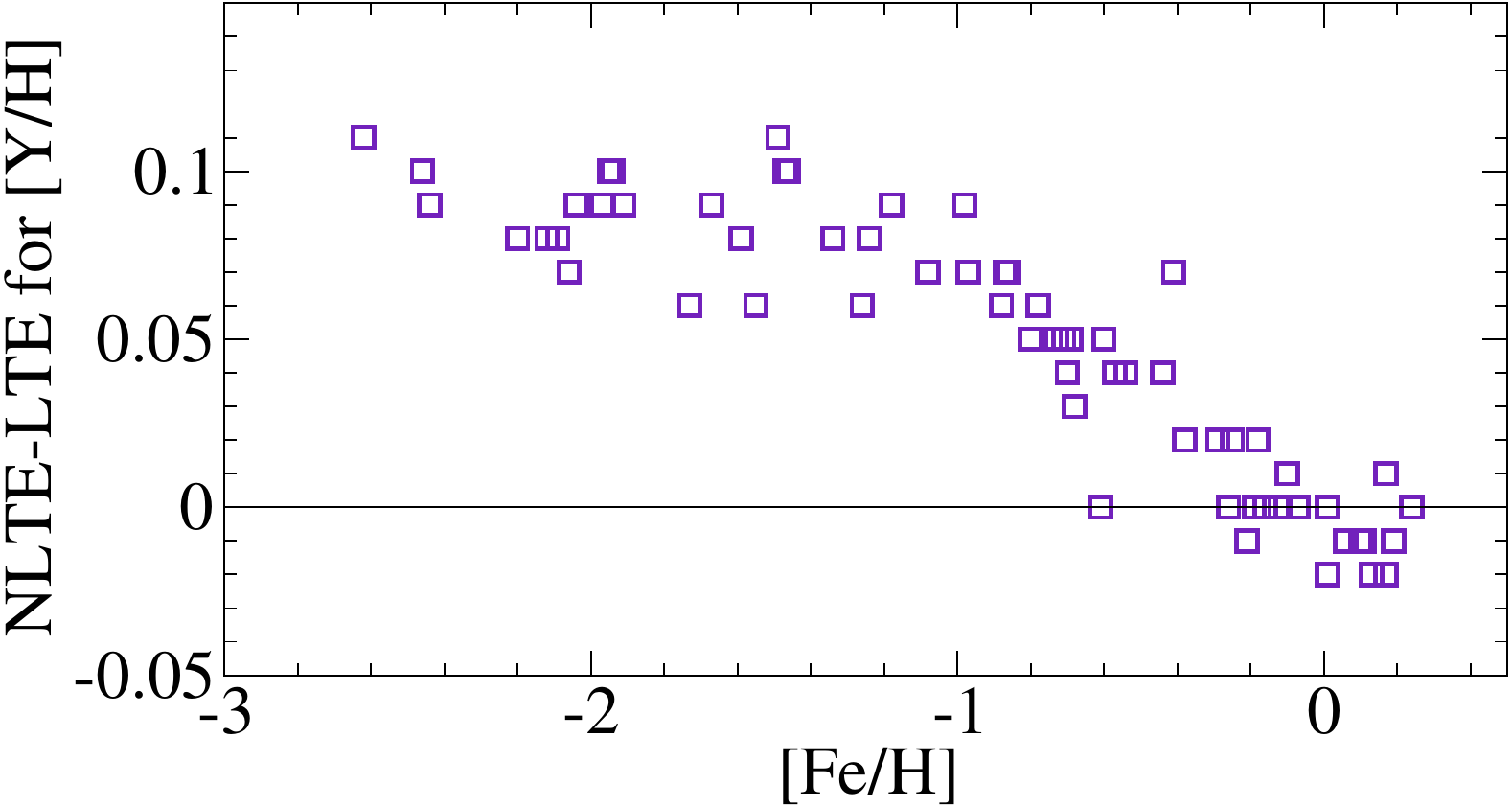}
 \caption{Differences in differential abundance [Y/H] between NLTE and LTE for 65 investigated stars.  }
 \label{NLTE-LTE}
 \end{center}
 \end{figure}

As mentioned, this study employs a line-by-line differential analysis method, comparing each line to the corresponding line in the Sun as a reference. In this context, we examine the influence of NLTE effects on the determination of average yttrium abundances [Y/H] and elemental ratios, which varies with the metallicity of a star. Figure~\ref{NLTE-LTE} displays the differences between the NLTE and LTE [Y/H] ratio for all investigated stars. The NLTE abundance corrections do not exceed 0.11~dex. The disparities between NLTE and LTE are negligible for stars with [Fe/H]$\sim$0.0, but they become more pronounced as the metallicity decreases.  NLTE abundance corrections for Y\ii\ depend on metallicity and neglecting NLTE effects distorce the Galactic trends.

Table~\ref{Sample} demonstrates that NLTE offers an additional benefit through the reduction of line-to-line scatter, which is observed in the majority of species and stars when compared to LTE. For example, for HD~94028 in LTE assumption [Y/Fe]~=~0.16$\pm$0.09, while significantly smaller statistical errors are achieved in NLTE, resulting in [Y/Fe]~=~0.26$\pm$0.06.

\begin{deluxetable*}{lcccccc}
\tablecaption{Characteristics of observed spectra. \label{tab_obs2}}
\tablewidth{0pt}
\tablehead{
\colhead{Star}&\colhead{Telescope/}     & \colhead{Spectral range}  & \colhead{Observing run}   &\colhead{$R$} & \colhead{$S/N$}  &\colhead{Run/Program ID}  \\
\colhead{}    &\colhead{Spectrograph }  & \colhead{\AA\,}           & \colhead{year/month/day}  &\colhead{}    & \colhead{}       & \colhead{}
}
\decimalcolnumbers
\startdata
  HD~29907   &  2  &   4590-6650  &  2001/03/08 & 51\,690   &   298  &  67.D-0086(A)   \\
  HD~31128   &  1  &   3781-6913  &  2008/01/16 & 115\,000  &   103  &  072.C-0488(E)  \\
  HD~59392   &  2  &   4590-6650  &  2020/09/01 & 51\,690   &   249  &  67.D-0086(A)   \\
  HD~97320   &  1  &   3781-6913  &  2009/03/20 & 115\,000  &   160  &  082.C-0212(B)  \\
  HD~193901  &  3  &   3772-7900  &  2021/08/23 & 140\,000  &   198  &  107.22QS.001   \\
  HD~298986  &  2  &   4590-6650  &  2001/03/08 & 51\,690   &   264  &  67.D-0086(A)   \\
  HD~102200  &  1  &   3781-6913  &  2007/03/10 & 115\,000  &   137  &  072.C-0488(E)  \\
  HD~3795    &  4  &   3527-9215  &  2017/12/03 & 48\,000   &   240  &  0100.A-9022(A) \\
  HD~32923   &  1  &   3781-6913  &  2004/12/24 & 115\,000  &   102  &  074.C-0364(A)  \\
  HD~40397   &  1  &   3781-6913  &  2016/02/04 & 115\,000  &   268  &  192.C-0852(A)  \\
  HD~64606   &  4  &   3527-9215  &  2004/03/10 & 48\,000   &   229  &  072.C-0033(A)  \\
  HD~69611   &  1  &   3781-6913  &  2016/03/06 & 115\,000  &   190  &  196.C-0042(A)  \\
  HD~114762  &  2  &   4785-6813  &  2001/03/07 & 107\,200  &   381  &  67.C-0206(A)   \\
  HD~201891  &  2  &   4785-6804  &  2012/10/18 & 66\,320   &   109  &  090.B-0605(A)  \\
\enddata
\tablecomments{{\bf Notes:}
Telescope/spectrograph: 1 = ESO-3.6/HARPS Echelle; 2 = ESO-VLT-U2/UVES; 3 = ESO-VLT-U1/ESPRESSO; 4 = MPG/ESO-2.2/FEROS}
\end{deluxetable*}

\section{Conclusions}\label{Sect:Conclusions}

For the first time, the departures from LTE for Y\ione\ -- Y\ii\ in the atmospheres of FGK stars across wide metallicity range are investigated. We present a new model atom of Y\ione\ -- Y\ii\  based on the most up-to-date atomic data including those for inelastic collisions with neutral hydrogen atoms and electrons. Collisional excitation and ionization by hydrogen atoms were considered for all levels. The calculated effective collision strengths for electron impact excitation of Y\ione\ and Y\ii\ are also presented in this study.

With the 1D MARCS solar model atmosphere, we obtain the solar NLTE abundance, log~$\epsilon_{\rm Y}$ = 2.21$\pm$0.05, from 15 lines of Y\ii\ and log~$\epsilon_{\rm Y}$ = 2.14$\pm$0.03 from one line of Y\ione. In LTE, the abundance difference between Y\ione\ and Y\ii\ amounts to -0.12~dex. Since,  we used only one Y\ione\ line, we adopted the solar NLTE abundance log~$\epsilon_{\rm Y}$ = 2.21$\pm$0.05 obtained from 15 lines of Y\ii\ only. The obtained result is in line with the 3D LTE result of \citet{2015A&A...573A..27G} and consistent within the error bars with the meteoritic yttrium abundance from \citet{2021SSRv..217...44L}.

Based on high-S/N, high-resolution spectra, the yttrium NLTE and LTE abundances were determined for 11 reference stars with
reliable atmospheric parameters, lieing in the following
ranges: 3930~K$\leq$~\Teff~$\leq$~6635~K, 1.1~$\leq$~log~$g$~$\leq$~4.6, and -2.55~$\leq$~[Fe/H]~$\leq$~+0.13. For seven reference stars, including the Sun, we can reach ionization equilibrium Y\ione\ -- Y\ii\ in NLTE.

In the models representing the atmospheres of F-G-K-type
stars of close-to-solar metallicity, the NLTE corrections for all Y\ii\ lines, which we analyzed, are positive and do not exceed 0.12~dex. The exceptions are Y\ii\ lines at 4374~\AA\ and 4883~\AA\ in solar atmosphere and the atmosphere of Procyon, respectively, for which $\Delta_{\rm NLTE}$~=~-0.01~dex.
In metal-poor stars, the NLTE corrections for all
Y\ii\ lines are still positive and stronger compared to solar metallicity stars, but do not exceed $\sim$0.20 dex. 
For Y\ione\ lines, NLTE effects are stronger compared to Y\ii\ 
and can reach up to $\sim$0.5 dex. In metal-poor stars, Y\ione\ lines are not available in the spectra. 
We provide a grid of the NLTE abundance corrections for Y\ii\ lines at 4398, 4883, 4900, and 5087~\AA.

We determined yttrium NLTE abundances for a sample of 65 FGK stars in the -2.62~$\leq$~[Fe/H]~$\leq$~+0.24 metallicity range.
We derive line-to-line differential abundances relative to the Sun using six lines of Y\ii\ at 4374, 4398, 4883, 4900, 5087, 5205~\AA. The NLTE leads to smaller line-to-line scatter for the most of the stars. The NLTE effects in Y\ii\ lines are important for stars with [Fe/H]~$\leq$~-1, and the NLTE--LTE abundance differences are increasing towards lower metallicity. 
 
Both thick and thin disk stars with -1.5~$\leq$~[Fe/H]~$\leq$~+0.24, show negative trends.
Metal-poor halo stars with -2.62$\leq$[Fe/H]$\leq$-1.46 show positive trend (Slope: +0.18$\pm$0.11) from [Y/Fe] $\approx$ -0.07 to 0.14~dex
The observed slope is in line with the $s+r$ model predicted by \citet{2020ApJ...900..179K}, however, their model predicts by $\sim$0.23~dex lower [Y/Fe].

 Our study has facilitated the execution of more precise analyses concerning Galactic trends and has enabled the presentation of a detailed comparison of trends in light neutron capture elements (Sr, Y, Zr) in NLTE. For Sr (Z = 38), Y (Z = 39), and Zr (Z = 40), in metal poor stars, the sequence 
[Sr/Fe]~$<$~[Y/Fe]~$<$~[Zr/Fe] applies. This observation could offer key insights into understanding the origin of light neutron capture elements.

This is the ﬁrst study of NLTE yttrium abundance trend for extensive metalicity range suitable for the Galactic chemical evolution research. 
We plan to use the current model atom for NLTE yttrium abundance determination in very metal-poor stars studied with LAMOST and Subaru. It will allow us to obtain [Y/Fe] vs. [Fe/H] trend for stars distributed across the metallicity down to -4.0 dex to provide observational constrains on GCE models. 

\acknowledgments
  
We thank the anonymous referee for valuable suggestions and comments.
This work was supported by the National Natural Science Foundation of China under grant Nos. 
11988101, 11890694, 12090044, 11833006, 12022304, 11973052, and National Key R$\&$D Program of China No. 2019YFA0405500. This research used the facilities of the Canadian Astronomy Data Centre operated by the National Research Council of Canada with the support of the Canadian Space Agency. We employed observations carried out with Shane/Hamilton spectrograph (the Lick Observatory). 

Based on observations collected at the European Southern Observatory under ESO programmes 67.C-0206(A), 67.D-0086(A), 072.C-0488(E), 072.C-0033(A), 074.C-0364(A), 076.C-0279(A), 078.D-0492(A), 080.D-0347(A), 082.C-0212(B), 090.B-0605(A), 0100.A-9022(A), 0103.D-0118(A), 0104.C-0863(A), 192.C-0852(A), 196.C-0042(A), 107.22QS.001.

Based on data obtained from the ESO Science Archive Facility with DOIs: 
https://doi.eso.org/10.18727/archive/21; https://doi.eso.org/10.18727/archive/24; https://doi.eso.org/10.18727/archive/33;
https://doi.eso.org/10.18727/archive/50.
 
We made use of the NIST, SIMBAD, and VALD databases.

\software{DETAIL \citep{detail}, SynthV\_NLTE \citep{2019ASPC..518..247T}, binmag \citep{binmag3,2018ascl.soft05015K}, MARCS \citep{2008A&A...486..951G}}.

\appendix

\section{Tables}

 \begin{deluxetable*}{ccccccccccccccc}
 \tablenum{A.1}
 \tablecaption{Effective collision strengths  (dimensionless) for selected temperatures in the range $T$=2000--20000 K, for excitation of the states of Y\ione\ by electrons. \label{ECS_Y1}}
\setlength{\tabcolsep}{2pt}
\tabletypesize{\tiny }
\tablewidth{0pt}
\tablehead{
\colhead{Transition} & \colhead{$J_i - J_j$} & \colhead{} & \colhead{} & \colhead{} & \colhead{} & \colhead{} & \colhead{} &
\colhead{} & \colhead{$T$ (K)} & \colhead{} & \colhead{} & \colhead{} & \colhead{} & \colhead{} \\
\colhead{} & \colhead{} & \colhead{2000} & \colhead{3000} & \colhead{4000} & \colhead{5000} & \colhead{6000} & \colhead{7000} & \colhead{8000} & \colhead{10000} &
\colhead{12000} & \colhead{14000} & \colhead{16000} & \colhead{18000} & \colhead{20000}
}
\decimalcolnumbers
\startdata
\multicolumn{15}c{Effective collision strengths calculated with experimental cross sections from \citet{1984JApSp..40..368K}} \\
a$^4$F -- y$^4$D$^{\circ}$ & 3/2 -- 1/2  & 4.31E-4  &  5.64E-4  &  7.13E-4  &  8.66E-4  &  1.02E-3  &  1.18E-3  &  1.33E-3  &  1.65E-3  &  1.96E-3  &  2.28E-3  &  2.60E-3  &  2.92E-3  &  3.25E-3  \\
a$^4$F -- y$^4$F$^{\circ}$ & 5/2 -- 5/2  & 6.11E-4  &  1.11E-3  &  1.65E-3  &  2.22E-3  &  2.81E-3  &  3.40E-3  &  3.99E-3  &  5.12E-3  &  6.19E-3  &  7.18E-3  &  8.09E-3  &  8.93E-3  &  9.71E-3  \\
a$^4$F -- y$^4$D$^{\circ}$ & 5/2 -- 5/2  & 7.12E-4  &  9.01E-4  &  1.12E-3  &  1.35E-3  &  1.58E-3  &  1.81E-3  &  2.04E-3  &  2.51E-3  &  2.98E-3  &  3.46E-3  &  3.93E-3  &  4.42E-3  &  4.90E-3  \\
a$^4$F -- y$^4$D$^{\circ}$ & 7/2 -- 5/2  & 3.46E-3  &  4.38E-3  &  5.44E-3  &  6.54E-3  &  7.67E-3  &  8.80E-3  &  9.94E-3  &  1.22E-2  &  1.45E-2  &  1.68E-2  &  1.92E-2  &  2.15E-2  &  2.39E-2  \\
a$^4$F -- x$^2$F$^{\circ}$ & 7/2 -- 5/2  & 2.76E-4  &  5.97E-4  &  9.80E-4  &  1.40E-3  &  1.85E-3  &  2.31E-3  &  2.79E-3  &  3.79E-3  &  4.83E-3  &  5.93E-3  &  7.06E-3  &  8.25E-3  &  9.48E-3  \\
a$^4$F -- y$^4$D$^{\circ}$ & 7/2 -- 7/2  & 2.77E-4  &  5.99E-4  &  9.82E-4  &  1.40E-3  &  1.85E-3  &  2.31E-3  &  2.79E-3  &  3.79E-3  &  4.84E-3  &  5.93E-3  &  7.07E-3  &  8.25E-3  &  9.48E-3  \\
a$^4$F -- z$^2$G$^{\circ}$ & 9/2 -- 7/2  & 1.18E-2  &  9.20E-3  &  8.90E-3  &  9.20E-3  &  9.73E-3  &  1.04E-2  &  1.11E-2  &  1.26E-2  &  1.42E-2  &  1.58E-2  &  1.74E-2  &  1.91E-2  &  2.08E-2  \\
a$^4$F -- y$^4$D$^{\circ}$ & 9/2 -- 7/2  & 2.53E-3  &  3.82E-3  &  5.18E-3  &  6.57E-3  &  7.97E-3  &  9.38E-3  &  1.08E-2  &  1.36E-2  &  1.65E-2  &  1.94E-2  &  2.23E-2  &  2.52E-2  &  2.82E-2  \\
z$^4$P$^{\circ}$ -- f$^4$D & 5/2 -- 7/2  & 2.00E-3  &  2.22E-3  &  2.61E-3  &  3.04E-3  &  3.51E-3  &  4.00E-3  &  4.51E-3  &  5.56E-3  &  6.62E-3  &  7.69E-3  &  8.76E-3  &  9.81E-3  &  1.08E-2  \\
\multicolumn{15}c{Effective collision strengths calculated with experimental cross sections from \citet{Smirnov2000}} \\
a$^2$D -- z$^2$D$^{\circ}$ & 3/2 -- 5/2  & 5.69E-3  &  9.14E-3  &  1.29E-2  &  1.70E-2  &  2.12E-2  &  2.56E-2  &  3.02E-2  &  3.94E-2  &  4.89E-2  &  5.83E-2  &  6.78E-2  &  7.71E-2  &  8.64E-2  \\
a$^2$D -- z$^2$D$^{\circ}$ & 3/2 -- 3/2  & 1.50E-2  &  2.75E-2  &  4.18E-2  &  5.76E-2  &  7.43E-2  &  9.17E-2  &  1.09E-1  &  1.45E-1  &  1.82E-1  &  2.18E-1  &  2.54E-1  &  2.90E-1  &  3.25E-1  \\
a$^2$D -- z$^2$F$^{\circ}$ & 3/2 -- 5/2  & 1.22E-2  &  2.38E-2  &  3.66E-2  &  5.02E-2  &  6.46E-2  &  7.96E-2  &  9.52E-2  &  1.28E-1  &  1.62E-1  &  1.97E-1  &  2.33E-1  &  2.70E-1  &  3.08E-1  \\
a$^2$D -- y$^2$D$^{\circ}$ & 3/2 -- 3/2  & 7.25E-2  &  1.26E-1  &  1.85E-1  &  2.45E-1  &  3.08E-1  &  3.73E-1  &  4.38E-1  &  5.73E-1  &  7.11E-1  &  8.51E-1  &  9.94E-1  &  1.14E-0  &  1.28E-0  \\
a$^2$D -- y$^2$P$^{\circ}$ & 3/2 -- 3/2  & 5.81E-3  &  1.11E-2  &  1.68E-2  &  2.28E-2  &  2.90E-2  &  3.53E-2  &  4.19E-2  &  5.56E-2  &  6.99E-2  &  8.48E-2  &  1.00E-1  &  1.16E-1  &  1.33E-1  \\
a$^2$D -- y$^2$F$^{\circ}$ & 3/2 -- 5/2  & 9.93E-2  &  1.54E-1  &  2.10E-1  &  2.69E-1  &  3.29E-1  &  3.92E-1  &  4.55E-1  &  5.87E-1  &  7.22E-1  &  8.61E-1  &  1.00E-0  &  1.15E-0  &  1.29E-0  \\
a$^2$D -- y$^2$P$^{\circ}$ & 3/2 -- 1/2  & 9.07E-3  &  1.75E-2  &  2.68E-2  &  3.66E-2  &  4.66E-2  &  5.69E-2  &  6.73E-2  &  8.85E-2  &  1.10E-1  &  1.33E-1  &  1.55E-1  &  1.79E-1  &  2.03E-1  \\
a$^2$D -- y$^2$D$^{\circ}$ & 3/2 -- 5/2  & 5.07E-3  &  9.23E-3  &  1.37E-2  &  1.84E-2  &  2.33E-2  &  2.84E-2  &  3.36E-2  &  4.42E-2  &  5.52E-2  &  6.64E-2  &  7.79E-2  &  8.95E-2  &  1.01E-1  \\
a$^2$D -- x$^2$P$^{\circ}$ & 3/2 -- 1/2  & 1.05E-2  &  2.24E-2  &  3.60E-2  &  5.05E-2  &  6.55E-2  &  8.09E-2  &  9.68E-2  &  1.30E-1  &  1.64E-1  &  1.99E-1  &  2.36E-1  &  2.73E-1  &  3.12E-1  \\
a$^2$D -- x$^2$P$^{\circ}$ & 3/2 -- 3/2  & 2.20E-3  &  3.72E-3  &  5.42E-3  &  7.26E-3  &  9.22E-3  &  1.13E-2  &  1.34E-2  &  1.79E-2  &  2.25E-2  &  2.74E-2  &  3.25E-2  &  3.78E-2  &  4.32E-2  \\
a$^2$D -- x$^2$F$^{\circ}$ & 3/2 -- 5/2  & 4.15E-3  &  4.65E-3  &  5.49E-3  &  6.49E-3  &  7.60E-3  &  8.79E-3  &  1.00E-2  &  1.26E-2  &  1.53E-2  &  1.80E-2  &  2.06E-2  &  2.33E-2  &  2.59E-2  \\
a$^2$D -- x$^2$D$^{\circ}$ & 3/2 -- 3/2  & 2.78E-3  &  3.89E-3  &  5.09E-3  &  6.34E-3  &  7.63E-3  &  8.95E-3  &  1.03E-2  &  1.30E-2  &  1.58E-2  &  1.87E-2  &  2.16E-2  &  2.45E-2  &  2.75E-2  \\
a$^2$D -- x$^2$D$^{\circ}$ & 3/2 -- 5/2  & 4.29E-4  &  7.88E-4  &  1.16E-3  &  1.55E-3  &  1.94E-3  &  2.34E-3  &  2.74E-3  &  3.56E-3  &  4.39E-3  &  5.25E-3  &  6.11E-3  &  6.99E-3  &  7.88E-3  \\
a$^2$D -- w$^2$D$^{\circ}$ & 3/2 -- 3/2  & 1.88E-3  &  2.96E-3  &  4.13E-3  &  5.33E-3  &  6.54E-3  &  7.75E-3  &  8.93E-3  &  1.12E-2  &  1.34E-2  &  1.54E-2  &  1.73E-2  &  1.91E-2  &  2.07E-2  \\
a$^2$D -- z$^2$D$^{\circ}$ & 5/2 -- 5/2  & 6.11E-2  &  9.85E-2  &  1.39E-1  &  1.83E-1  &  2.29E-1  &  2.78E-1  &  3.27E-1  &  4.28E-1  &  5.32E-1  &  6.35E-1  &  7.39E-1  &  8.41E-1  &  9.43E-1  \\
a$^2$D -- z$^2$F$^{\circ}$ & 5/2 -- 5/2  & 1.88E-3  &  3.67E-3  &  5.64E-3  &  7.74E-3  &  9.96E-3  &  1.23E-2  &  1.47E-2  &  1.98E-2  &  2.51E-2  &  3.05E-2  &  3.62E-2  &  4.19E-2  &  4.77E-2  \\
a$^2$D -- z$^2$F$^{\circ}$ & 5/2 -- 7/2  & 1.64E-2  &  3.18E-2  &  4.82E-2  &  6.51E-2  &  8.23E-2  &  9.98E-2  &  1.18E-1  &  1.54E-1  &  1.91E-1  &  2.29E-1  &  2.68E-1  &  3.08E-1  &  3.48E-1  \\
a$^2$D -- y$^2$D$^{\circ}$ & 5/2 -- 3/2  & 2.20E-2  &  3.85E-2  &  5.62E-2  &  7.48E-2  &  9.40E-2  &  1.14E-1  &  1.34E-1  &  1.75E-1  &  2.17E-1  &  2.60E-1  &  3.04E-1  &  3.49E-1  &  3.94E-1  \\
a$^2$D -- y$^2$P$^{\circ}$ & 5/2 -- 3/2  & 1.24E-2  &  2.38E-2  &  3.61E-2  &  4.89E-2  &  6.22E-2  &  7.59E-2  &  9.00E-2  &  1.20E-1  &  1.50E-1  &  1.83E-1  &  2.16E-1  &  2.50E-1  &  2.86E-1  \\
a$^2$D -- y$^2$F$^{\circ}$ & 5/2 -- 5/2  & 4.09E-2  &  6.33E-2  &  8.66E-2  &  1.11E-1  &  1.36E-1  &  1.62E-1  &  1.88E-1  &  2.43E-1  &  2.99E-1  &  3.57E-1  &  4.16E-1  &  4.76E-1  &  5.37E-1  \\
a$^2$D -- y$^2$D$^{\circ}$ & 5/2 -- 5/2  & 8.52E-2  &  1.55E-1  &  2.31E-1  &  3.11E-1  &  3.93E-1  &  4.79E-1  &  5.67E-1  &  7.47E-1  &  9.33E-1  &  1.12E-0  &  1.32E-0  &  1.52E-0  &  1.72E-0  \\
a$^2$D -- y$^2$F$^{\circ}$ & 5/2 -- 7/2  & 1.75E-1  &  2.47E-1  &  3.21E-1  &  3.99E-1  &  4.78E-1  &  5.61E-1  &  6.45E-1  &  8.21E-1  &  1.00E-0  &  1.19E-0  &  1.39E-0  &  1.59E-0  &  1.79E-0  \\
a$^2$D -- x$^2$P$^{\circ}$ & 5/2 -- 3/2  & 3.30E-2  &  5.58E-2  &  8.15E-2  &  1.09E-1  &  1.39E-1  &  1.70E-1  &  2.02E-1  &  2.69E-1  &  3.40E-1  &  4.14E-1  &  4.91E-1  &  5.71E-1  &  6.53E-1  \\
a$^2$D -- x$^2$F$^{\circ}$ & 5/2 -- 5/2  & 5.19E-4  &  5.82E-4  &  6.87E-4  &  8.13E-4  &  9.53E-4  &  1.10E-3  &  1.26E-3  &  1.58E-3  &  1.92E-3  &  2.26E-3  &  2.59E-3  &  2.92E-3  &  3.25E-3  \\
a$^2$D -- x$^2$D$^{\circ}$ & 5/2 -- 3/2  & 7.86E-4  &  1.10E-3  &  1.44E-3  &  1.80E-3  &  2.16E-3  &  2.54E-3  &  2.92E-3  &  3.70E-3  &  4.50E-3  &  5.31E-3  &  6.13E-3  &  6.97E-3  &  7.81E-3  \\
a$^2$D -- x$^2$F$^{\circ}$ & 5/2 -- 7/2  & 8.14E-3  &  1.10E-2  &  1.42E-2  &  1.75E-2  &  2.10E-2  &  2.45E-2  &  2.82E-2  &  3.58E-2  &  4.36E-2  &  5.15E-2  &  5.93E-2  &  6.70E-2  &  7.46E-2  \\
a$^2$D -- x$^2$D$^{\circ}$ & 5/2 -- 5/2  & 2.13E-3  &  3.90E-3  &  5.78E-3  &  7.70E-3  &  9.65E-3  &  1.16E-2  &  1.36E-2  &  1.77E-2  &  2.19E-2  &  2.61E-2  &  3.04E-2  &  3.48E-2  &  3.93E-2  \\
a$^2$D -- w$^2$P$^{\circ}$ & 5/2 -- 3/2  & 3.27E-4  &  6.24E-4  &  9.67E-4  &  1.34E-3  &  1.73E-3  &  2.13E-3  &  2.55E-3  &  3.41E-3  &  4.31E-3  &  5.24E-3  &  6.19E-3  &  7.16E-3  &  8.16E-3  \\
a$^2$D -- w$^2$F$^{\circ}$ & 5/2 -- 7/2  & 7.47E-2  &  7.99E-2  &  8.32E-2  &  8.59E-2  &  8.86E-2  &  9.11E-2  &  9.36E-2  &  9.81E-2  &  1.02E-1  &  1.06E-1  &  1.10E-1  &  1.13E-1  &  1.17E-1  \\
a$^2$F -- x$^2$P$^{\circ}$ & 5/2 -- 3/2  & 4.33E-4  &  7.66E-4  &  1.16E-3  &  1.60E-3  &  2.09E-3  &  2.62E-3  &  3.19E-3  &  4.41E-3  &  5.74E-3  &  7.18E-3  &  8.71E-3  &  1.03E-2  &  1.20E-2  \\
a$^2$F -- x$^2$F$^{\circ}$ & 5/2 -- 5/2  & 2.13E-3  &  2.45E-3  &  2.98E-3  &  3.61E-3  &  4.32E-3  &  5.10E-3  &  5.92E-3  &  7.67E-3  &  9.52E-3  &  1.14E-2  &  1.33E-2  &  1.53E-2  &  1.72E-2  \\
a$^2$F -- x$^2$D$^{\circ}$ & 5/2 -- 3/2  & 1.34E-3  &  1.93E-3  &  2.58E-3  &  3.28E-3  &  4.02E-3  &  4.80E-3  &  5.61E-3  &  7.30E-3  &  9.08E-3  &  1.09E-2  &  1.29E-2  &  1.48E-2  &  1.69E-2  \\
a$^2$F -- w$^2$D$^{\circ}$ & 5/2 -- 3/2  & 1.67E-3  &  2.69E-3  &  3.82E-3  &  5.02E-3  &  6.26E-3  &  7.50E-3  &  8.75E-3  &  1.12E-2  &  1.36E-2  &  1.59E-2  &  1.81E-2  &  2.01E-2  &  2.21E-2  \\
a$^2$F -- w$^2$F$^{\circ}$ & 5/2 -- 7/2  & 9.91E-3  &  1.07E-2  &  1.13E-2  &  1.18E-2  &  1.22E-2  &  1.27E-2  &  1.32E-2  &  1.41E-2  &  1.49E-2  &  1.57E-2  &  1.64E-2  &  1.71E-2  &  1.78E-2  \\
a$^2$F -- x$^2$F$^{\circ}$ & 7/2 -- 7/2  & 3.94E-3  &  5.49E-3  &  7.23E-3  &  9.12E-3  &  1.11E-2  &  1.33E-2  &  1.55E-2  &  2.03E-2  &  2.53E-2  &  3.05E-2  &  3.57E-2  &  4.10E-2  &  4.62E-2  \\
a$^2$F -- x$^2$D$^{\circ}$ & 7/2 -- 5/2  & 1.69E-3  &  3.18E-3  &  4.80E-3  &  6.51E-3  &  8.30E-3  &  1.02E-2  &  1.21E-2  &  1.61E-2  &  2.03E-2  &  2.48E-2  &  2.94E-2  &  3.42E-2  &  3.91E-2  \\
b$^2$D -- x$^2$D$^{\circ}$ & 3/2 -- 3/2  & 5.94E-4  &  8.55E-4  &  1.15E-3  &  1.46E-3  &  1.80E-3  &  2.14E-3  &  2.51E-3  &  3.27E-3  &  4.08E-3  &  4.91E-3  &  5.78E-3  &  6.68E-3  &  7.60E-3  \\
b$^2$D -- w$^2$D$^{\circ}$ & 3/2 -- 3/2  & 1.23E-3  &  1.99E-3  &  2.83E-3  &  3.72E-3  &  4.64E-3  &  5.57E-3  &  6.50E-3  &  8.33E-3  &  1.01E-2  &  1.18E-2  &  1.35E-2  &  1.50E-2  &  1.65E-2  \\
b$^2$D -- w$^2$F$^{\circ}$ & 3/2 -- 5/2  & 7.07E-5  &  1.71E-4  &  2.98E-4  &  4.43E-4  &  6.00E-4  &  7.66E-4  &  9.41E-4  &  1.31E-3  &  1.71E-3  &  2.13E-3  &  2.58E-3  &  3.04E-3  &  3.52E-3  \\
b$^2$D -- x$^2$D$^{\circ}$ & 5/2 -- 5/2  & 5.77E-5  &  1.09E-4  &  1.64E-4  &  2.23E-4  &  2.84E-4  &  3.48E-4  &  4.14E-4  &  5.52E-4  &  6.98E-4  &  8.51E-4  &  1.01E-3  &  1.17E-3  &  1.34E-3  \\
b$^2$D -- w$^2$D$^{\circ}$ & 5/2 -- 5/2  & 1.43E-3  &  2.46E-3  &  3.57E-3  &  4.71E-3  &  5.88E-3  &  7.07E-3  &  8.28E-3  &  1.07E-2  &  1.31E-2  &  1.55E-2  &  1.78E-2  &  2.00E-2  &  2.22E-2  \\
b$^2$D -- w$^2$P$^{\circ}$ & 5/2 -- 3/2  & 2.08E-4  &  4.07E-4  &  6.43E-4  &  9.05E-4  &  1.19E-3  &  1.48E-3  &  1.80E-3  &  2.46E-3  &  3.17E-3  &  3.92E-3  &  4.71E-3  &  5.53E-3  &  6.38E-3  \\
b$^2$D -- w$^2$F$^{\circ}$ & 5/2 -- 7/2  & 8.41E-2  &  9.10E-2  &  9.58E-2  &  1.00E-1  &  1.04E-1  &  1.09E-1  &  1.13E-1  &  1.20E-1  &  1.27E-1  &  1.34E-1  &  1.41E-1  &  1.47E-1  &  1.53E-1  \\
a$^2$G -- z$^2$H$^{\circ}$ & 9/2 -- 11/2 & 3.02E-4  &  1.34E-3  &  3.14E-3  &  5.56E-3  &  8.43E-3  &  1.16E-2  &  1.50E-2  &  2.21E-2  &  2.93E-2  &  3.63E-2  &  4.29E-2  &  4.92E-2  &  5.51E-2  \\
a$^2$G -- y$^2$G$^{\circ}$ & 9/2 -- 9/2  & 7.25E-4  &  2.14E-3  &  4.05E-3  &  6.22E-3  &  8.50E-3  &  1.08E-2  &  1.31E-2  &  1.75E-2  &  2.17E-2  &  2.57E-2  &  2.95E-2  &  3.31E-2  &  3.65E-2  \\
a$^2$G -- z$^2$G$^{\circ}$ & 7/2 -- 7/2  & 5.52E-3  &  9.70E-3  &  1.42E-2  &  1.90E-2  &  2.38E-2  &  2.86E-2  &  3.34E-2  &  4.27E-2  &  5.15E-2  &  5.99E-2  &  6.81E-2  &  7.59E-2  &  8.36E-2  \\
a$^2$G -- z$^2$H$^{\circ}$ & 7/2 -- 9/2  & 2.29E-4  &  9.96E-4  &  2.33E-3  &  4.14E-3  &  6.35E-3  &  8.85E-3  &  1.16E-2  &  1.74E-2  &  2.34E-2  &  2.93E-2  &  3.50E-2  &  4.05E-2  &  4.56E-2  \\
a$^2$G -- y$^2$G$^{\circ}$ & 7/2 -- 7/2  & 9.69E-4  &  2.41E-3  &  4.21E-3  &  6.20E-3  &  8.28E-3  &  1.04E-2  &  1.24E-2  &  1.65E-2  &  2.03E-2  &  2.39E-2  &  2.73E-2  &  3.05E-2  &  3.36E-2  \\\hline
\enddata
\end{deluxetable*}

\begin{longrotatetable}
 \begin{deluxetable*}{ccccccccccccccccc}
  \tablenum{A.2}
 \tablecaption{Effective collision strengths for selected temperatures in the range $T$=2000--28000 K, for excitation of the states of Y\ii\ by electrons. \label{ECS_Y2}}
\setlength{\tabcolsep}{2pt}
\tabletypesize{\tiny }
\tablewidth{0pt}
\tablehead{
\colhead{Transition} & \colhead{$J_i - J_j$} & \colhead{} & \colhead{} & \colhead{} & \colhead{} & \colhead{} & \colhead{} &
\colhead{} & \colhead{$T$ (K)} & \colhead{} & \colhead{} & \colhead{} & \colhead{} & \colhead{} & \colhead{} & \colhead{} \\
\colhead{} & \colhead{} & \colhead{2000} & \colhead{3000} & \colhead{4000} & \colhead{5000} & \colhead{6000} & \colhead{7000} & \colhead{8000} & \colhead{10000} &
\colhead{12000} & \colhead{14000} & \colhead{16000} & \colhead{18000} & \colhead{20000} & \colhead{24000} & \colhead{28000}
}
\decimalcolnumbers
\startdata
\multicolumn{17}c{Effective collision strengths calculated with experimental cross sections from \citet{2002OptSp..93..351S}} \\
y$^3$P$^{\circ}$ -- 7s $^3$D& 2 -- 3  &  2.42E-08 &  2.01E-06  & 2.07E-05 & 8.93E-05 & 2.46E-04 & 5.21E-04 & 9.33E-04 & 2.19E-03 & 4.02E-03 & 6.35E-03 & 9.12E-03 & 1.23E-02 & 1.57E-02 & 2.32E-02 & 3.13E-02  \\
z$^3$P$^{\circ}$ -- g$^3$D  & 0 -- 1  &  6.25E-17 &  1.31E-12  & 2.14E-10 & 4.88E-09 & 4.08E-08 & 1.91E-07 & 6.20E-07 & 3.36E-06 & 1.08E-05 & 2.55E-05 & 4.96E-05 & 8.47E-05 & 1.32E-04 & 2.63E-04 & 4.44E-04  \\
z$^3$P$^{\circ}$ -- g$^3$D  & 1 -- 2  &  1.96E-16 &  4.08E-12  & 6.67E-10 & 1.52E-08 & 1.26E-07 & 5.91E-07 & 1.92E-06 & 1.04E-05 & 3.32E-05 & 7.84E-05 & 1.53E-04 & 2.60E-04 & 4.05E-04 & 8.08E-04 & 1.36E-03  \\
z$^3$P$^{\circ}$ -- g$^3$D  & 2 -- 3  &  3.67E-16 &  7.43E-12  & 1.20E-09 & 2.71E-08 & 2.25E-07 & 1.05E-06 & 3.39E-06 & 1.83E-05 & 5.84E-05 & 1.38E-04 & 2.68E-04 & 4.57E-04 & 7.11E-04 & 1.42E-03 & 2.39E-03  \\
z$^3$P$^{\circ}$ -- e$^3$P  & 1 -- 2  &  1.47E-18 &  1.08E-13  & 3.31E-11 & 1.10E-09 & 1.18E-08 & 6.63E-08 & 2.46E-07 & 1.61E-06 & 5.87E-06 & 1.52E-05 & 3.16E-05 & 5.70E-05 & 9.24E-05 & 1.97E-04 & 3.47E-04  \\
z$^3$P$^{\circ}$ -- e$^3$P  & 0 -- 1  &  2.16E-19 &  1.80E-14  & 5.97E-12 & 2.07E-10 & 2.30E-09 & 1.31E-08 & 4.95E-08 & 3.30E-07 & 1.21E-06 & 3.16E-06 & 6.60E-06 & 1.19E-05 & 1.93E-05 & 4.12E-05 & 7.29E-05  \\
z$^3$D$^{\circ}$ -- f$^3$P  & 2 -- 1  &  1.51E-17 &  4.43E-13  & 8.74E-11 & 2.22E-09 & 1.99E-08 & 9.82E-08 & 3.31E-07 & 1.89E-06 & 6.26E-06 & 1.51E-05 & 3.00E-05 & 5.18E-05 & 8.13E-05 & 1.64E-04 & 2.80E-04  \\
z$^3$D$^{\circ}$ -- f$^3$P  & 3 -- 2  &  2.67E-17 &  7.24E-13  & 1.37E-10 & 3.40E-09 & 3.01E-08 & 1.46E-07 & 4.89E-07 & 2.76E-06 & 9.07E-06 & 2.18E-05 & 4.30E-05 & 7.41E-05 & 1.16E-04 & 2.34E-04 & 3.96E-04  \\
z$^3$P$^{\circ}$ -- e$^3$P  & 1 -- 1  &  1.16E-18 &  9.65E-14  & 3.19E-11 & 1.11E-09 & 1.23E-08 & 7.02E-08 & 2.65E-07 & 1.76E-06 & 6.49E-06 & 1.69E-05 & 3.53E-05 & 6.37E-05 & 1.04E-04 & 2.21E-04 & 3.90E-04  \\
z$^1$D$^{\circ}$ -- e$^3$F  & 2 -- 2  &  3.18E-18 &  1.95E-13  & 5.55E-11 & 1.76E-09 & 1.83E-08 & 1.00E-07 & 3.66E-07 & 2.33E-06 & 8.32E-06 & 2.12E-05 & 4.37E-05 & 7.79E-05 & 1.25E-04 & 2.62E-04 & 4.58E-04  \\
z$^3$P$^{\circ}$ -- e$^3$P  & 2 -- 2  &  6.33E-18 &  4.63E-13  & 1.43E-10 & 4.74E-09 & 5.10E-08 & 2.86E-07 & 1.06E-06 & 6.95E-06 & 2.53E-05 & 6.55E-05 & 1.36E-04 & 2.46E-04 & 3.99E-04 & 8.49E-04 & 1.50E-03  \\
z$^3$P$^{\circ}$ -- e$^3$P  & 1 -- 0  &  1.13E-18 &  9.18E-14  & 2.99E-11 & 1.03E-09 & 1.13E-08 & 6.42E-08 & 2.41E-07 & 1.60E-06 & 5.85E-06 & 1.52E-05 & 3.17E-05 & 5.70E-05 & 9.24E-05 & 1.96E-04 & 3.44E-04  \\
z$^3$F$^{\circ}$ -- e$^3$F  & 3 -- 3  &  1.28E-18 &  1.06E-13  & 3.51E-11 & 1.22E-09 & 1.35E-08 & 7.68E-08 & 2.89E-07 & 1.92E-06 & 7.05E-06 & 1.83E-05 & 3.83E-05 & 6.91E-05 & 1.12E-04 & 2.39E-04 & 4.21E-04  \\
z$^3$F$^{\circ}$ -- e$^3$F  & 2 -- 2  &  1.18E-18 &  7.23E-14  & 2.06E-11 & 6.51E-10 & 6.78E-09 & 3.71E-08 & 1.36E-07 & 8.64E-07 & 3.09E-06 & 7.87E-06 & 1.62E-05 & 2.89E-05 & 4.65E-05 & 9.75E-05 & 1.70E-04  \\
z$^3$P$^{\circ}$ -- e$^3$S  & 2 -- 1  &  1.90E-17 &  1.18E-12  & 3.33E-10 & 1.05E-08 & 1.09E-07 & 5.93E-07 & 2.16E-06 & 1.37E-05 & 4.86E-05 & 1.23E-04 & 2.53E-04 & 4.49E-04 & 7.19E-04 & 1.50E-03 & 2.59E-03  \\
y$^3$P$^{\circ}$ -- g$^3$D  & 1 -- 2  &  2.04E-16 &  3.05E-12  & 4.23E-10 & 8.69E-09 & 6.77E-08 & 3.01E-07 & 9.41E-07 & 4.82E-06 & 1.49E-05 & 3.42E-05 & 6.51E-05 & 1.09E-04 & 1.67E-04 & 3.26E-04 & 5.39E-04  \\
z$^1$D$^{\circ}$ -- e$^3$G  & 2 -- 3  &  8.58E-19 &  7.45E-14  & 2.50E-11 & 8.73E-10 & 9.72E-09 & 5.59E-08 & 2.12E-07 & 1.43E-06 & 5.29E-06 & 1.39E-05 & 2.93E-05 & 5.33E-05 & 8.73E-05 & 1.88E-04 & 3.36E-04  \\
z$^3$D$^{\circ}$ -- e$^3$F  & 2 -- 3  &  6.16E-18 &  3.65E-13  & 1.02E-10 & 3.18E-09 & 3.29E-08 & 1.79E-07 & 6.52E-07 & 4.14E-06 & 1.47E-05 & 3.75E-05 & 7.71E-05 & 1.37E-04 & 2.21E-04 & 4.63E-04 & 8.08E-04  \\
z$^3$D$^{\circ}$ -- e$^3$F  & 1 -- 2  &  2.50E-18 &  1.54E-13  & 4.37E-11 & 1.38E-09 & 1.44E-08 & 7.89E-08 & 2.88E-07 & 1.84E-06 & 6.57E-06 & 1.68E-05 & 3.45E-05 & 6.16E-05 & 9.91E-05 & 2.08E-04 & 3.63E-04  \\
z$^3$D$^{\circ}$ -- e$^3$F  & 3 -- 4  &  7.91E-18 &  4.38E-13  & 1.18E-10 & 3.62E-09 & 3.69E-08 & 1.99E-07 & 7.18E-07 & 4.51E-06 & 1.60E-05 & 4.04E-05 & 8.28E-05 & 1.47E-04 & 2.36E-04 & 4.93E-04 & 8.59E-04  \\
z$^3$F$^{\circ}$ -- e$^3$G  & 4 -- 5  &  8.87E-18 &  8.00E-13  & 2.75E-10 & 9.76E-09 & 1.10E-07 & 6.34E-07 & 2.41E-06 & 1.63E-05 & 6.05E-05 & 1.59E-04 & 3.34E-04 & 6.04E-04 & 9.85E-04 & 2.11E-03 & 3.73E-03  \\
z$^3$D$^{\circ}$ -- f$^3$D  & 3 -- 1  &  5.64E-19 &  7.34E-14  & 3.03E-11 & 1.20E-09 & 1.45E-08 & 8.80E-08 & 3.48E-07 & 2.50E-06 & 9.66E-06 & 2.61E-05 & 5.64E-05 & 1.04E-04 & 1.73E-04 & 3.81E-04 & 6.88E-04  \\
z$^3$D$^{\circ}$ -- f$^3$D  & 3 -- 3  &  9.21E-19 &  1.25E-13  & 5.33E-11 & 2.16E-09 & 2.66E-08 & 1.65E-07 & 6.60E-07 & 4.82E-06 & 1.89E-05 & 5.18E-05 & 1.13E-04 & 2.10E-04 & 3.49E-04 & 7.73E-04 & 1.40E-03  \\
z$^3$D$^{\circ}$ -- e$^3$G  & 3 -- 3  &  1.37E-18 &  1.19E-13  & 3.98E-11 & 1.39E-09 & 1.55E-08 & 8.92E-08 & 3.38E-07 & 2.28E-06 & 8.48E-06 & 2.23E-05 & 4.71E-05 & 8.56E-05 & 1.40E-04 & 3.03E-04 & 5.42E-04  \\
z$^1$D$^{\circ}$ -- e$^3$D  & 2 -- 2  &  2.87E-19 &  3.84E-14  & 1.61E-11 & 6.42E-10 & 7.80E-09 & 4.77E-08 & 1.90E-07 & 1.37E-06 & 5.31E-06 & 1.44E-05 & 3.12E-05 & 5.80E-05 & 9.65E-05 & 2.14E-04 & 3.90E-04  \\
z$^3$F$^{\circ}$ -- e$^3$D  & 3 -- 3  &  1.68E-19 &  3.02E-14  & 1.46E-11 & 6.39E-10 & 8.26E-09 & 5.27E-08 & 2.16E-07 & 1.62E-06 & 6.46E-06 & 1.78E-05 & 3.90E-05 & 7.28E-05 & 1.22E-04 & 2.70E-04 & 4.92E-04  \\
z$^3$F$^{\circ}$ -- e$^3$D  & 2 -- 1  &  1.11E-18 &  1.17E-13  & 4.33E-11 & 1.60E-09 & 1.85E-08 & 1.09E-07 & 4.20E-07 & 2.89E-06 & 1.09E-05 & 2.88E-05 & 6.09E-05 & 1.11E-04 & 1.82E-04 & 3.92E-04 & 6.97E-04  \\
z$^3$F$^{\circ}$ -- e$^3$D  & 3 -- 2  &  7.10E-19 &  9.49E-14  & 3.97E-11 & 1.59E-09 & 1.93E-08 & 1.18E-07 & 4.69E-07 & 3.38E-06 & 1.31E-05 & 3.57E-05 & 7.74E-05 & 1.44E-04 & 2.39E-04 & 5.31E-04 & 9.68E-04  \\
y$^3$P$^{\circ}$ -- e$^3$P  & 2 -- 2  &  1.19E-18 &  8.75E-14  & 2.70E-11 & 8.99E-10 & 9.68E-09 & 5.43E-08 & 2.02E-07 & 1.33E-06 & 4.84E-06 & 1.26E-05 & 2.62E-05 & 4.73E-05 & 7.69E-05 & 1.64E-04 & 2.91E-04  \\
z$^3$F$^{\circ}$ -- e$^3$D  & 4 -- 3  &  7.40E-19 &  1.33E-13  & 6.45E-11 & 2.82E-09 & 3.64E-08 & 2.32E-07 & 9.53E-07 & 7.16E-06 & 2.85E-05 & 7.87E-05 & 1.72E-04 & 3.21E-04 & 5.37E-04 & 1.19E-03 & 2.17E-03  \\
y$^3$P$^{\circ}$ -- e$^3$G  & 2 -- 3  &  4.14E-19 &  3.61E-14  & 1.21E-11 & 4.24E-10 & 4.72E-09 & 2.72E-08 & 1.03E-07 & 6.96E-07 & 2.59E-06 & 6.82E-06 & 1.44E-05 & 2.62E-05 & 4.30E-05 & 9.30E-05 & 1.67E-04  \\
z$^3$D$^{\circ}$ -- e$^3$D  & 3 -- 3  &  7.14E-19 &  1.28E-13  & 6.22E-11 & 2.72E-09 & 3.51E-08 & 2.24E-07 & 9.21E-07 & 6.91E-06 & 2.76E-05 & 7.60E-05 & 1.66E-04 & 3.11E-04 & 5.19E-04 & 1.16E-03 & 2.10E-03  \\
z$^3$D$^{\circ}$ -- e$^3$D  & 1 -- 1  &  6.08E-19 &  6.41E-14  & 2.38E-11 & 8.81E-10 & 1.02E-08 & 5.99E-08 & 2.31E-07 & 1.59E-06 & 5.98E-06 & 1.58E-05 & 3.35E-05 & 6.11E-05 & 1.00E-04 & 2.16E-04 & 3.85E-04  \\
z$^3$D$^{\circ}$ -- e$^3$D  & 2 -- 2  &  4.01E-19 &  5.37E-14  & 2.25E-11 & 8.98E-10 & 1.09E-08 & 6.69E-08 & 2.66E-07 & 1.92E-06 & 7.45E-06 & 2.03E-05 & 4.39E-05 & 8.16E-05 & 1.36E-04 & 3.02E-04 & 5.50E-04  \\
z$^3$D$^{\circ}$ -- e$^3$D  & 2 -- 1  &  5.50E-19 &  5.81E-14  & 2.15E-11 & 7.97E-10 & 9.20E-09 & 5.42E-08 & 2.09E-07 & 1.44E-06 & 5.42E-06 & 1.43E-05 & 3.04E-05 & 5.54E-05 & 9.07E-05 & 1.96E-04 & 3.49E-04  \\
y$^3$P$^{\circ}$ -- e$^3$S  & 1 -- 1  &  2.39E-18 &  1.48E-13  & 4.21E-11 & 1.33E-09 & 1.38E-08 & 7.52E-08 & 2.74E-07 & 1.74E-06 & 6.20E-06 & 1.58E-05 & 3.24E-05 & 5.77E-05 & 9.25E-05 & 1.93E-04 & 3.36E-04  \\
z$^1$F$^{\circ}$ -- f$^3$D  & 3 -- 3  &  2.91E-19 &  3.96E-14  & 1.69E-11 & 6.86E-10 & 8.45E-09 & 5.23E-08 & 2.10E-07 & 1.53E-06 & 6.03E-06 & 1.65E-05 & 3.60E-05 & 6.71E-05 & 1.12E-04 & 2.48E-04 & 4.50E-04  \\
y$^3$P$^{\circ}$ -- e$^3$S  & 2 -- 1  &  6.46E-18 &  4.02E-13  & 1.14E-10 & 3.59E-09 & 3.73E-08 & 2.04E-07 & 7.42E-07 & 4.72E-06 & 1.68E-05 & 4.27E-05 & 8.78E-05 & 1.56E-04 & 2.51E-04 & 5.24E-04 & 9.09E-04  \\
z$^3$D$^{\circ}$ -- e$^3$D  & 3 -- 2  &  5.23E-19 &  6.99E-14  & 2.93E-11 & 1.17E-09 & 1.42E-08 & 8.71E-08 & 3.46E-07 & 2.50E-06 & 9.71E-06 & 2.64E-05 & 5.73E-05 & 1.06E-04 & 1.77E-04 & 3.93E-04 & 7.18E-04  \\
y$^3$P$^{\circ}$ -- e$^3$D  & 1 -- 2  &  1.12E-19 &  1.50E-14  & 6.29E-12 & 2.52E-10 & 3.06E-09 & 1.88E-08 & 7.46E-08 & 5.39E-07 & 2.10E-06 & 5.71E-06 & 1.24E-05 & 2.30E-05 & 3.84E-05 & 8.54E-05 & 1.56E-04  \\
z$^1$F$^{\circ}$ -- e$^3$D  & 3 -- 2  &  1.49E-19 &  2.00E-14  & 8.37E-12 & 3.35E-10 & 4.07E-09 & 2.50E-08 & 9.93E-08 & 7.17E-07 & 2.79E-06 & 7.61E-06 & 1.65E-05 & 3.07E-05 & 5.12E-05 & 1.14E-04 & 2.08E-04  \\
\multicolumn{17}c{Effective collision strengths calculated with experimental cross sections from \citet{Smirnov2001}}  \\
 a$^1$S --  y$^1$P$^{\circ}$ &  0–1   &  5.40E-19 &  1.19E-13  & 6.31E-11 & 2.88E-09 & 3.82E-08 & 2.48E-07 & 1.03E-06 & 7.86E-06 & 3.16E-05 & 8.75E-05 & 1.92E-04 & 3.58E-04 & 5.99E-04 & 1.33E-03 & 2.41E-03  \\
 a$^1$D --  y$^1$P$^{\circ}$ &  2–1   &  1.45E-18 &  3.22E-13  & 1.70E-10 & 7.78E-09 & 1.03E-07 & 6.71E-07 & 2.78E-06 & 2.12E-05 & 8.54E-05 & 2.37E-04 & 5.19E-04 & 9.71E-04 & 1.62E-03 & 3.61E-03 & 6.56E-03  \\
 z$^1$D$^{\circ}$ -- g$^1$D  &  2–2   &  1.11E-18 &  1.42E-13  & 5.71E-11 & 2.21E-09 & 2.62E-08 & 1.57E-07 & 6.14E-07 & 4.32E-06 & 1.65E-05 & 4.40E-05 & 9.39E-05 & 1.72E-04 & 2.82E-04 & 6.10E-04 & 1.09E-03  \\
 z$^3$P$^{\circ}$ -- f$^1$D  &  2–2   &  5.16E-18 &  4.51E-13  & 1.51E-10 & 5.25E-09 & 5.82E-08 & 3.33E-07 & 1.25E-06 & 8.35E-06 & 3.06E-05 & 7.94E-05 & 1.66E-04 & 2.98E-04 & 4.83E-04 & 1.03E-03 & 1.81E-03  \\
 z$^3$F$^{\circ}$ -- g$^1$D  &  2–2   &  8.28E-19 &  1.05E-13  & 4.25E-11 & 1.64E-09 & 1.95E-08 & 1.17E-07 & 4.57E-07 & 3.22E-06 & 1.23E-05 & 3.28E-05 & 7.00E-05 & 1.28E-04 & 2.11E-04 & 4.56E-04 & 8.10E-04  \\
 z$^1$P$^{\circ}$ -- g$^1$D  &  1–2   &  7.60E-19 &  9.67E-14  & 3.90E-11 & 1.51E-09 & 1.79E-08 & 1.07E-07 & 4.20E-07 & 2.95E-06 & 1.13E-05 & 3.01E-05 & 6.43E-05 & 1.18E-04 & 1.93E-04 & 4.18E-04 & 7.44E-04  \\
 z$^1$D$^{\circ}$ -- f$^1$D  &  2–2   &  1.64E-18 &  1.44E-13  & 4.82E-11 & 1.68E-09 & 1.86E-08 & 1.06E-07 & 4.01E-07 & 2.67E-06 & 9.79E-06 & 2.54E-05 & 5.30E-05 & 9.55E-05 & 1.55E-04 & 3.29E-04 & 5.79E-04  \\
 z$^3$D$^{\circ}$ -- g$^1$D  &  1–2   &  5.31E-19 &  6.76E-14  & 2.72E-11 & 1.05E-09 & 1.25E-08 & 7.51E-08 & 2.94E-07 & 2.07E-06 & 7.88E-06 & 2.11E-05 & 4.50E-05 & 8.25E-05 & 1.36E-04 & 2.93E-04 & 5.22E-04  \\
 z$^1$P$^{\circ}$ -- f$^1$S  &  1–0   &  3.97E-18 &  4.14E-13  & 1.53E-10 & 5.63E-09 & 6.49E-08 & 3.82E-07 & 1.47E-06 & 1.01E-05 & 3.78E-05 & 9.96E-05 & 2.10E-04 & 3.81E-04 & 6.22E-04 & 1.33E-03 & 2.36E-03  \\
 z$^3$F$^{\circ}$ -- f$^1$D  &  2–2   &  1.56E-18 &  1.36E-13  & 4.57E-11 & 1.59E-09 & 1.76E-08 & 1.01E-07 & 3.81E-07 & 2.53E-06 & 9.30E-06 & 2.42E-05 & 5.04E-05 & 9.08E-05 & 1.47E-04 & 3.13E-04 & 5.51E-04  \\
 z$^1$P$^{\circ}$ -- f$^1$D  &  1–2   &  1.92E-18 &  1.68E-13  & 5.63E-11 & 1.96E-09 & 2.18E-08 & 1.24E-07 & 4.69E-07 & 3.12E-06 & 1.15E-05 & 2.98E-05 & 6.22E-05 & 1.12E-04 & 1.82E-04 & 3.86E-04 & 6.80E-04  \\
 z$^3$D$^{\circ}$ -- f$^1$S  &  1–0   &  6.88E-19 &  7.19E-14  & 2.65E-11 & 9.77E-10 & 1.13E-08 & 6.63E-08 & 2.55E-07 & 1.75E-06 & 6.56E-06 & 1.73E-05 & 3.65E-05 & 6.63E-05 & 1.08E-04 & 2.32E-04 & 4.11E-04  \\
 a$^1$D --  z$^1$F$^{\circ}$ &  2–3   &  1.44E-19 &  6.77E-14  & 5.15E-11 & 2.92E-09 & 4.47E-08 & 3.22E-07 & 1.44E-06 & 1.23E-05 & 5.32E-05 & 1.56E-04 & 3.58E-04 & 6.94E-04 & 1.20E-03 & 2.80E-03 & 5.32E-03  \\
 z$^1$F$^{\circ}$ -- e$^1$G  &  3–4   &  3.93E-18 &  3.59E-13  & 1.23E-10 & 4.37E-09 & 4.91E-08 & 2.84E-07 & 1.08E-06 & 7.35E-06 & 2.74E-05 & 7.24E-05 & 1.53E-04 & 2.79E-04 & 4.58E-04 & 9.92E-04 & 1.77E-03  \\
 b$^1$D --  y$^1$P$^{\circ}$ &  2–1   &  1.83E-19 &  4.05E-14  & 2.15E-11 & 9.85E-10 & 1.31E-08 & 8.52E-08 & 3.54E-07 & 2.71E-06 & 1.09E-05 & 3.04E-05 & 6.68E-05 & 1.25E-04 & 2.10E-04 & 4.69E-04 & 8.56E-04  \\
 z$^1$D$^{\circ}$ -- e$^1$D  &  2–2   &  1.19E-19 &  2.25E-14  & 1.12E-11 & 4.92E-10 & 6.39E-09 & 4.10E-08 & 1.69E-07 & 1.27E-06 & 5.10E-06 & 1.41E-05 & 3.11E-05 & 5.85E-05 & 9.84E-05 & 2.22E-04 & 4.09E-04  \\
 a$^1$S --  z$^1$P$^{\circ}$ &  0–1   &  3.51E-20 &  2.07E-14  & 1.77E-11 & 1.07E-09 & 1.72E-08 & 1.28E-07 & 5.92E-07 & 5.24E-06 & 2.33E-05 & 6.99E-05 & 1.63E-04 & 3.19E-04 & 5.55E-04 & 1.31E-03 & 2.49E-03  \\
 z$^1$F$^{\circ}$ -- f$^1$D  &  3–2   &  3.38E-18 &  2.96E-13  & 9.93E-11 & 3.47E-09 & 3.85E-08 & 2.20E-07 & 8.31E-07 & 5.55E-06 & 2.04E-05 & 5.30E-05 & 1.11E-04 & 2.00E-04 & 3.25E-04 & 6.92E-04 & 1.22E-03  \\
 z$^3$D$^{\circ}$ -- e$^1$D  &  1–2   &  4.01E-20 &  7.63E-15  & 3.78E-12 & 1.67E-10 & 2.17E-09 & 1.39E-08 & 5.72E-08 & 4.32E-07 & 1.73E-06 & 4.82E-06 & 1.06E-05 & 1.99E-05 & 3.36E-05 & 7.57E-05 & 1.40E-04  \\
 z$^3$D$^{\circ}$ -- e$^1$D  &  2–2   &  1.19E-19 &  2.26E-14  & 1.12E-11 & 4.95E-10 & 6.43E-09 & 4.12E-08 & 1.70E-07 & 1.28E-06 & 5.14E-06 & 1.43E-05 & 3.14E-05 & 5.91E-05 & 9.95E-05 & 2.24E-04 & 4.15E-04  \\
 a$^3$D --  z$^1$P$^{\circ}$ &  1–1   &  1.69E-20 &  9.93E-15  & 8.50E-12 & 5.17E-10 & 8.29E-09 & 6.18E-08 & 2.85E-07 & 2.52E-06 & 1.12E-05 & 3.37E-05 & 7.84E-05 & 1.54E-04 & 2.68E-04 & 6.32E-04 & 1.20E-03  \\
 a$^3$D --  z$^1$P$^{\circ}$ &  2–1   &  3.20E-20 &  1.88E-14  & 1.61E-11 & 9.79E-10 & 1.57E-08 & 1.17E-07 & 5.40E-07 & 4.78E-06 & 2.13E-05 & 6.38E-05 & 1.49E-04 & 2.92E-04 & 5.07E-04 & 1.20E-03 & 2.28E-03  \\
 y$^3$P$^{\circ}$ -- e$^1$F  &  2–3   &  1.66E-19 &  2.44E-14  & 1.06E-11 & 4.34E-10 & 5.35E-09 & 3.30E-08 & 1.32E-07 & 9.52E-07 & 3.69E-06 & 9.96E-06 & 2.14E-05 & 3.95E-05 & 6.53E-05 & 1.43E-04 & 2.57E-04  \\
 y$^1$P$^{\circ}$ -- h$^1$D  &  1–2   &  1.79E-15 &  1.17E-11  & 1.07E-09 & 1.73E-08 & 1.15E-07 & 4.55E-07 & 1.31E-06 & 6.00E-06 & 1.73E-05 & 3.79E-05 & 6.99E-05 & 1.15E-04 & 1.74E-04 & 3.34E-04 & 5.50E-04  \\
 a$^3$D --  z$^1$D$^{\circ}$ &  1–2   &  4.10E-21 &  4.10E-15  & 4.55E-12 & 3.22E-10 & 5.70E-09 & 4.55E-08 & 2.20E-07 & 2.06E-06 & 9.49E-06 & 2.89E-05 & 6.79E-05 & 1.34E-04 & 2.33E-04 & 5.46E-04 & 1.03E-03  \\
 a$^3$D --  z$^1$D$^{\circ}$ &  2–2   &  7.91E-21 &  7.91E-15  & 8.77E-12 & 6.21E-10 & 1.10E-08 & 8.77E-08 & 4.23E-07 & 3.98E-06 & 1.83E-05 & 5.58E-05 & 1.31E-04 & 2.58E-04 & 4.49E-04 & 1.05E-03 & 1.99E-03  \\
 a$^3$F --  z$^1$F$^{\circ}$ &  4–3   &  2.71E-20 &  1.27E-14  & 9.70E-12 & 5.51E-10 & 8.43E-09 & 6.08E-08 & 2.73E-07 & 2.33E-06 & 1.01E-05 & 2.96E-05 & 6.80E-05 & 1.32E-04 & 2.28E-04 & 5.36E-04 & 1.02E-03  \\
 a$^1$D --  z$^1$D$^{\circ}$ &  2–2   &  2.84E-20 &  2.84E-14  & 3.16E-11 & 2.24E-09 & 3.96E-08 & 3.16E-07 & 1.53E-06 & 1.43E-05 & 6.60E-05 & 2.01E-04 & 4.73E-04 & 9.32E-04 & 1.62E-03 & 3.81E-03 & 7.19E-03  \\
 z$^1$F$^{\circ}$ -- e$^1$D  &  3–2   &  3.80E-19 &  7.23E-14  & 3.59E-11 & 1.58E-09 & 2.06E-08 & 1.32E-07 & 5.45E-07 & 4.12E-06 & 1.66E-05 & 4.61E-05 & 1.02E-04 & 1.91E-04 & 3.22E-04 & 7.29E-04 & 1.35E-03  \\
 a$^3$F --  z$^1$P$^{\circ}$ &  2–1   &  9.84E-21 &  5.79E-15  & 4.96E-12 & 3.02E-10 & 4.86E-09 & 3.63E-08 & 1.67E-07 & 1.48E-06 & 6.63E-06 & 1.99E-05 & 4.64E-05 & 9.13E-05 & 1.59E-04 & 3.77E-04 & 7.20E-04  \\
 a$^3$P --  z$^1$F$^{\circ}$ &  2–3   &  4.68E-21 &  2.20E-15  & 1.68E-12 & 9.55E-11 & 1.46E-09 & 1.06E-08 & 4.75E-08 & 4.05E-07 & 1.76E-06 & 5.18E-06 & 1.19E-05 & 2.32E-05 & 4.01E-05 & 9.43E-05 & 1.80E-04  \\
 b$^1$D --  z$^1$F$^{\circ}$ &  2–3   &  1.18E-20 &  5.56E-15  & 4.25E-12 & 2.41E-10 & 3.70E-09 & 2.67E-08 & 1.20E-07 & 1.03E-06 & 4.46E-06 & 1.31E-05 & 3.01E-05 & 5.87E-05 & 1.02E-04 & 2.39E-04 & 4.56E-04  \\
\multicolumn{17}c{Effective collision strengths calculated with experimental cross sections from \citet{2002DokPh..47...34S}} \\
 a$^3$D -- y$^3$P$^{\circ}$  & 1 -- 1 &  1.11E-19 & 3.35E-14   & 2.03E-11 & 1.00E-09 & 1.40E-08 & 9.43E-08 & 4.02E-07 & 3.18E-06 & 1.31E-05 & 3.72E-05 & 8.29E-05 & 1.57E-04 & 2.66E-04 & 6.05E-04 & 1.12E-03  \\
 a$^3$D -- y$^3$P$^{\circ}$  & 2 -- 2 &  2.90E-20 & 1.59E-14   & 1.31E-11 & 7.74E-10 & 1.22E-08 & 8.98E-08 & 4.09E-07 & 3.56E-06 & 1.56E-05 & 4.63E-05 & 1.07E-04 & 2.09E-04 & 3.63E-04 & 8.55E-04 & 1.63E-03  \\
 a$^3$D -- y$^3$P$^{\circ}$  & 1 -- 0 &  3.09E-20 & 1.78E-14   & 1.50E-11 & 9.04E-10 & 1.44E-08 & 1.06E-07 & 4.87E-07 & 4.26E-06 & 1.88E-05 & 5.59E-05 & 1.29E-04 & 2.53E-04 & 4.38E-04 & 1.03E-03 & 1.95E-03  \\
 a$^3$D -- y$^3$P$^{\circ}$  & 2 -- 1 &  3.96E-19 & 1.19E-13   & 7.24E-11 & 3.58E-09 & 4.99E-08 & 3.36E-07 & 1.43E-06 & 1.13E-05 & 4.68E-05 & 1.32E-04 & 2.95E-04 & 5.61E-04 & 9.49E-04 & 2.16E-03 & 3.99E-03  \\
 a$^3$D -- y$^3$P$^{\circ}$  & 3 -- 2 &  1.24E-19 & 6.80E-14   & 5.59E-11 & 3.32E-09 & 5.23E-08 & 3.84E-07 & 1.75E-06 & 1.52E-05 & 6.70E-05 & 1.99E-04 & 4.59E-04 & 8.96E-04 & 1.55E-03 & 3.67E-03 & 6.99E-03  \\
 a$^1$S -- z$^3$D$^{\circ}$  & 0 -- 1 &  9.15E-21 & 3.71E-15   & 2.61E-12 & 1.41E-10 & 2.07E-09 & 1.46E-08 & 6.39E-08 & 5.28E-07 & 2.24E-06 & 6.47E-06 & 1.46E-05 & 2.81E-05 & 4.79E-05 & 1.10E-04 & 2.04E-04  \\
 a$^3$D -- z$^3$D$^{\circ}$  & 2 -- 3 &  2.27E-20 & 1.19E-14   & 9.57E-12 & 5.60E-10 & 8.74E-09 & 6.38E-08 & 2.89E-07 & 2.49E-06 & 1.09E-05 & 3.19E-05 & 7.33E-05 & 1.42E-04 & 2.45E-04 & 5.72E-04 & 1.08E-03  \\
 a$^3$D -- z$^3$D$^{\circ}$  & 1 -- 2 &  1.73E-20 & 8.49E-15   & 6.61E-12 & 3.79E-10 & 5.84E-09 & 4.23E-08 & 1.90E-07 & 1.62E-06 & 7.05E-06 & 2.07E-05 & 4.74E-05 & 9.20E-05 & 1.59E-04 & 3.70E-04 & 6.99E-04  \\
 a$^3$D -- z$^3$D$^{\circ}$  & 3 -- 3 &  9.01E-20 & 4.73E-14   & 3.80E-11 & 2.22E-09 & 3.47E-08 & 2.54E-07 & 1.15E-06 & 9.89E-06 & 4.31E-05 & 1.27E-04 & 2.92E-04 & 5.66E-04 & 9.75E-04 & 2.28E-03 & 4.30E-03  \\
 a$^3$D -- z$^3$D$^{\circ}$  & 1 -- 1 &  9.53E-20 & 3.87E-14   & 2.72E-11 & 1.46E-09 & 2.16E-08 & 1.52E-07 & 6.66E-07 & 5.50E-06 & 2.34E-05 & 6.75E-05 & 1.53E-04 & 2.93E-04 & 5.00E-04 & 1.15E-03 & 2.13E-03  \\
 a$^3$D -- z$^3$D$^{\circ}$  & 2 -- 2 &  7.21E-20 & 3.54E-14   & 2.76E-11 & 1.58E-09 & 2.44E-08 & 1.76E-07 & 7.94E-07 & 6.78E-06 & 2.94E-05 & 8.64E-05 & 1.98E-04 & 3.84E-04 & 6.62E-04 & 1.54E-03 & 2.92E-03  \\
 a$^3$D -- z$^3$D$^{\circ}$  & 2 -- 1 &  5.42E-20 & 2.20E-14   & 1.55E-11 & 8.32E-10 & 1.23E-08 & 8.62E-08 & 3.79E-07 & 3.13E-06 & 1.33E-05 & 3.84E-05 & 8.68E-05 & 1.67E-04 & 2.84E-04 & 6.52E-04 & 1.21E-03  \\
 a$^3$D -- z$^3$D$^{\circ}$  & 3 -- 2 &  4.84E-20 & 2.38E-14   & 1.85E-11 & 1.06E-09 & 1.64E-08 & 1.19E-07 & 5.33E-07 & 4.56E-06 & 1.98E-05 & 5.81E-05 & 1.33E-04 & 2.58E-04 & 4.45E-04 & 1.04E-03 & 1.96E-03  \\
 a$^3$D -- z$^3$F$^{\circ}$  & 3 -- 4 &  6.73E-17 & 5.36E-12   & 1.67E-09 & 5.53E-08 & 5.90E-07 & 3.29E-06 & 1.21E-05 & 7.89E-05 & 2.86E-04 & 7.39E-04 & 1.54E-03 & 2.78E-03 & 4.52E-03 & 9.68E-03 & 1.72E-02  \\
 a$^3$D -- z$^3$F$^{\circ}$  & 2 -- 3 &  3.68E-20 & 2.66E-14   & 2.52E-11 & 1.63E-09 & 2.72E-08 & 2.09E-07 & 9.83E-07 & 8.99E-06 & 4.10E-05 & 1.25E-04 & 2.95E-04 & 5.87E-04 & 1.03E-03 & 2.49E-03 & 4.83E-03  \\
 a$^3$D -- z$^3$F$^{\circ}$  & 1 -- 2 &  9.82E-21 & 8.22E-15   & 8.34E-12 & 5.61E-10 & 9.63E-09 & 7.53E-08 & 3.59E-07 & 3.34E-06 & 1.54E-05 & 4.72E-05 & 1.12E-04 & 2.22E-04 & 3.91E-04 & 9.42E-04 & 1.82E-03  \\
 a$^3$D -- z$^3$F$^{\circ}$  & 2 -- 2 &  4.65E-21 & 3.90E-15   & 3.95E-12 & 2.66E-10 & 4.56E-09 & 3.57E-08 & 1.70E-07 & 1.58E-06 & 7.30E-06 & 2.24E-05 & 5.30E-05 & 1.05E-04 & 1.85E-04 & 4.46E-04 & 8.62E-04  \\
 a$^3$D -- z$^3$F$^{\circ}$  & 3 -- 3 &  1.42E-20 & 1.02E-14   & 9.69E-12 & 6.26E-10 & 1.05E-08 & 8.04E-08 & 3.78E-07 & 3.46E-06 & 1.58E-05 & 4.81E-05 & 1.14E-04 & 2.26E-04 & 3.98E-04 & 9.60E-04 & 1.86E-03  \\
 a$^3$D -- z$^3$F$^{\circ}$  & 3 -- 2 &  2.05E-21 & 1.71E-15   & 1.74E-12 & 1.17E-10 & 2.01E-09 & 1.57E-08 & 7.50E-08 & 6.98E-07 & 3.21E-06 & 9.85E-06 & 2.33E-05 & 4.64E-05 & 8.17E-05 & 1.97E-04 & 3.80E-04  \\
 a$^1$D -- z$^3$D$^{\circ}$  & 2 -- 2 &  1.47E-20 & 7.23E-15   & 5.63E-12 & 3.23E-10 & 4.98E-09 & 3.61E-08 & 1.62E-07 & 1.39E-06 & 6.03E-06 & 1.77E-05 & 4.06E-05 & 7.88E-05 & 1.36E-04 & 3.17E-04 & 6.00E-04  \\
 a$^1$D -- z$^3$D$^{\circ}$  & 2 -- 1 &  1.38E-20 & 5.61E-15   & 3.94E-12 & 2.12E-10 & 3.14E-09 & 2.20E-08 & 9.67E-08 & 8.00E-07 & 3.40E-06 & 9.82E-06 & 2.22E-05 & 4.27E-05 & 7.30E-05 & 1.68E-04 & 3.12E-04  \\
 a$^1$D -- z$^3$F$^{\circ}$  & 2 -- 3 &  1.78E-21 & 1.29E-15   & 1.22E-12 & 7.87E-11 & 1.32E-09 & 1.01E-08 & 4.76E-08 & 4.35E-07 & 1.99E-06 & 6.06E-06 & 1.43E-05 & 2.85E-05 & 5.02E-05 & 1.21E-04 & 2.35E-04  \\
 a$^1$D -- z$^3$F$^{\circ}$  & 2 -- 2 &  5.93E-21 & 4.97E-15   & 5.04E-12 & 3.40E-10 & 5.83E-09 & 4.56E-08 & 2.18E-07 & 2.03E-06 & 9.34E-06 & 2.86E-05 & 6.79E-05 & 1.35E-04 & 2.38E-04 & 5.73E-04 & 1.11E-03  \\
 a$^3$D -- z$^3$P$^{\circ}$  & 1 -- 2 &  9.33E-21 & 3.80E-15   & 2.67E-12 & 1.44E-10 & 2.13E-09 & 1.49E-08 & 6.57E-08 & 5.43E-07 & 2.31E-06 & 6.66E-06 & 1.51E-05 & 2.89E-05 & 4.93E-05 & 1.13E-04 & 2.12E-04  \\
 a$^1$S -- z$^3$P$^{\circ}$  & 0 -- 1 &  3.21E-20 & 7.74E-15   & 4.19E-12 & 1.93E-10 & 2.58E-09 & 1.68E-08 & 6.99E-08 & 5.36E-07 & 2.17E-06 & 6.06E-06 & 1.34E-05 & 2.53E-05 & 4.27E-05 & 9.66E-05 & 1.79E-04  \\
 a$^3$D -- z$^3$P$^{\circ}$  & 2 -- 2 &  1.00E-19 & 4.08E-14   & 2.87E-11 & 1.54E-09 & 2.28E-08 & 1.60E-07 & 7.04E-07 & 5.82E-06 & 2.47E-05 & 7.14E-05 & 1.61E-04 & 3.10E-04 & 5.29E-04 & 1.22E-03 & 2.28E-03  \\
 a$^3$D -- z$^3$P$^{\circ}$  & 3 -- 2 &  2.56E-19 & 1.04E-13   & 7.34E-11 & 3.95E-09 & 5.84E-08 & 4.10E-07 & 1.80E-06 & 1.49E-05 & 6.33E-05 & 1.83E-04 & 4.14E-04 & 7.94E-04 & 1.35E-03 & 3.12E-03 & 5.83E-03  \\
 a$^3$D -- z$^3$P$^{\circ}$  & 1 -- 1 &  2.05E-19 & 4.93E-14   & 2.67E-11 & 1.23E-09 & 1.64E-08 & 1.07E-07 & 4.46E-07 & 3.42E-06 & 1.38E-05 & 3.87E-05 & 8.55E-05 & 1.61E-04 & 2.72E-04 & 6.17E-04 & 1.14E-03  \\
 a$^3$D -- z$^3$P$^{\circ}$  & 2 -- 1 &  5.53E-19 & 1.33E-13   & 7.21E-11 & 3.33E-09 & 4.43E-08 & 2.89E-07 & 1.20E-06 & 9.23E-06 & 3.73E-05 & 1.04E-04 & 2.31E-04 & 4.36E-04 & 7.36E-04 & 1.67E-03 & 3.09E-03  \\
 a$^3$D -- z$^3$P$^{\circ}$  & 1 -- 0 &  2.01E-20 & 8.07E-15   & 5.73E-12 & 3.14E-10 & 4.72E-09 & 3.38E-08 & 1.51E-07 & 1.29E-06 & 5.66E-06 & 1.68E-05 & 3.91E-05 & 7.70E-05 & 1.35E-04 & 3.23E-04 & 6.28E-04  \\
 a$^1$D -- z$^3$P$^{\circ}$  & 2 -- 2 &  2.13E-20 & 8.70E-15   & 6.12E-12 & 3.30E-10 & 4.88E-09 & 3.42E-08 & 1.51E-07 & 1.25E-06 & 5.30E-06 & 1.53E-05 & 3.46E-05 & 6.65E-05 & 1.14E-04 & 2.62E-04 & 4.90E-04  \\
 a$^3$F -- z$^3$D$^{\circ}$  & 3 -- 3 &  8.76E-21 & 4.61E-15   & 3.71E-12 & 2.17E-10 & 3.40E-09 & 2.48E-08 & 1.13E-07 & 9.72E-07 & 4.25E-06 & 1.25E-05 & 2.88E-05 & 5.60E-05 & 9.67E-05 & 2.26E-04 & 4.29E-04  \\
 a$^3$F -- z$^3$D$^{\circ}$  & 2 -- 1 &  3.90E-20 & 1.58E-14   & 1.12E-11 & 6.02E-10 & 8.89E-09 & 6.25E-08 & 2.75E-07 & 2.28E-06 & 9.68E-06 & 2.80E-05 & 6.36E-05 & 1.22E-04 & 2.09E-04 & 4.81E-04 & 8.96E-04  \\
 a$^3$F -- z$^3$D$^{\circ}$  & 4 -- 3 &  3.99E-20 & 2.10E-14   & 1.69E-11 & 9.90E-10 & 1.55E-08 & 1.13E-07 & 5.13E-07 & 4.43E-06 & 1.94E-05 & 5.72E-05 & 1.31E-04 & 2.56E-04 & 4.41E-04 & 1.03E-03 & 1.96E-03  \\
 a$^3$F -- z$^3$D$^{\circ}$  & 3 -- 2 &  3.46E-20 & 1.70E-14   & 1.33E-11 & 7.62E-10 & 1.18E-08 & 8.52E-08 & 3.84E-07 & 3.29E-06 & 1.43E-05 & 4.21E-05 & 9.66E-05 & 1.88E-04 & 3.24E-04 & 7.59E-04 & 1.44E-03  \\
 a$^3$F -- z$^3$F$^{\circ}$  & 4 -- 4 &  8.99E-21 & 6.65E-15   & 6.35E-12 & 4.12E-10 & 6.88E-09 & 5.28E-08 & 2.48E-07 & 2.26E-06 & 1.02E-05 & 3.09E-05 & 7.25E-05 & 1.43E-04 & 2.49E-04 & 5.93E-04 & 1.13E-03  \\
 a$^3$F -- z$^3$F$^{\circ}$  & 2 -- 2 &  8.24E-22 & 6.91E-16   & 7.02E-13 & 4.73E-11 & 8.12E-10 & 6.36E-09 & 3.04E-08 & 2.83E-07 & 1.31E-06 & 4.02E-06 & 9.53E-06 & 1.90E-05 & 3.35E-05 & 8.09E-05 & 1.57E-04  \\
 a$^3$F -- z$^3$F$^{\circ}$  & 3 -- 3 &  4.01E-21 & 2.90E-15   & 2.75E-12 & 1.78E-10 & 2.98E-09 & 2.29E-08 & 1.08E-07 & 9.90E-07 & 4.52E-06 & 1.38E-05 & 3.27E-05 & 6.52E-05 & 1.15E-04 & 2.78E-04 & 5.41E-04  \\
 a$^3$F -- z$^3$F$^{\circ}$  & 4 -- 3 &  1.88E-21 & 1.36E-15   & 1.29E-12 & 8.36E-11 & 1.40E-09 & 1.08E-08 & 5.07E-08 & 4.65E-07 & 2.12E-06 & 6.49E-06 & 1.54E-05 & 3.06E-05 & 5.40E-05 & 1.31E-04 & 2.54E-04  \\
 a$^3$P -- y$^3$P$^{\circ}$  & 2 -- 2 &  6.96E-21 & 3.83E-15   & 3.16E-12 & 1.88E-10 & 2.96E-09 & 2.19E-08 & 9.98E-08 & 8.71E-07 & 3.85E-06 & 1.14E-05 & 2.65E-05 & 5.20E-05 & 9.05E-05 & 2.15E-04 & 4.12E-04  \\
 a$^3$P -- y$^3$P$^{\circ}$  & 1 -- 0 &  2.14E-21 & 1.23E-15   & 1.04E-12 & 6.30E-11 & 1.00E-09 & 7.45E-09 & 3.42E-08 & 3.00E-07 & 1.33E-06 & 3.97E-06 & 9.22E-06 & 1.81E-05 & 3.14E-05 & 7.42E-05 & 1.41E-04  \\
 a$^3$P -- z$^3$D$^{\circ}$  & 2 -- 3 &  9.36E-21 & 4.93E-15   & 3.98E-12 & 2.33E-10 & 3.65E-09 & 2.67E-08 & 1.21E-07 & 1.05E-06 & 4.60E-06 & 1.36E-05 & 3.13E-05 & 6.09E-05 & 1.05E-04 & 2.47E-04 & 4.70E-04  \\
 a$^3$P -- z$^3$D$^{\circ}$  & 1 -- 2 &  7.12E-21 & 3.51E-15   & 2.74E-12 & 1.58E-10 & 2.44E-09 & 1.77E-08 & 7.97E-08 & 6.84E-07 & 2.98E-06 & 8.79E-06 & 2.02E-05 & 3.94E-05 & 6.81E-05 & 1.60E-04 & 3.04E-04  \\
 a$^3$P -- z$^3$D$^{\circ}$  & 1 -- 1 &  4.15E-21 & 1.69E-15   & 1.19E-12 & 6.44E-11 & 9.53E-10 & 6.70E-09 & 2.95E-08 & 2.45E-07 & 1.05E-06 & 3.03E-06 & 6.89E-06 & 1.33E-05 & 2.27E-05 & 5.24E-05 & 9.80E-05  \\
 a$^3$P -- z$^3$D$^{\circ}$  & 2 -- 1 &  1.10E-20 & 4.48E-15   & 3.16E-12 & 1.71E-10 & 2.52E-09 & 1.78E-08 & 7.82E-08 & 6.49E-07 & 2.77E-06 & 8.03E-06 & 1.82E-05 & 3.51E-05 & 6.02E-05 & 1.39E-04 & 2.60E-04  \\\hline
\enddata
\end{deluxetable*}
\end{longrotatetable}

\startlongtable
 \begin{deluxetable*}{cccccccccccc}
   \tablenum{A.3}
 \tablecaption{Atmospheric parameters of the sample stars and yttrium LTE and NLTE abundances. \label{Sample}}
\setlength{\tabcolsep}{2pt}
\tabletypesize{\scriptsize }
\tablewidth{20pt}
\tablehead{
\colhead{$\#$} & \colhead{Star} & \colhead{ \Teff } & \colhead{log$g$} & \colhead{[Fe/H]}  & \colhead{ $\xi_t$ } & \colhead{Pop. } & \colhead{[Y/Fe]$_{\rm LTE}$} & \colhead{ $\sigma_{\rm LTE}$} & \colhead{[Y/Fe]$_{\rm NLTE}$} & \colhead{ $\sigma_{\rm NLTE}$} & \colhead{$N_{\rm lines}$}  \\
\colhead{}   & \colhead{}     & \colhead{K}       & \colhead{CGS}    & \colhead{dex}     & \colhead{ \kms }    & \colhead{}   & \colhead{}                & \colhead{}                   & \colhead{}                    & \colhead{}  & \colhead{}
}
\decimalcolnumbers
\startdata
1   & HD~19373    &  6045  &  4.24   &    0.10       &  1.20   & Thin disk     &  -0.11  & 0.05 &  -0.11   &  0.05  & 4 \\
2   & HD~22484    &  6000  &  4.07   &    0.01       &  1.10   & Thin disk     &  -0.06  & 0.06 &  -0.08   &  0.05  & 3 \\
3   & HD~24289    &  5980  &  3.71   &   -1.94       &  1.10   & Halo          &  -0.03  & 0.08 &  0.07    &  0.07  & 4 \\
4   & HD~30562    &  5900  &  4.08   &    0.17       &  1.30   & Thin disk     &  -0.05  & 0.05 &  -0.07   &  0.05  & 3 \\
5   & HD~30743    &  6450  &  4.20   &   -0.44       &  1.80   & Thin disk     &  0.04   & 0.05 &  0.08    &  0.05  & 3 \\
6   & HD~34411    &  5850  &  4.23   &    0.01       &  1.20   & Thin disk     &  -0.08  & 0.05 &  -0.08   &  0.05  & 3 \\
7   & HD~43318    &  6250  &  3.92   &   -0.19       &  1.70   & Thin disk     &  -0.02  & 0.02 &  -0.02   &  0.02  & 2 \\
8   & HD~45067    &  5960  &  3.94   &   -0.16       &  1.50   & Thin disk     &  -0.04  & 0.04 &  -0.04   &  0.04  & 5 \\
9   & HD~45205    &  5790  &  4.08   &   -0.87       &  1.10   & Thick disk    &  -0.04  & 0.02 &  0.03    &  0.04  & 3 \\
10  & HD~52711    &  5900  &  4.33   &   -0.21       &  1.20   & Thin disk     &  -0.04  & 0.02 &  -0.04   &  0.02  & 4 \\
11  & HD~58855    &  6410  &  4.32   &   -0.29       &  1.60   & Thin disk     &  0.06   & 0.02 &  0.08    &  0.00  & 3 \\
12  & HD~59374    &  5850  &  4.38   &   -0.88       &  1.20   & Thick disk    &  -0.02  & 0.06 &  0.04    &  0.06  & 5 \\
13  & HD~59984    &  5930  &  4.02   &   -0.69       &  1.40   & Thin disk     &  -0.17  & 0.05 &  -0.12   &  0.04  & 4 \\
14  & HD~62301    &  5840  &  4.09   &   -0.70       &  1.30   & Thick disk    &  -0.12  & 0.06 &  -0.08   &  0.05  & 5 \\
15  & HD~64090    &  5400  &  4.70   &   -1.73       &  0.70   & Halo          &  -0.07  & 0.06 &  -0.01   &  0.03  & 3 \\
16  & HD~69897    &  6240  &  4.24   &   -0.25       &  1.40   & Thin disk     &  -0.05  & 0.06 &  -0.03   &  0.05  & 4 \\
17  & HD~74000    &  6225  &  4.13   &   -1.97       &  1.30   & Halo          &  0.07   & 0.05 &  0.16    &  0.02  & 4 \\
18  & HD~76932    &  5870  &  4.10   &   -0.98       &  1.30   & Thick disk    &  0.11   & 0.05 &  0.20    &  0.03  & 4 \\
19  & HD~82943    &  5970  &  4.37   &    0.19       &  1.20   & Thin disk     &  -0.20  & 0.01 &  -0.21   &  0.02  & 4 \\
20  & HD~89744    &  6280  &  3.97   &    0.13       &  1.70   & Thin disk     &  0.05   & 0.07 &  0.03    &  0.04  & 3 \\
21  & HD~90839    &  6195  &  4.38   &   -0.18       &  1.40   & Thin disk     &  0.03   & 0.04 &  0.05    &  0.03  & 4 \\
22  & HD~92855    &  6020  &  4.36   &   -0.12       &  1.30   & Thin disk     &  -0.05  & 0.02 &  -0.05   &  0.02  & 2 \\
23  & HD~94028    &  5970  &  4.33   &   -1.47       &  1.30   & Thick disk    &  0.16   & 0.09 &  0.26    &  0.06  & 6 \\
24  & HD~99984    &  6190  &  3.72   &   -0.38       &  1.80   & Thin disk     &  0.07   & 0.04 &  0.09    &  0.04  & 4 \\
25  & HD~100563   &  6460  &  4.32   &    0.06       &  1.60   & Thin disk     &  0.04   & 0.02 &  0.03    &  0.04  & 3 \\
26  & HD~102870   &  6170  &  4.14   &    0.11       &  1.50   & Thin disk     &  -0.06  & 0.06 &  -0.07   &  0.06  & 4 \\
27  & HD~103095   &  5130  &  4.66   &   -1.26       &  0.90   & Halo          &  -0.02  & 0.03 &  0.04    &  0.01  & 3 \\
28  & HD~105755   &  5800  &  4.05   &   -0.73       &  1.20   & Thin disk     &  -0.07  & 0.03 &  -0.02   &  0.02  & 4 \\
29  & HD~106516   &  6300  &  4.44   &   -0.73       &  1.50   & Thick disk    &  -0.03  & 0.05 &  0.02    &  0.04  & 3 \\
30  & HD~108177   &  6100  &  4.22   &   -1.67       &  1.10   & Halo          &  0.00   & 0.03 &  0.09    &  0.04  & 6 \\
31  & HD~110897   &  5920  &  4.41   &   -0.57       &  1.20   & Thin disk     &  -0.11  & 0.05 &  -0.07   &  0.04  & 4 \\
32  & HD~114710   &  6090  &  4.47   &    0.06       &  1.10   & Thin disk     &  0.10   & 0.04 &  0.09    &  0.04  & 4 \\
33  & HD~115617   &  5490  &  4.40   &   -0.10       &  1.10   & Thin disk     &  -0.18  & 0.05 &  -0.17   &  0.06  & 3 \\
34  & HD~134088   &  5730  &  4.46   &   -0.80       &  1.10   & Thick disk    &  -0.04  & 0.03 &  0.01    &  0.05  & 3 \\
35  & HD~134169   &  5890  &  4.02   &   -0.78       &  1.20   & Thin disk     &  -0.12  & 0.04 &  -0.06   &  0.03  & 4 \\
36  & HD~138776   &  5650  &  4.30   &    0.24       &  1.30   & Thin disk     &  -0.24  & 0.05 &  -0.24   &  0.05  & 3 \\
37  & HD~142091   &  4810  &  3.12   &   -0.07       &  1.20   & Thin disk     &  -0.26  & 0.02 &  -0.26   &  0.02  & 2 \\
38  & HD~142373   &  5830  &  3.96   &   -0.54       &  1.40   & Thin disk     &  -0.17  & 0.07 &  -0.13   &  0.07  & 4 \\
39  & BD+07 4841  &  6130  &  4.15   &   -1.46       &  1.30   & Halo          &  0.19   & 0.05 &  0.29    &  0.05  & 6 \\
40  & BD+09 0352  &  6150  &  4.25   &   -2.09       &  1.30   & Halo          &  -0.03  & 0.14 &  0.05    &  0.13  & 2 \\
41  & BD+24 1676  &  6210  &  3.90   &   -2.44       &  1.50   & Halo          &  -0.12  & 0.14 &  -0.04   &  0.11  & 2 \\
42  & BD+29 2091  &  5860  &  4.67   &   -1.91       &  0.80   & Halo          &  -0.19  & 0.02 &  -0.10   &  0.03  & 4 \\
43  & BD+37 1458  &  5500  &  3.70   &   -1.95       &  1.00   & Halo          &  0.14   & 0.06 &  0.24    &  0.08  & 4 \\
44  & BD+66 0268  &  5300  &  4.72   &   -2.06       &  0.60   & Halo          &  -0.13  & 0.04 &  -0.06   &  0.04  & 3 \\
45  & BD-04 3208  &  6390  &  4.08   &   -2.20       &  1.40   & Halo          &  -0.05  & 0.03 &  0.03    &  0.03  & 1 \\
46  & BD-13 3442  &  6400  &  3.95   &   -2.62       &  1.40   & Halo          &  0.05   & 0.03 &  0.16    &  0.03  & 1 \\
47  & G090-003    &  6007  &  3.90   &   -2.04       &  1.30   & Halo          &  -0.13  & 0.04 &  -0.04   &  0.06  & 4 \\
48  & HD~29907    &  5500  &  4.64   &   -1.55       &  0.60   & Halo          &  -0.09  & 0.07 &  -0.03   &  0.07  & 3 \\
49  & HD~31128    &  5980  &  4.49   &   -1.49       &  1.20   & Thick disc    &  0.03   & 0.05 &  0.14    &  0.04  & 5 \\
50  & HD~59392    &  6010  &  4.02   &   -1.59       &  1.40   & Halo          &  0.14   & 0.05 &  0.22    &  0.04  & 4 \\
51  & HD~97320    &  6110  &  4.27   &   -1.18       &  1.40   & Thick disc    &  0.05   & 0.07 &  0.14    &  0.06  & 5 \\
52  & HD~193901   &  5780  &  4.46   &   -1.08       &  0.90   & Halo          &  -0.12  & 0.04 &  -0.05   &  0.03  & 5 \\
53  & HD~298986   &  6130  &  4.30   &   -1.34       &  1.40   & Halo          &  -0.12  & 0.05 &  -0.03   &  0.05  & 4 \\
54  & HD~102200   &  6115  &  4.20   &   -1.24       &  1.40   & Thick disc    &  -0.04  & 0.06 &  0.04    &  0.05  & 6 \\
55  & HD~3795     &  5475  &  3.85   &   -0.61       &  1.00   & Thick disc    &  0.06   & 0.08 &  0.06    &  0.07  & 4 \\
56  & HD~32923    &  5710  &  4.03   &   -0.26       &  1.20   & Thin disc     &  0.03   & 0.05 &  0.03    &  0.04  & 4 \\
57  & HD~40397    &  5550  &  4.39   &   -0.17       &  1.00   & Thick disc    &  -0.02  & 0.06 &  -0.01   &  0.07  & 3 \\
58  & HD~64606    &  5280  &  4.63   &   -0.68       &  1.00   & Thick disc    &  -0.16  & 0.09 &  -0.14   &  0.08  & 4 \\
59  & HD~69611    &  5940  &  4.17   &   -0.60       &  1.20   & Thick disc    &  -0.04  & 0.04 &  0.01    &  0.03  & 4 \\
60  & HD~114762   &  5930  &  4.18   &   -0.71       &  1.20   & Thick disc    &  -0.17  & 0.04 &  -0.12   &  0.04  & 4 \\
61  & HD~201891   &  5900  &  4.29   &   -0.97       &  1.20   & Thick disc    &  -0.19  & 0.06 &  -0.13   &  0.05  & 4 \\\hline
\enddata
\end{deluxetable*}

\bibliography{yttrium_mod}
\bibliographystyle{aasjournal}

\end{document}